%% file: main.tex
\documentclass[12pt]{article}
\pdfoutput=1
\usepackage{ifpdf}
\usepackage[a4paper,top=2.7cm,width=17cm, bottom=2.7cm, left=2cm]{geometry}
\usepackage{graphicx}
\usepackage{microtype}
\usepackage[utf8x]{inputenc}
\usepackage[T1]{fontenc}
\usepackage{ae}
\usepackage{aecompl}
\usepackage{amsmath}
\usepackage{amssymb}
\usepackage{mathdots}
\usepackage{setspace}
\usepackage[nosort]{cite}
\def\bfseries{\fontseries \bfdefault \selectfont \boldmath}

\usepackage[english]{babel}
\usepackage{datetime}
\shortdate 
\usepackage{bm}
\usepackage{subfigure}
\usepackage[normalem]{ulem} 
\usepackage{soul} 
\usepackage[titletoc,title]{appendix}
\usepackage{authblk}
\usepackage{xcolor,soul}
\usepackage[colorlinks=true,linktocpage=true,linkcolor=blue,citecolor=blue]{hyperref}
\usepackage{titlesec}
\titleformat*{\section}{\large\bfseries}
\titleformat*{\subsection}{\bfseries}
\titleformat*{\subsubsection}{\bfseries}

\usepackage{hyperref}
\hypersetup{
    colorlinks,
    citecolor=black,
    filecolor=black,
    linkcolor=black,
    urlcolor=black
}

\newcommand{\rf}[1]{Eq.~(\ref{#1})}
\newcommand{\rfn}[1]{~(\ref{#1})}
\newcommand{\rfs}[1]{Section~\ref{#1}}

\newcommand{\rff}[1]{Fig.~\ref{#1}}
\newcommand{\rfc}[1]{Ref.~\cite{#1}}

\newcommand{\p}{\partial}

\newcommand{\D}{{\mathcal D}}
\newcommand{\f}[2]{\frac{#1}{#2}}
\newcommand{\pd}[2]{\frac{\partial{#1}}{\partial{#2}}}

\newcommand{\sym}{${\mathcal N}=4$}
\newcommand{\symm}{${\mathcal N}=4$ SYM}

\newcommand{\bea}{\begin{eqnarray}}
\newcommand{\eea}{\end{eqnarray}}
\newcommand{\be}{\begin{eqnarray}}
\newcommand{\ee}{\end{eqnarray}}
\newcommand{\bel}[1]{\begin{eqnarray}\label{#1}}

\newcommand{\nn}{\nonumber}
\newcommand{\sr}{\eta_s}
\newcommand{\edens}{\mathcal{E}}
\newcommand{\PP}{\mathcal{P}}
\newcommand{\pa}{\mathcal{A}}
\newcommand{\pL}{\mathcal{P}_L}
\newcommand{\pT}{\mathcal{P}_T}
\newcommand{\paz}{{\pa_\star}}

\newcommand{\dndy}{\f{dN}{dy}}

\newcommand{\taur}{\tau_{R}}
\newcommand{\LL}{\mathcal{L}}
\newcommand{\MM}{\mathcal{M}}
\newcommand{\xa}{\Xi_1}
\newcommand{\xb}{\Xi_2}
\newcommand{\tpi}{\tau_{\Pi}}

\def\pimunu{{\pi^{\mu \nu}}}
\def\sigmamunu{{\sigma^{\mu \nu}}}

\begin{document}

\title{{\bf 
Hydrodynamic Attractors\\ 
in Ultrarelativistic Nuclear Collisions
}}

\renewcommand\Authfont{\scshape\small}

\author[1]{Jakub Jankowski\thanks{Jakub.Jankowski@uwr.edu.pl}}
\author[2,3]{Micha\l{} Spali\'nski\thanks{Michal.Spalinski@ncbj.gov.pl}}

\affil[1]{Institute of Theoretical Physics, University of Wroc{\l}aw, Pl. Maxa Borna 9, 50-204 Wroc{\l}aw, Poland}
\affil[2]{Physics Department, University of Bia\l{}ystok, 15-245 Bia\l{}ystok, Poland}
\affil[3]{National Center for Nuclear Research, 00-681 Warszawa, Poland}

\vspace{15pt}

\date{}

\maketitle

\thispagestyle{empty}

\begin{abstract} 

    One of the many physical questions that have emerged from studies of
    heavy-ion collisions at RHIC and the LHC concerns the validity of
    hydrodynamic modelling at the very early stages, when the Quark-Gluon Plasma
    system produced is still far from isotropy. In this article we review the
    idea of far-from-equilibrium hydrodynamic attractors as a way to understand
    how the complexity of initial states of nuclear matter is reduced so that a
    hydrodynamic description can be effective.

\end{abstract}

\newpage
\tableofcontents
\newpage

\input{intro}

%

\input{hic}

%
\input{mis}

%
\input{attractors}

%
\input{kinetic}

%
\input{sym}

%
\input{phase}

%
\input{prehydro}

%
\input{beyond}

%
\input{outlook}

%
  

{\bf Acknowledgements }
We would like to thank many of our colleagues with whom we have had countless 
  discussions about attractors over the years. This includes especially  Wojciech Florkowski, Michał Heller and Viktor Svensson.
  JJ is supported by the National Science Centre, Poland, under grant 2018/29/B/ST2/02457.
  MS is supported by the National Science Centre, Poland, under grants 2018/29/B/ST2/02457 and 2021/41/B/ST2/02909.
  For the purpose of Open Access, the author has applied a CC-BY public copyright licence to any Author Accepted Manuscript (AAM) version arising from this submission.
%

\bibliography{bibdynam}
\bibliographystyle{elsarticle-num}

\end{document}

%% file: intro.tex
\section{Introduction}
\label{sec:intro}

The theory of the strong nuclear interactions, Quantum Chromodynamics, is
beautiful on many levels, one being the simplicity of its formulation. This
simplicity hides a richness of phenomena which remains beyond reach even now,
after decades of research. While the spectrum of hadronic states is often viewed
as a problem solved at least at some level by lattice calculations, the
collective properties which are relevant for the physics of hadronic matter at
finite temperature and density are far less well understood. The main motivation
for this review article comes from studies of Quark-Gluon Plasma (QGP) created
in ultrarelativistic nuclear
collisions~\cite{Kolb:2003dz,Heinz:2013th,Busza:2018rrf,Schenke:2021mxx}. Such
enquiries are of great intrinsic interest, as they address the nature of
Yang-Mills theory itself.  They also have wide ranging implications in diverse
areas of physics, such as nonequilibrium statistical physics, nuclear physics
and astrophysics~\cite{Lovato:2022vgq,Sorensen:2023zkk,Achenbach:2023pba}. 

The heavy-ion collision (HIC) programme is an experimental study of the
properties of strongly-interacting matter, currently pursued at RHIC and the
LHC. Extracting physical properties of QCD matter from collider data is a
formidable challenge.  A crucial element of the analysis is the application of
hydrodynamic models; usually these are variants of the Mueller-Israel-Stewart
theory (MIS)~\cite{Muller:1967zza,Israel:1976tn,Israel:1979wp}. The traditional
formulation of relativistic hydrodynamics assumes that the system under
consideration is approximately in a state of local thermodynamic equilibrium.
The leading order description is then the theory of perfect fluids, and
dissipative effects are accounted for by augmenting the perfect fluid evolution
by adding terms involving gradients of the hydrodynamic variables.  In order to
explain the observed signs of fluidity ({\it e.g.} elliptic flow), the
hydrodynamic stage of simulations has to begin at rather early times, after an
interval of less than 1~fm/c, when the system is still very anisotropic.  Even
though the drop of QGP is not at all close to local equilibrium, hydrodynamic
modelling is very
successful~\cite{Heinz:2002un,Romatschke:2007mq,Teaney:2003kp,Schenke:2021mxx}.
This is a puzzle which touches on foundational questions in
relativistic fluid dynamics.  The discovery of far-from-equilibrium attractors
is a possible resolution of this puzzle~\cite{Heller:2015dha}. 

Theoretical analysis of HIC began already in the 1970s (see e.g.
Ref.~\cite{Yagi:2005yb}). A
crucially important step was made in the seminal work of 
Bjorken~\cite{Bjorken:1982qr}, who pointed out that in a certain kinematic
regime one should expect the initial conditions, as well as subsequent dynamics,
to be approximately invariant under Lorentz boosts along the collision axis.
Supplemented with the assumption of conformal invariance, this has opened the
door to analytic calculations in a situation where one might have thought
numerical computations were the only possible approach. The results of these
calculations have limited applicability due to the strong symmetry assumptions
explained in more detail in the following Section, but they have led to a wealth
of insights. One of them, which has emerged in the past few years, is the notion
of hydrodynamic attractors. 

The term ``attractor'' has a number of meanings. 
In the present context it is
best to think of hydrodynamic attractors as submanifolds of the phase space of
the theory under consideration which are approached asymptotically in the course
of dissipative evolution. The appearance of such attractors at late time is
entirely expected, but it was found that in some cases this attractor extends to
early times, when the system is very far from equilibrium~\cite{Heller:2015dha}.
Such far-from-equilibrium attractors have been identified in many model
systems~\cite{Romatschke:2017vte,Strickland:2018ayk,Almaalol:2020rnu,Noronha:2021syv}
and it is essentially clear that their origin at early times is kinematical:
they arise due to the strong longitudinal expansion~\cite{Blaizot:2017ucy}.
This effect appears in any theory or model of equilibration, be it a
hydrodynamic or kinetic theory model, or presumably QCD itself. Since attractor
behaviour eliminates much of the complexity of initial states as well as of the
dynamics, it may be feasible to match the attractor of QCD to the attractor of a
much simpler phenomenological model, such as the widely used MIS model of
hydrodynamics.
This provides a possible explanation of the success of hydrodynamic simulations
in the description of heavy-ion collisions.  At present, one cannot claim this
with a high degree of certainty, since this explanation relies on studies
involving rather strong symmetry assumptions.  They are valid to some degree at
the early stages of QGP evolution, but it is not yet known to what extent their
violation affects the robustness of hydrodynamic attractors. 
Nevertheless, we regard this possibility with a degree of confidence. 

In this article we review the theoretical underpinnings of hydrodynamic
attractors as well as some applications which are directly relevant to
phenomenological studies.  We hope that our article will be somewhat
complementary to existing reviews, such as
Refs.~\cite{Florkowski:2017olj,Berges:2020fwq,Soloviev:2021lhs}.  We begin, in
\rfs{sec:HIC}, with a brief account of the physical setting of heavy-ion
collisions and the emergence of boost-invariance, a symmetry property which
plays a crucial role in the entire picture. In \rfs{sec:MIS} we emphasise the
conceptual difference between hydrodynamics, understood as an asymptotic
statement about equilibrating systems, and hydrodynamic models which provide a
dynamical description with appropriate asymptotics. Attractors are then
introduced, first in the context of hydrodynamic models in \rfs{sec:attractors},
and then in the framework of kinetic theory in \rfs{sec:attractorKT}. In
\rfs{sec:attractorHolo} we turn to the example of \sym\ supersymmetric
Yang-Mills theory (SYM), which has historically played a crucial role in the
paradigm shift which occurred over the last decade, having provided (thanks to
the AdS/CFT correspondence) a theoretical laboratory based on first principles
where the transition to hydrodynamic behaviour could be investigated. In
\rfs{sec:PhaseSpace} we describe the phase space approach to attractors, this
time aiming for a treatment independent of any special choice of variables. Such
an approach is potentially useful in identifying attractors without relying on
simplifying symmetry assumptions.  \rfs{sec:prehydro} reviews some recent
quantitative applications of attractors to the modelling of heavy ion
collisions.  In \rfs{sec:beyond} we summarise what has been learnt from studies
of conformal Bjorken flow and review some results concerning attractors in
models where some of the symmetry assumptions have been relaxed, specifically by
incorporating the breaking of conformal symmetry or the inclusion of transverse
dynamics.  Finally, \rfs{sec:Out} offers some opinions on research directions
one can envisage following from the developments discussed in this review.

%% file: hic.tex
\section{Heavy Ion Collisions}
\label{sec:HIC}

Although heavy ions are collided at a wide range of collision energies, the
concept of a hydrodynamic attractor has emerged from attempts to understand the
behaviour of hadronic matter at highest available energy densities.
In this section we will review some of the relevant kinematics as 
well as the idea of boost-invariance, which plays a key role at early stages of the
collision.

\subsection{The spacetime picture of the collision}
\label{subsect:general}

\begin{figure}[t]
\begin{center}
\includegraphics[height = .4\textheight]{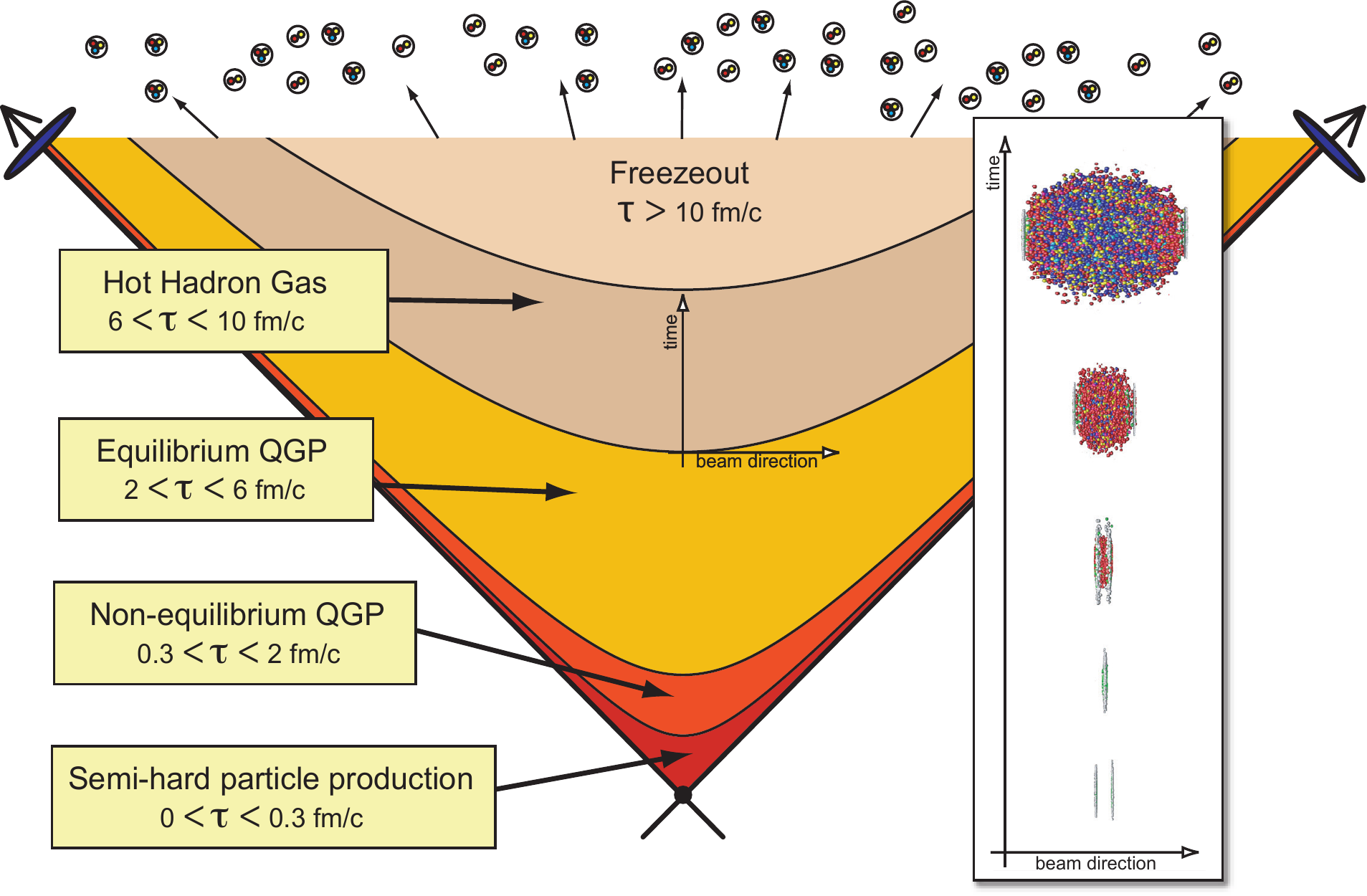} 
\caption{Schematic picture of an ultrarelativistic heavy ion collision with estimated time scales. Figure taken from Ref. \cite{Strickland:2014pga}.}
\label{fig:timescalesHIC}
\end{center}
\end{figure}

In all collision systems and for all collision energies the relevant physics is
a challenge for existing theoretical techniques and eludes a direct treatment
based on QCD. A number of approximations and model approaches have emerged, each
taking advantage of special circumstances arising at various stages of evolution.  Those
different theoretical patches merge together into a coherent picture
consisting of following phases (see e.g.~\cite{Strickland:2014pga}):
\begin{itemize}
  \item {\bf collective state formation} ($0\leq\tau\lesssim 0.3~\rm fm/c$)\\
      gluon dominated, governed by semi-hard particle scattering; 
  \item {\bf pre-hydrodynamic collective flow} ($0.3\lesssim\tau\lesssim  2~\rm fm/c$)\\ 
      highly anisotropic QGP flow with large pressure gradients;
  \item {\bf hydrodynamic evolution} ($2\lesssim\tau\lesssim~6~\rm fm/c$) \\
      leading up to the QCD crossover followed by hadronisation;
  \item {\bf hot hadron gas } ($6\lesssim\tau\lesssim10~\rm fm/c$)\\ 
      expanding gas of hadrons exhibiting re-scattering processes; 
  \item {\bf freeze-out} ($\tau\gtrsim~10~\rm fm/c$)\\ 
      free-streaming gas of non-interacting hadrons. 
\end{itemize}
The sequence of events defined above is schematically pictured in
Fig.~\ref{fig:timescalesHIC}. The hyperbolae represent surfaces of constant
proper time $\tau\equiv\sqrt{t^2-z^2}$, while the nuclei move in the $z$
direction almost along the light cones.  In phenomenological computations,
hydrodynamic models are successfully used already at times around
$\tau\lesssim1$~fm/c, where the system is still highly anisotropic. Therefore,
our focus in this review is on the first two stages, the goal being to
understand how it is that the prehydrodynamic stage of evolution can be
described by fluid-dynamical models.

\subsection{The initial stages}
\label{subsec:initial_state}

The initial state of a heavy ion collision remains the most uncertain element of
the theoretical picture described in Sec.~\ref{subsect:general}, as it is the
domain of non-perturbative quantum field theory. Nevertheless, crucial insights
into the relevant physics were formulated already in the early 1980s.  Two heavy
ions approaching one another at ultrarelativistic velocity are highly Lorentz
contracted along the direction of motion, with the factor
$\gamma=1/\sqrt{1-v^2/c^2}\sim100$ typical for RHIC conditions and more than $1000$ for
the LHC. The ultrarelativistic nature of the collisions has critically important
consequences for the physics of the subsequent evolution. 

The fundamental observations originate in
Refs.~\cite{Gottfried:1974yp,Low:1977gq,Anishetty:1980zp} and rely on the notion
of {\em nuclear transparency}, which states that the highly Lorentz-contracted
nuclei essentially pass through each other, creating a central fragmentation
region of energy density high enough for a deconfined state of QCD matter to
form. The contracted nuclei are treated as if they were of infinite transverse
extent, with no dynamics in the transverse plane. The baryon number of the
colliding nuclei is carried away from this region by the receding projectiles,
leaving behind a drop of approximately baryon-neutral plasma.

The physical picture developed in Ref.~\cite{Bjorken:1982qr} envisages matter
moving essentially along the collision axis, which we take to be the z-axis,
with velocity $v=z/t$ in the centre of mass frame, in a manner reminiscent of
the Hubble expansion of the Universe. This assumption is equivalent to
invariance under boosts along the collision axis and it can be tested
experimentally. It implies that the number of charged particles per unit
rapidity $dN_{\rm ch}/d\eta$ is independent of rapidity in the region
$\eta\approx0$. Experimental data from PHOBOS~\cite{Back:2002wb,PHOBOS:2002whd}
shown in \rff{fig:dNchTutorial} demonstrate the emergence of a central plateau
region with increasing collision energy in the range $\sqrt{s}=19.6-200$~GeV in
the Au-Au system. In consequence, at earliest times, longitudinal expansion
dominates the dynamics and the transverse flow builds up only somewhat later.
This effect is strongest for central collisions.

\begin{figure}
\begin{center}
\includegraphics[width =.9\textwidth]{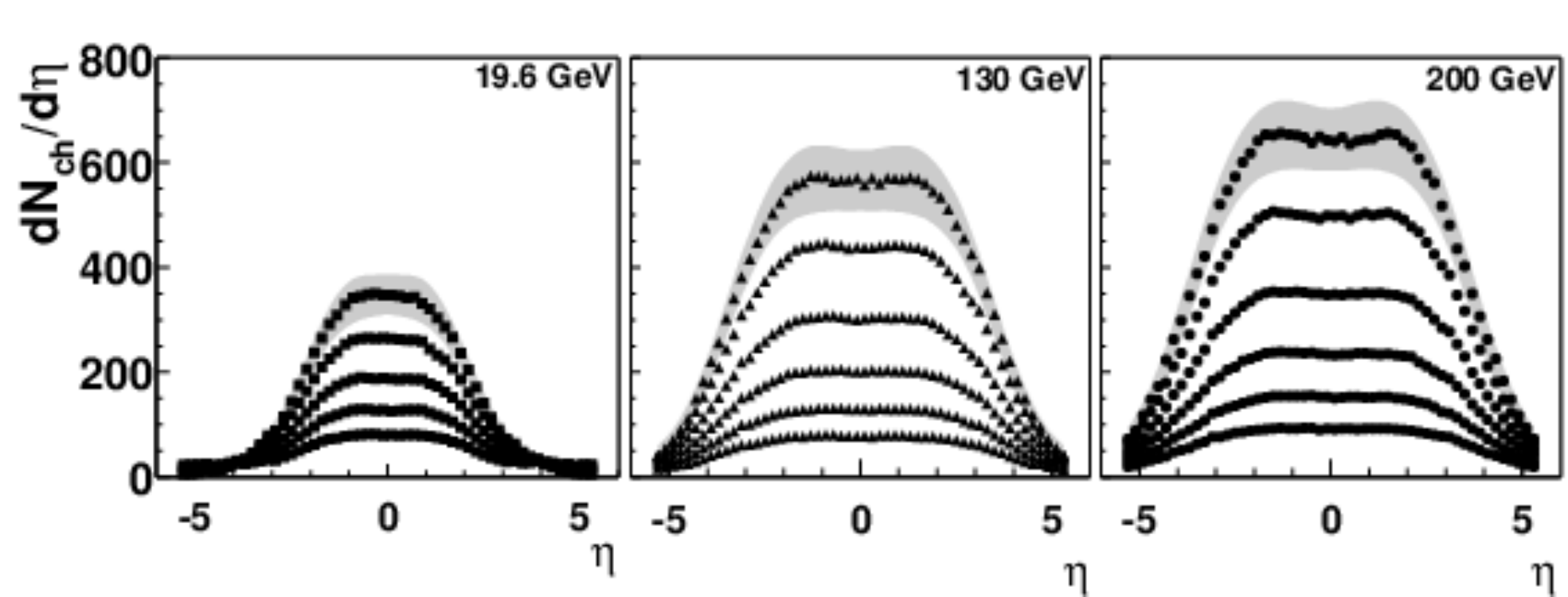}
\caption{Emergence of a central plateau region in the charge particle production
rate in Au-Au collisions for increasing collision energy
$\sqrt{s}=19.6,~130~{\rm and}~200~\rm GeV$. The grey band represents most central
collisions in the $0-6\%$ centrality bin. The plots are taken from
Refs.~\cite{Back:2002wb,PHOBOS:2002whd}.
}
\label{fig:dNchTutorial}
\end{center}
\end{figure}


This idealised picture can be expressed as a set of symmetry assumptions which
define {\em Bjorken flow}. To do this, it is very convenient to use the proper
time $\tau$ and spacetime rapidity $\sr={\rm arctanh}(z/t)$
coordinates\footnote{It is easy to check that for boost-invariant flow 
 $\eta_s = \eta$.}. In terms of these, the Minkowski metric takes the form:
\begin{equation}
    ds^2 =-dt^2+dz^2+dx_\perp^2 =-d\tau^2 +\tau^2 d\sr^2 +dx_\perp^2~,
\end{equation}
where $x_\perp=(x,y)$ are coordinates in the transverse plane. The physical
idealisations sketched in the previous paragraphs translate to the statement that
the physics is independent of spacetime rapidity $\sr$ as well as the
coordinates in the transverse plane. The components of the relativistic flow
velocity assume the form $(u^\mu) = (1,0,0,0)$, with $u_\mu u^\mu=-1$.  

The fundamental local observable which will be the focus of our considerations
is the energy-momentum tensor.  Under the symmetry assumptions stated above it
can be expressed in terms of three functions of the proper time $\tau$: 
\begin{equation}
    T^{\mu}_{\nu}= {\rm diag}\left\{-\edens(\tau),\pL(\tau),\pT(\tau),\pT(\tau)\right\}~,
    \label{eq:Tmn}
\end{equation}
where $\edens$ is the energy density in the local rest-frame, and the
eigenvalues $\pL, \pT$ are referred to as the longitudinal and transverse
pressures. The form of Eq.~(\ref{eq:Tmn}) does not rely on the applicability of
a hydrodynamic description, as it is determined only by the symmetry assumptions
reviewed above. 

One can parametrise the eigenvalues $\pL, \pT$
as 
\begin{equation}
    \pL=\PP\left(1-\frac{2}{3}\mathcal{A}\right), \qquad 
    \pT=\PP\left(1+\frac{1}{3}\mathcal{A}\right)~,
    \label{eq:AdefNC}
\end{equation}
where
\be
\PP \equiv \frac{1}{3}\left(\pL + 2\pT\right)
\ee
is naturally interpreted as the average pressure, while $\pa$ reflects the pressure anisotropy
\be
\pa  \equiv \f{\pL-\pT}{\PP}.
\ee
The pressure anisotropy is a measure of distance from equilibrium, or more
precisely, from spatial isotropy, which is a necessary condition for equilibrium
in the absence of external fields.

\subsection{Conformal symmetry}
\label{sec:confsym}

Since QCD at high energies is approximately scale invariant, it is natural to
impose conformal symmetry to simplify the mathematical description. This is a
very powerful assumption which requires tracelessness of the energy momentum
tensor $T^\mu_\mu=0$, and implies that $\PP=\edens/3$. The energy-momentum tensor
for conformal Bjorken flow can thus be expressed in terms of two functions of
proper time, $\edens$ and $\pa$.
Conservation of the energy momentum tensor 
\begin{equation}
    \label{eq:cons}
    \nabla_\mu T^{\mu\nu} = 0
\end{equation}
relates the pressure anisotropy to the logarithmic derivative of the energy density:
\begin{equation}
    \pa(\tau) = 6\left(1+\frac{3}{4}\tau\partial_\tau\ln\edens\right)~.  
    \label{eq:Adef3}
\end{equation}

For conformal systems it is also very convenient to introduce the concept of
{\it effective temperature} $T(\tau)$, defined by
\begin{equation}
    \edens = C_e T^4~,
    \label{eq:Tdef}
\end{equation}
where $C_e$ is a constant which depends on the number of degrees of
freedom. This equation has the form of a conformal equation of state, so that
in an equilibrium state $T$ is the thermodynamic temperature. Away from
equilibrium \rf{eq:Tdef} defines $T$ as equal to the temperature of an equilibrium state
with the same energy density. 

At asymptotically late times the system approaches local thermodynamic
equilibrium, so the pressure anisotropy tends to zero and the energy-momentum
tensor in~\rf{eq:Tmn} approaches the perfect-fluid form. The way this happens is
determined by the microscopic dynamics which governs the evolution of the
pressure anisotropy.  Once $\pa(\tau)$ is known, the energy density is
determined by \rf{eq:Adef3} up to a single integration constant which sets the
scale. In this sense, for Bjorken flow the dynamics is captured by the
pressure anisotropy.  In the late time limit, if we set $\pa\approx 0$, then
\rf{eq:cons} determines the 
effective temperature
\begin{equation}
    T = \f{\Lambda}{(\Lambda\tau)^{1/3}} 
\label{eq:bjorken}
\end{equation}
where $\Lambda$ is the integration constant containing information about the
initial condition.  This is a consequence of local equilibrium and the
conservation of energy-momentum, so it is valid regardless of any dynamical
details.

\subsection{Prehydrodynamic evolution and the hydrodynamic attractor}
\label{subsec:prehydro}


\begin{figure}[t]
\begin{center}
\includegraphics[height = .3\textheight]{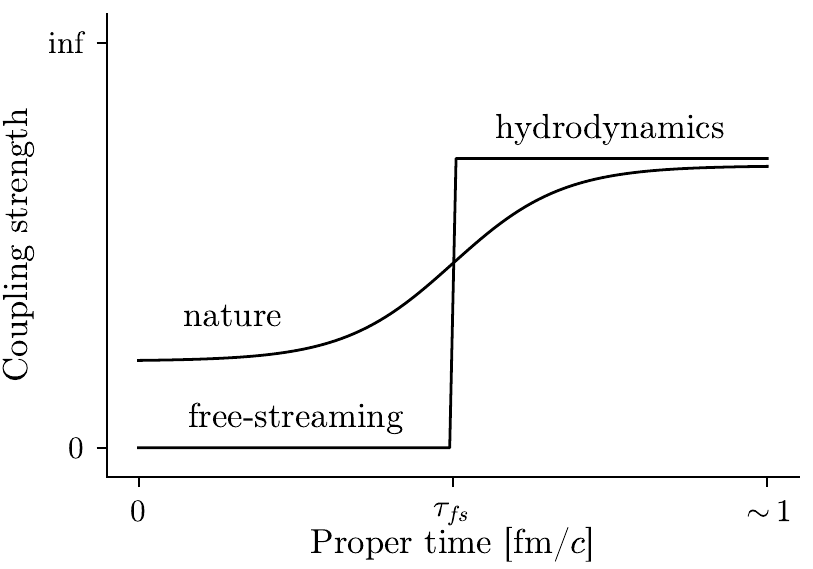} 
\caption{Schematic picture of a coupling evolution and the transition to hydrodynamics. Figure taken from Ref. \cite{Moreland:2018gsh}.}
\label{fig:hydrodynamisation}
\end{center}
\end{figure}


The early stages of QGP dynamics are not well understood at this time. At a
qualitative level one may say that the longitudinally expanding, approximately
boost-invariant initial state begins to build up transverse pressure and evolves
toward local thermal equilibrium. The main challenge is to understand how this
state becomes amenable to a description in terms of hydrodynamics. Consistency
with observation suggests that this happens on a timescale of about
$0.3\leq\tau\leq1~\rm fm/c$, when the system is still very anisotropic, and
hydrodynamics in the usual sense would not be expected to apply.  And yet, one
has to accept as fact that hydrodynamic simulations capture many essential
features of QGP dynamics. 

An important point is that gluon self-interactions are not only responsible for
asymptotic freedom, but also for their proliferation, which leads to a dense
medium. Attempts to describe it in terms of quasiparticles require parameter
values such that the mean free path of constituents cannot be large compared to
their de Broglie wavelength~\cite{Busza:2018rrf}. This implies that despite the
weakness of parton interactions at small distances, strong collective effects
should be expected and are seen as playing a key role in the thermalisation
process~\cite{Blaizot:1987nc,Baier:2000sb}.  The precise way this plays out is
still the subject of current  research, but it is feasible that following a
regime where a field-theoretical description is necessary, the system enters a
stage which can be described by approximately free-streaming quasiparticles (for
recent reviews please see e.g.
Refs.~\cite{Schlichting:2019abc,Berges:2020fwq,Schenke:2021mxx}).  The simplest
way to model this situation is to adopt a "step-function approach" and assume
that particles free stream for some time $\tau_{\rm fs}$, and at that point the
description switches to hydrodynamic evolution at a time when the expanding
plasma system is still far from equilibrium. This is schematically depicted in
Fig.~\ref{fig:hydrodynamisation}.

The successful application of hydrodynamic models in such far-from-equilibrium
situations implies that the complexity of initial states is rapidly reduced
within a very short interval of proper-time.  Since this happens for all initial
states, the system can be said to reach a far-from-equilibrium {\em hydrodynamic
attractor}.  In the context of boost-invariant flow this implies that any
potentially complex dynamics of the pressure anisotropy should give way to
universal features already at very early times, very far from the perfect fluid
domain.  Thus, hydrodynamic attractors enter the picture as an interface to the
hydrodynamic stage. In principle, this attractor could describe free streaming
at the very earliest times, but it is not known whether this is the case or not.
At present we have to resort to various models and uncontrolled approximations,
some of which (such as kinetic theory) imply free streaming, while others do
not.  

In the next seven Sections we will review the early-time dynamics and the
appearance of far-from-equilibrium attractors in various model systems.  We will
also address the important issue of relaxing some of the symmetry assumptions
which we have described in this Section.

%% file: mis.tex
\section{Hydrodynamic models of equilibration}
\label{sec:MIS}

The appearance of attractors at the early stages of QGP dynamics can be
understood most easily in the context of what we refer to here as {\em hydrodynamic
models of equilibration}. This Section reviews the necessary conceptual
framework by clarifying the relationship between hydrodynamic behaviour and this
simplest class of models where its emergence can be studied.  Since this review
is focused on attractors, the aim of this section is not to introduce the
subject of relativistic hydrodynamics, which is well covered by the
existing sources (see e.g.
~\cite{Romatschke:2009im,Florkowski:2017olj,Romatschke:2017ejr}), but rather to
present a perspective which is useful for understanding hydrodynamic attractors.

\subsection{Conservation laws}

Hydrodynamic behaviour follows from conservation laws, the
most fundamental ones being those which express spacetime symmetries. In the
relativistic setting they take the form of the conservation law of the
energy-momentum tensor: 
\bel{eq:conservation}
\nabla_\mu T^{\mu\nu} = 0~.
\ee
In the context of a microscopic theory, such as a quantum field theory,
$T^{\mu\nu}$ above would refer to the expectation value of the energy-momentum
operator in some state, while in a kinetic theory model this would be a suitable
moment of the distribution function (see \rfs{sec:attractorKT}). When the system
is in local equilibrium, this quantity can be expressed in the perfect fluid
form, which is just a constant boost of its value at equilibrium:
\bel{eq:perfect}
    T_{\mu\nu}= \edens u_\mu u_\nu + \PP \Delta_{\mu\nu}\,,
\ee
where $\Delta_{\mu\nu}=g_{\mu\nu}+u_\mu u_\nu$ and $u$ is the boost parameter --
the relativistic velocity. Throughout this review, the metric $g$ is assumed to
be that of flat Minkowski space. The quantities $\edens$ and $\PP$ are scalars
which can be interpreted as the energy density and pressure in the local rest
frame. They are usually expressed in terms of the local effective temperature
$T$ through equations of state. The effective temperature and flow velocity are
then referred to as the {\em hydrodynamic variables}.  Due to the normalisation
condition of the four-velocity ($u\cdot u = -1$), there are four independent
variables. 

If the system is not in global equilibrium, the four hydrodynamic variables
are no longer constant and energy momentum tensor 
will depart from the perfect fluid form
\bel{eq:dissipative}
T_{\mu\nu}= \edens u_\mu u_\nu + \PP \Delta_{\mu\nu} + \pi^{\mu\nu}.
\ee
The correction $\pi^{\mu\nu}$ appearing above will be referred to as the
dissipative tensor.  This tensor vanishes unless the hydrodynamic variables vary
in spacetime, so one expects that sufficiently close to equilibrium it can be
expressed as a series of terms involving derivatives of the hydrodynamic
variables; this series is referred to as the {\em hydrodynamic gradient
expansion}\footnote{Unless explicitly indicated otherwise, we use the terms {\em
gradient} and {\em derivative} to mean derivatives with respect to the spacetime
variables, as opposed to purely spacial derivatives.}.  The gradient expansion
provides an asymptotic description of a given flow sufficiently close to
equilibrium. This asymptotic behaviour is strongly constrained by symmetries and
is thus common to many microscopic systems.  

The definition of the hydrodynamic variables is physically
unambiguous only in global equilibrium.  In general, one can redefine them
according to 
 \be
 \label{eq:frame}
 \edens = \tilde{\edens} + \delta\edens,\quad u^\mu= \tilde{u}^\mu+\delta u^\mu ~.
 \ee
In the context of the gradient expansion the delta-terms appearing above can be
thought of as being of order one or higher. Up to some finite order such
redefinitions can be used to impose so-called hydrodynamic {\em frame
conditions} which eliminate some components of the energy-momentum tensor. A
very convenient requirement of this type is the Landau condition
\begin{equation}
\label{eq:landauc}
    u_\mu \pi^{\mu\nu} = 0\,.
\end{equation}
Unless stated otherwise, in this review we will be assuming that this choice has been made. 

\subsection{Modelling hydrodynamics}

The basic idea of hydrodynamic models is to adopt the hydrodynamic
variables $(u^\mu)$
and $T$ as independent classical fields in an effective description of the dynamics
of the energy-momentum tensor. Hydrodynamic models then view the conservation
equations \rf{eq:conservation} not as a statement about the expectation value of
energy-momentum in a microscopic theory, but rather as a set of four evolution
equations which determine the dynamics of the four hydrodynamic variables.
With the energy-momentum tensor in the form given in \rf{eq:perfect}, this leads
to the relativistic theory of perfect fluids.  

In order to
incorporate dissipation one needs to express the dissipative tensor
$\pi^{\mu\nu}$ in \rf{eq:dissipative} in terms of
the hydrodynamic variables and their gradients. It is natural to do this by
using the gradient expansion, which from this perspective is the most general
parametrisation of near-equilibrium behaviour, including all the terms allowed
by symmetries. 

In conformal theories it is very convenient to express gradients in terms of
{\em Weyl-covariant derivative} $\mathcal{D}_\mu$ which differs from the ordinary derivative by
terms involving the four-velocity $u$ and its gradient. Its general definition and properties can be found in
Ref.~\cite{Loganayagam:2008is} (see also the appendix E of
Ref.~\cite{Florkowski:2017olj} for a brief summary).  
The simplest possibility is to set
\bel{eq:navstokes}
\pi_{\mu\nu} = - \eta \sigma_{\mu\nu} \equiv - \eta \left(\D_{\mu} u_\nu +
    \D_\nu u_\mu\right) =
- \eta \left(\p_\mu u_\nu +
\partial_\nu u_\mu-\frac{2}{3}\Delta_{\mu\nu}\partial_\alpha u^\alpha\right)~,
\ee
which is the unique term of first order in gradients which is consistent with
Lorentz and conformal invariance. The coefficient $\eta$ appearing here is the
shear viscosity, which is a scalar function of the effective temperature.  The
resulting model is the relativistic generalisation of Navier-Stokes theory.  In
contrast to non-relativistic case, this theory is acausal, because it possesses
solutions which propagate at arbitrarily large velocities. In consequence, this  
theory is also
unstable~\cite{Hiscock:1983zz,Hiscock:1985zz,Romatschke:2009im,Gavassino:2021owo}. 

To obtain a consistent and practically useful dynamical model one needs to
provide a prescription for augmenting the conservation equations
\rf{eq:conservation} in such a way as to be able to calculate the time evolution
of arbitrary initial data.  This prescription has to guarantee stability under
perturbations of equilibrium, as well as causality of propagation.  It must also
ensure the correct asymptotic behaviour as equilibrium is approached, which is
given by \rf{eq:navstokes}.  These requirements are very strong, and precious
few examples exist where they have been proved to be satisfied (see
Refs.~\cite{Bemfica:2017wps,Bemfica:2019cop,Bemfica:2020xym,Bemfica:2019knx,Bemfica:2020zjp}).
In the remainder of this Section we review the most widely-used approaches,
where they can be satisfied at least at the linearised level.

\subsection{The MIS approach}

The MIS approach~\cite{Israel:1976tn,Israel:1979wp} does not assume an explicit
form of the dissipative tensor in terms of gradients of the hydrodynamic
variables. Instead, it posits a separate set of partial differential equations
for the dissipative tensor. These are formulated in such a way as to possess 
asymptotic solutions in the form of the gradient expansion parametrised in
terms of some finite number of scalar parameters. 

In the simplest variant of MIS theory the dissipative tensor satisfies equations
of the form of a relaxation equation
\begin{equation}
   (\tpi \D  +1)\pi_{\mu\nu}= - \eta \sigma^{\mu\nu} + \dots
   \label{eq:MISpi}
\end{equation}
where $\D\equiv u^\mu\D_\mu$. The properties of the Weyl-covariant derivative
ensure that the Landau condition is preserved under time evolution. One may also
include additional terms in this equation, as discussed below.  As written, this
model guarantees stability as well as causality at the linearised level, as long as
the relaxation time is large enough, satisfying the bound (see e.g.
Ref.\cite{Romatschke:2009im,Florkowski:2017olj})
\bel{eq:causal}
T \tau_\pi > 2 \eta/s~. 
\ee
Causality and stability at the nonlinear level are much more challenging to
establish, as discussed e.g. in Ref.~\cite{Bemfica:2020zjp}.

The solution to the relaxation equation \rfn{eq:MISpi} can be formally expanded in gradients:
\be
\label{eq:pigradshort}
\pimunu &=& -\eta \sigmamunu +  \tpi  \D\left(\eta\sigmamunu\right) +\ldots 
\ee
where the ellipsis denotes terms of third and higher orders. 
The leading term is of the Navier-Stokes form given in \rf{eq:navstokes}. 
The second and higher order terms are affected by the 
precise set of terms chosen for the right hand side of \rf{eq:MISpi}. 
In order to view a hydrodynamic model of equilibration as an effective
description of some underlying theory, one needs to have a means of matching the
two. This can be done using the gradient expansion which, as a perturbative
series around the state of global equilibrium, can be computed in any dynamical
theory -- at least in principle.  This circumstance makes it possible to match
parameters by comparing terms of the gradient expansion calculated in a
microscopic theory with analogous terms calculated in a hydrodynamic
model~\cite{Bhattacharyya:2007vjd}. For this to be generally possible at a given order in
the gradient expansion, the series in \rf{eq:pigradshort} would have to
include all terms allowed by Lorentz (and conformal) symmetry at this order. 
\rf{eq:MISpi} can match any microscopic model to first order in gradients, but
if one wishes to have the option to match to second order, additional terms are
needed. In Ref.~\cite{Baier:2007ix} the complete
set of second order terms which are consistent with
Lorentz and conformal covariance was determined. They can be matched by the
gradient expansion of the following relaxation equation
\be
\label{eq:brsss}
\left(\tau_\pi \D + 1 \right) \pi^{\mu\nu}
= -\eta \sigma^{\mu\nu}  
+ \lambda_1 {\pi^{\langle\mu}}_\lambda \pi^{\nu\rangle\lambda} +
\lambda_2 {\pi^{\langle\mu}}_\lambda \omega^{\nu\rangle\lambda} +
\lambda_3 {\omega^{\langle\mu}}_\lambda \omega^{\nu\rangle\lambda}\,.
\ee
Here $\lambda_1, \lambda_2,\lambda_3$ are additional transport coefficients
which guarantee
matching to second order in gradients~\footnote{We have omitted terms which
vanish in a flat metric background.}, 
\be
\omega^{\mu\nu} = \f{1}{2} 
\left(\D^\mu u^\nu - \D^\nu u^\mu \right),
\ee
is the kinetic vorticity,
and 
the angular brackets are defined as 
\be
{}^{\langle}   A^{ \mu\nu \rangle}   \equiv A^{\langle \mu\nu \rangle} = \f{1}{2}   \Delta^{\mu\alpha} \Delta^{\nu \beta} \left( A_{\alpha \beta} + A_{\beta\alpha} \right)
- \f{1}{3}  \Delta^{\mu\nu} \Delta^{\alpha \beta} A_{\alpha \beta} .
\ee
In the remainder of this review when talking about MIS theory we will have in
mind the above form of the relaxation equations, sometimes referred to as the
BRSSS equations. 
It is worth pointing out that while \rf{eq:brsss} is general enough so that
its gradient expansion includes all the terms in \rf{eq:pigradshort} with
arbitrary coefficients, it is not unique~\cite{Baier:2007ix}.

Finally, we note that while MIS theory is the most widely-used framework for
building models of hydrodynamics, other approaches exist, such as
anisotropic hydrodynamics (for a review and references see e.g.
Ref.~\cite{Florkowski:2017olj}).

\subsection{Lessons from linear response}
\label{sec:linresponse}

Important insights into nonequilibrium dynamics follow from linearisation around
the state of global equilibrium. 
For our purposes it is enough to consider here the state of homogeneous equilibrium (non-rotating, without any external fields).
The hydrodynamic variables which solve the
linearised equations are then 
proportional to the harmonic factor $\exp
\left(-i \omega(k) t + i \vec{k}\cdot\vec{x}\right)$.  
The dispersion relations
which define the different solutions (modes) fall into two categories: the
{\em hydrodynamic modes} whose frequency vanishes with at long wavelengths, 
$\lim_{k\rightarrow0} \omega(k) = 0$, and the {\em nonhydrodynamic modes} which are
gapped: $\lim_{k\rightarrow0} \omega(k) \neq 0$.  This gap -- the frequency at
vanishing wave vector $k$ -- sets the asymptotic damping rate of the transient modes.  The damping
of the hydrodynamic modes diminishes with $k$, so modes of long
wavelengths are weakly damped. 

For example, linearisation of the evolution equations of MIS theory reveals a
set of hydrodynamic sound and shear modes\footnote{The radius of convergence of the series expansions of $\omega(k)$  is set by  singularities in the complexified $k$ place which
reflect mode
collisions~\cite{Withers:2018srf,Grozdanov:2019uhi,Heller:2020hnq,Heller:2020jif}.}
\be
    \omega_{\rm shear} &=& -i \frac{\eta}{s T}k^2 + O(k^4)~,
\\
    \omega_{\rm sound} &=& \pm \f{1}{\sqrt{3}} k -\frac{2i}{3}\frac{\eta}{s T} k^2  + O(k^3)~,
\ee
as well a some nonhydrodynamic modes which are damped
regardless of wavelength: their dispersion relation is $\omega = - i/\tau_\Pi +
O(k^2)$. In the limit when the relaxation time vanishes, the nonhydrodynamic modes
decouple and this theory reduces to Navier-Stokes theory. A calculation of the
velocity of sound (see e.g. Refs.~\cite{Romatschke:2009im,Florkowski:2017olj})
gives
\be
v = \f{1}{\sqrt{3}} \sqrt{1 + 4 \f{\eta/s}{T\tpi}} .
\ee
The condition
\rf{eq:causal} provides a limit on how small the relaxation time can be without
violating causality.  Thus, a natural way to think of nonhydrodynamic modes is
to view them as a regulator~\cite{Spalinski:2016fnj} (somewhat in the spirit of a
``UV-completion'' of quantum field theories), with the relaxation time
playing the role of a regulator parameter. 

This happens not just in MIS-type theories, but in many other hydrodynamic
models which are causal at least at the linear level, such as
BDNK~\cite{Bemfica:2017wps,Bemfica:2019knx,Kovtun:2019hdm,Noronha:2021syv} and
HJSW~\cite{Heller:2014wfa}. Indeed, recent
results~\cite{Heller:2022ejw,Gavassino:2023myj} strongly suggest that the
presence of nonhydrodynamic modes is a necessary condition for causality.  The
existence of hydrodynamic modes follows from conservation laws, while the
nonhydrodynamic modes are required to maintain causality.  The nonhydrodynamic
modes account for transient behaviour, while the long-lived hydrodynamic modes
express a measure of universality in the approach to equilibrium.

\subsection{Modeling the non-hydrodynamic sector}
\label{sec:HJSW}

The appearance of nonhydrodynamic modes in models of relativistic hydrodynamics
mirrors the structure of microscopic theories. However, the analysis of
linearised perturbations of microscopic models reveals a much more complicated
picture than the simple nonhydrodynamic sector of MIS theory.  This happens in
models of kinetic theory~\cite{Romatschke:2015gic}, as well as strongly coupled
field theories described using methods based on the AdS/CFT
correspondence~\cite{Kovtun:2004de}. In both these cases there is an infinite
number of nonhydrodynamic modes, and in the latter case they are not purely
decaying. Sufficiently close to equilibrium the details of this sector are not
relevant, as the near-equilibrium physics is captured by the hydrodynamic
modes~\cite{Bantilan:2022ech}. However, in practice models of hydrodynamics are
often used further away from equilibrium, so models with different
nonhydrodynamic sectors will a priori lead to different results.  In such
situations one is really probing the physics of the regulator.

This raises the question whether it is possible to engineer hydrodynamic models
which mimic nontrivial nonhydrodynamic sectors. An example of such a model was
put forward in Ref.~\cite{Heller:2014wfa} and will be referred to as the HJSW
model (see also  \cite{Florkowski:2017olj,Aniceto:2015mto,Gavassino:2022roi}).
The motivation behind its formulation was to mimic the behaviour of strongly
coupled $\mathcal{N}=4$ SYM theory, where the least-damped transient modes
depend very weakly on momentum (a phenomenon known as ultralocality).
This leads to an evolution equation for the dissipative tensor of the form 
\begin{equation}
    \label{eq:hjsw}
    \left(\frac{1}{T}\mathcal{D}\right)^2\pi_{\mu\nu}
    +2\Omega_I\frac{1}{T}\mathcal{D}\pi_{\mu\nu}+|\Omega|^2\pi_{\mu\nu}=-\eta|\Omega|^2\sigma_{\mu\nu}-C_\sigma\frac{1}{T}\mathcal{D}\left(\eta\sigma_{\mu\nu}\right)+\dots~.
\end{equation}
This equation is a replacement for the MIS/BRSSS relaxation equation,
\rf{eq:MISpi}. The parameters $\eta, \Omega_R, \Omega_I, C_\sigma$ play the same
role as the transport coefficients appearing in \rf{eq:MISpi}, and $|\Omega|^2 =
\Omega_R^2+ \Omega_I^2$. The term with parameter $C_\sigma$ was introduced to
broaden the domain where the theory is stable and causal at the linearised
level.  The physical meaning of these parameters is
partially revealed by formally expanding \rf{eq:hjsw} in gradients, which yields
\rf{eq:pigradshort} with the identification 
\be
\tau_\Pi =  \frac{2\Omega_I - C_\sigma}{|\Omega|^2} \f{1}{T}
\ee
and $\eta$ retaining its meaning as the shear viscosity. 

Further insight is gained by calculating the dispersion relations for linear perturbations of equilibrium.
Apart from the standard hydrodynamic modes we see nonhydrodynamic modes
\be
\omega_\pm(k) = - i \Omega_I \pm \Omega_R + O(k)
\ee
whose relaxation rate is set by $\Omega_I$. In contrast to MIS theory, these
modes are not purely decaying: they also oscillate with frequency set by
$\Omega_R$. This captures the patterns of least-damped quasinormal mode of the
black brane appearing in the dual description of \symm\ theory (see
\rfs{sec:attractorHolo}), where  $\Omega_{R} \approx 9.8$ and $\Omega_{I}
\approx 8.6$.  Of course, in the spirit of hydrodynamics, \rf{eq:hjsw} could in
principle apply to any theory with a similar pattern of nonhydrodynamic modes.
Thus, at least at the level of the gradient expansion, this model contains
Navier-Stokes theory in the near-equilibrium limit -- just like MIS -- but
provides a different regulator sector. 
The assumption of ultralocality which has lead to \rf{eq:hjsw} is a useful
simplification, but it is not strictly obeyed in SYM, and can be avoided at the
level of hydrodynamic models~\cite{Heller:2021yjh}. 

The idea of including nonhydrodynamic modes in a deliberate manner has also been
the founding concept of the Hydro+ programme~\cite{Stephanov:2017ghc}, which is
being actively developed in connection with the search for signals of a critical
point in the QCD phase diagram through heavy-ion collisions. Recent work
developing this circle of ideas  includes
Refs.~\cite{Abbasi:2020xli,Ke:2022tqf}.

\subsection{General frames}
\label{sec:gf}

Another interesting class of hydrodynamic models was discovered quite recently
by Disconzi, Bemfica, Noronha and Kovtun in
Refs.~\cite{Bemfica:2017wps,Bemfica:2019knx,Kovtun:2019hdm}.  These models,
usually referred to by the acronym BDNK,  deviate from the MIS approach in that
they do not introduce additional hydrodynamic fields beyond those already
present in Navier-Stokes theory and rely only on the conservation equations to
provide the dynamics.  

The basic insight of BDNK was to recognise that the Landau condition,
\rf{eq:landauc}, is not a fundamental requirement, but rather one of many
ways of pinning down the definition of the hydrodynamic variables
off-equilibrium. So instead of \rf{eq:navstokes}, at first order in
gradients one could adopt the following form of the dissipative tensor:
\be
\label{eq:bdnk}
\pi_{\mu\nu} = \tau^{\mu\nu} + \mathcal{C}\left(u_\mu u_\nu+ \frac{1}{3}\Delta_{\mu\nu} \right)+ \mathcal{Q}_\mu u_\nu + \mathcal{Q}_\nu u_\mu
\ee
where
\be
\tau^{\mu\nu} = - \eta \sigma^{\mu\nu},
\quad
\mathcal{Q}^\mu
= -\tau_\psi\Delta^{\mu\lambda}\D_{\lambda}\edens, 
\quad
\mathcal{C}
= -\tau_\phi \mathcal{D}\edens
\label{eq:bdnk2}
\ee
where $\tau_\phi, \tau_\psi$ are new transport coefficients. These  additional
terms in \rf{eq:bdnk} (relative to \rf{eq:navstokes}) could be removed using the
frame freedom \rf{eq:frame}, which would amount to imposing the Landau
condition. No new dynamical fields are introduced: $\tau^{\mu\nu},
\mathcal{Q}^\mu, \mathcal{C}$ are expressed explicitly in terms of the basic
hydrodynamical variables $\edens, u^\mu$. Nevertheless, this theory is causal
and stable~\cite{Bemfica:2020zjp} for suitable choices of parameters, because
relaxing the Landau frame condition  introduces a nonhydrodynamic sector.  The
structure of this sector turns out to be the same as in MIS theories, but the
evolution equations are different and lead to the same physics only close to
equilibrium~\cite{Bantilan:2022ech}.  Models of this type are the subject of a
number of interesting recent
studies~\cite{Hoult:2020eho,Pandya:2021ief,Pandya:2022pif,Pandya:2022sff,Gavassino:2023odx,Gavassino:2023qwl}.

The general-frame concept can be taken further in the spirit of the MIS
approach~\cite{Noronha:2021syv}. The basic idea is to replace \rf{eq:bdnk2} by a
set of relaxation equations\footnote{The published version of
Ref.~\cite{Noronha:2021syv} presents a rather general implementation of this
idea. A conformal implementation, similar to what we review here, can be found
in the original arXiv.org (v1) submission. The relaxation equations \rfn{eq:nss}
contain only the conceptually essential terms; the original reference contains
some additional contributions motivated by entropy considerations.}
\be
\tau_\pi 
\mathcal{D}\pi^{\mu\nu}
+ \pi^{\mu\nu}
&=& - \eta \sigma^{\mu\nu}
\nn\\
\tau_Q
\D\mathcal{Q}^{\mu}
+ \mathcal{Q}^\mu
&=& -\tau_\psi \Delta^{\mu\lambda}\D_{\lambda}\edens
\nn\\
\tau_C
\mathcal{D}\mathcal{C}
+ \mathcal{C}
&=& -\tau_\phi \mathcal{D}\edens
\label{eq:nss}
\ee
with additional transport coefficients $\tau_\pi, \tau_Q, \tau_C$.  The
resulting model has more degrees of freedom and a nonhydrodynamic sector which
is larger than in either MIS or BDNK, and is of some practical as well as
conceptual significance~\cite{Gavassino:2020ubn,Gavassino:2021cli}.  We will return
to it briefly in \rfs{sec:attractors}, since it offers some additional insights
into attractor behaviour.

%% file: attractors.tex
\section{Attractors in hydrodynamic models}
\label{sec:attractors}

Hydrodynamic attractors were first identified in hydrodynamic models, and
subsequently studied in other models of equilibration such as kinetic theory and
strongly coupled theories amenable to studies based on the AdS/CFT
correspondence.  This Section reviews attractors arising in hydrodynamic
models of Bjorken flow and introduces a number of concepts which will be used in
the remainder of this article.

\subsection{Bjorken flow in MIS theory}

We now turn to the description of Bjorken flow in MIS theory, specifically 
the BRSSS version~\cite{Baier:2007ix}.  As reviewed in \rfs{sec:HIC},
the dynamics of the energy-momentum tensor in this case is captured by the
pressure anisotropy $\pa(\tau)$ and the energy density $\edens(\tau)$, or
equivalently the effective temperature $T(\tau)$.  Conservation of the
energy-momentum tensor reduces 
to \rf{eq:Adef3}, which can be written  as the evolution equation for the 
effective temperature:
\be 
\label{eq:consbjorken} 
\tau\p_\tau \log T(\tau) = -\f{1}{3} + \f{1}{18} \pa(\tau)~,  
\ee
while the MIS relaxation equation~\rf{eq:brsss} becomes an evolution equation
for the pressure anisotropy. To write it down most explicitly one needs to take
full advantage of the constraints of conformal symmetry.   

Conformal symmetry implies that the energy scale is set by the local effective
temperature. The transport coefficients are then determined by dimensional
analysis~\footnote{The other transport coefficients appearing in  \rf{eq:brsss}
    are similarly constrained, but are not relevant for Bjorken flow.}
\begin{equation}
\label{eq:tcs}
\tau_\pi=\frac{C_{\tau\Pi}}{T}, \qquad \eta=C_\eta s~,\qquad \lambda_1=\frac{C_{\lambda_1}}{T \eta}~,
\end{equation}
where $s=4\edens/3T$ is the entropy density,  
up to dimensionless constants $C_{\tau\Pi},C_\eta,C_{\lambda_1}$. These
constants can be fitted to experiment, or matched to an underlying microscopic
theory in cases where an explicit calculation of the gradient expansion is
feasible. An example of such a calculation for the cases of \symm\ was carried
out in Ref.~\cite{Bhattacharyya:2007vjd,Baier:2007ix} using the AdS/CFT correspondence, with
the result
\begin{equation}
C_{\tau\Pi}=\frac{2-\log 2 }{2\pi}~,  \qquad C_\eta=\frac{1}{4\pi}~, \qquad C_{\lambda_1}=\frac{1}{2\pi}~.
\end{equation}
These values provide a useful point of reference as well as an order of
magnitude estimate which is sometimes used in hydrodynamic simulations. 

Once the transport coefficients are written in the form \rf{eq:tcs}, the
MIS/BRSSS relaxation equation can be written in the form
\be
\label{eq:misbjorken} 
    C_{\tau\Pi} \left(\tau \pa'(\tau) + \f{2}{9} \pa^2(\tau)\right) = 8 C_\eta -\tau T(\tau) \left( \pa(\tau) + \frac{C_\lambda}{12 C_\eta}  \pa(\tau)^2\right).
\ee
The system of two coupled ordinary differential equations, \rf{eq:consbjorken}
and \rf{eq:misbjorken} determines the dynamics of Bjorken flow in MIS theory.

\subsection{Late time asymptotics of Bjorken flow}

The evolution equations,
\rf{eq:consbjorken} and \rf{eq:misbjorken}, can be combined to give a single ODE which
determines the dynamics of the effective temperature
\be
    \label{eq:MISTeom}
    C_{\tau\Pi}\tau T'' &+&
    \frac{3}{2}\tau\left(\frac{C_{\lambda_1}\tau}{C_\eta}+\frac{2C_{\tau\Pi}}{T}\right){T'}^2+
    \left(\frac{11C_{\tau\Pi}}{3}+\frac{(C_\eta+C_{\lambda_1})\tau
    T}{C_\eta}\right)T'+\nn\\
    &+& \frac{(2C_\eta+C_{\lambda_1})T^2}{6C_\eta}-
    \frac{4(C_\eta-C_{\tau\Pi})T}{9\tau}=0~.
    \ee
It is easy to see that at large proper-times this equation has an asymptotic 
solution of the form
\begin{equation}
    T(\tau)=\frac{\Lambda}{(\Lambda\tau)^\frac{1}{3}}
    \left(1-\frac{2C_\eta}{3(\Lambda\tau)^\frac{2}{3}}+
    O\left(\f{1}{(\Lambda\tau)^{4/3}}\right) \right)~,
    \label{eq:MISTasym}
\end{equation}
where $\Lambda$ is an integration constant.  Since the initial value problem for
\rf{eq:MISTeom} allows for the choice of two integration constants, namely the
initial temperature and its derivative, it is clear that the asymptotic solution
\rf{eq:MISTasym} contains only half the information encoded in the initial
state. This is a consequence of dissipation. A complete solution would require
augmenting this result with additional terms which depend on the remaining
initial data, but vanish faster than any power of proper time. We will return to
this important point below in \rfs{sec:late}. 

Quite generally, dissipation implies an effective loss of information:
specifically, a partial loss of memory of the initial state of the system. The initial
state can be far from equilibrium and may be characterised by many parameters.
On the other hand, the final state of equilibrium is characterised by very few
parameters. The asymptotic late-time behaviour of the system will thus be
partially independent of the initial state.  This process of ``information
loss'' can be studied using modern asymptotic methods. Furthermore, it lies at
the heart of the idea of hydrodynamic attractors, which -- as we will discuss in
detail -- is fundamentally the observation that generic initial states evolve
into a region of phase space which can be effectively covered by a subset of all
possible initial conditions.

\subsection{Universal variables}

In the case of conformal Bjorken flow it is possible to make the notion of
information loss described above even sharper by using suitable variables which
are correlated in a universal way: variables in which the asymptotic behaviour
near equilibrium is completely independent of initial conditions.  This is not a
typical situation and is only possible due to the very strong symmetry
assumptions.

Conformal symmetry suggests using the dimensionless pressure anisotropy $\pa$
and introducing the dimensionless variable $w\equiv \tau T$. At late times, when
the temperature follows \rf{eq:bjorken},
$w\sim\tau^{2/3}$, so that it can be thought of as a ``clock variable'': the
proper time in units of local effective temperature. Since the relaxation time
$\tau_\Pi \sim 1/T$, one also has $w\sim\tau/\tau_\Pi$, so one can think of this
variable as the proper time in units of the relaxation time. 
Using these dimensionless variables, the conservation equation~\eqref{eq:consbjorken} can be
written as
\be 
\label{eq:consw} 
\f{d\log T}{d \log w} = \f{\pa - \ 6}{\pa + 12}~, 
\ee
and the MIS equation \rf{eq:misbjorken} takes the form 
\begin{equation}
    \label{eq:MISAeom}
C_{\tau\Pi}\left(1+\frac{\mathcal{A}(w)}{12}\right) \mathcal{A'}(w)+\left(\frac{C_{\tau\Pi}}{3w}+\frac{C_{\lambda_1}}{8C_\eta}\right)\mathcal{A}(w)^2=\frac{3}{2}\left(\frac{8C_\eta}{w}-\mathcal{A}(w)\right)~.
\end{equation}
The remarkable point here is that \rf{eq:MISAeom} is a self-contained equation
which can be solved independently of the conservation law \rf{eq:consw}.  Once
solutions $\pa(w)$ are found, they can be used in \rf{eq:consw} to determine the
corresponding evolution of the effective temperature. 

For a perfect fluid $\pa=0$ and either \rf{eq:consbjorken} or \rf{eq:consw}
suffices to determine the solution, leading to Bjorken's \rf{eq:bjorken}.
However, for dissipative systems one must also specify a nontrivial solution of
\rf{eq:MISAeom}, which depends on the microscopic dynamics of the plasma through
the transport coefficients, as well as on the initial state of the system.  If a
solution $\pa(w)$ of \rf{eq:MISAeom} is given, one can integrate
\rf{eq:consw} to solve for the effective temperature as a function of $w$:
\begin{equation}
  \label{eq.evol}
   T(w) =  \Phi_\pa (w, w_0)  T(w_0)~,
\end{equation}
for some initial condition 
$T(w_0)$, with the function $\Phi_\pa$ being
\begin{equation}
  \label{eq.phidef}
  \Phi_\pa (w, w_0) = \exp\left(\int_{w_0}^{w} \f{dx}{x}
  \f{\pa(x) - \ 6}{\pa(x) + 12} \right)~.
\end{equation}
The subscript $\pa$ which appears above indicates the functional dependence of this
quantity on the pressure anisotropy.

\subsection{The hydrodynamic attractor}

It is straightforward to solve \rf{eq:MISAeom} numerically. As expected, at late
times all solutions tend to zero as equilibrium is approached. However, a rather striking
picture emerges when studying the behaviour of solutions obtained by setting
initial conditions at a sequence of diminishing initial values of
$w$, as seen in \rff{fig:MISattractor}. It is evident that
the solution curves rapidly approach a distinguished locus, which is referred to
as a far-from-equilibrium attractor~\cite{Heller:2015dha}. This attractor curve is 
determined uniquely by this procedure, and will be denoted by $\pa_\star$. 
It is the extension of the hydrodynamic attractor expected near equilibrium into
the early-time, nonequilibrium region. 


\begin{figure}
\begin{center}
    \includegraphics[height =.55\textwidth]{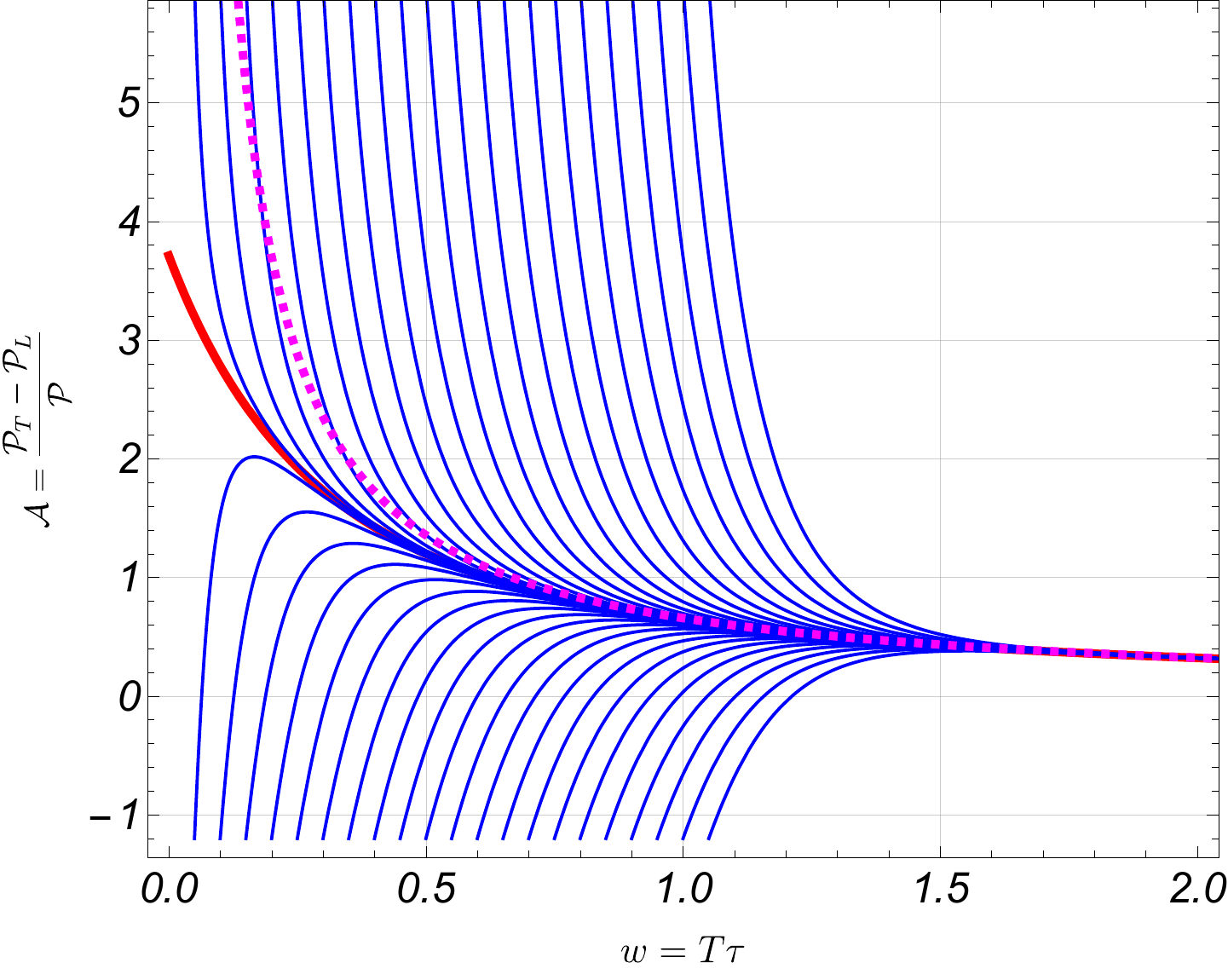}
\caption{Some solutions of \rf{eq:MISAeom} (blue lines) plotted together with
the attractor (red line); the dashed magenta line represents second order viscous hydrodynamics.}
\label{fig:MISattractor}
\end{center}
\end{figure}


It is physically important that solutions initialised off  the attractor
approach it rapidly while the pressure anisotropy is high and the system is
still far from equilibrium.  This fact leads to a potential explanation of the
early thermalisation puzzle, as we will argue in the following.  Note also that
solutions which start out below the attractor are initially driven {\em away}
from equilibrium toward the attractor. As discussed further below, this is a
consequence of the strong longitudinal expansion. 

The emerging picture is that for a given range of initial conditions,
apart from an initial transient, the function $\pa(w)$ quickly approaches a
universal attractor $\pa_\star(w)$ which is determined by the
microscopic theory under consideration. We
assume that the physically interesting range of initial conditions is in the
basin of attraction of this unique attractor.  This suggests that it should be a
good approximation to replace the form of the pressure anisotropy $\pa(w)$, as
it appears in \rf{eq:consw}, by the attractor $\pa_\star(w)$:
\begin{equation}
  \label{eq.approx}
   T(w) \approx  \Phi_{\pa_\star} (w, w_0)  T(w_0)~.
\end{equation}
Within such an approximation, the temperature at late times is determined by the
temperature at early times {\em alone}: the remaining dependence on the initial
state is neglected by assuming that the effective dynamics of the system is
captured by its attractor, apart from a negligible initial transient\footnote{
    An example which bears some similarity to what is considered here is the
    idea of an inflationary attractor in cosmology, which also captures the
    effective loss of information about the pre-inflationary features of our
Universe (see e.g.~\cite{Mukhanov:2005sc}).}.

The attractor apparent in the pressure anisotropy $\pa(w)$ is particularly
striking, but it is a manifestation of an intrinsic feature of this dynamical  system, as
well as many other like it, however one chooses to describe them. It also has
implications for other observables,  such as the speed of sound away from
equilibrium~\cite{Cartwright:2022hlg}.

The qualitative picture seen in \rff{fig:MISattractor} is typical of Bjorken
flow in many models of equilibration, including various extensions of MIS
theory, anisotropic hydrodynamics~\cite{Strickland:2017kux,Alqahtani:2022xvo},
kinetic theory as well as strongly coupled \symm\ theory.  Before reviewing some
of them, we will try to understand the features seen in this plot in a
quantitative way, using asymptotic methods to extract the relevant physics from
\rf{eq:MISAeom}.

\subsection{Early time behaviour}
\label{sec:MISearly}

As it is clear from \rff{fig:MISattractor}, at small
values of $w$ generic solutions are divergent, apart from the attractor which is
regular there. It is straightforward to check that if we assume that the
pressure anisotropy approaches a finite, constant value $\pa_\pm$ as $w
\rightarrow {0}$,
then the only possible values consistent with \rf{eq:MISAeom} are
\begin{equation}
    \pa_\pm = \pm 6\sqrt{C_\eta/C_{\tau\Pi}}~.
\end{equation}
The negative option is unstable, it acts as a repulsor; we will not discuss it 
further here. The positive value
provides the initial condition which can be used to determine the attractor
numerically. 

The early-time behaviour of regular solutions of \rf{eq:MISAeom} can be studied
analytically through a convergent series expansion in powers of
$w$~\cite{Heller:2015dha,Kurkela:2019set,Aniceto:2022dnm} 
\begin{equation}
    \paz(w)=\sum_{n=0}^\infty c_n w^n=6\sqrt{\frac{C_\eta}{C_{\tau\Pi}}}-    \frac{9(C_{\lambda_1}+2\sqrt{C_\eta C_{\tau\Pi}})}{C_{\tau\Pi}(2 C_{\tau\Pi} + 9 \sqrt{C_\eta C_{\tau\Pi}})}w+\cdots
\end{equation}
In the following we will denote the attractor solution by
$\pa_\star$. 
The remaining solutions of \rf{eq:MISAeom} diverge at $w=0$, but are seen to
approach the attractor rapidly. 
From a physical perspective it is very important to understand how exactly this
happens and what is the reason for it. One can look for solutions of
the form
\begin{equation}
    \pa(w) = \paz(w) + \delta\pa(w)
\end{equation}
where $\delta\pa(w)$ is dominant for $w$ approaching zero.  
The equation of motion \rf{eq:MISAeom} then takes the approximate form
\begin{equation}
    w \delta \pa'(w)+4 \delta\pa(w)=0~,
\end{equation}
which gives $\delta\pa\sim w^{-4}$. This result is independent of the transport
coefficients, which suggests a kinematic origin of this phenomenon.  More
specifically, the physical mechanism behind it can be identified with the strong
longitudinal expansion of the system.  The implications of this fact will be
discussed in \rfs{sec:beyond}.

\subsection{Late time behaviour}
\label{sec:late}

At large values of $w$, all solutions plotted in \rff{fig:MISattractor} approach
the curve corresponding to the leading order of the gradient expansion. This
can be seen directly in \rf{eq:MISAeom} by noting that as $w \rightarrow
\infty$ both terms on the left hand side of are subdominant, so that the leading
asymptotic behaviour is 
\begin{equation}
    \pa(w)\sim\frac{8 C_\eta}{w}\,.
    \label{eq:MISlatetimeL}
\end{equation}
Just as the late-time solution of \rf{eq:bjorken}, this implies a loss of initial
state information, because \rf{eq:MISAeom} which governs the dynamics of the
pressure anisotropy requires an initial condition, so a general solution would
contain a single integration constant. This information is completely absent
from the asymptotic solution \rf{eq:MISlatetimeL}, which is completely
universal, identical for all initial conditions. 

As an aside, it is amusing to note that the leading asymptotic behaviour of the
pressure anisotropy can be made not just independent of the initial conditions,
but even across different theories, which at this order differ only by the value
of $\eta/s$. Indeed, defining $\tilde{w}\equiv w/8C_\eta$,
the asymptotics of the pressure anisotropy in any conformal theory are simply
$\pa\sim 1/\tilde{w}$~\cite{Heller:2016rtz}.  This observation has found applications in situations
where the late-time behaviours of different theories are compared. 

The leading asymptotic behaviour of the pressure anisotropy captured by
\rf{eq:MISlatetimeL} is corrected by an infinite series of subleading terms:
\begin{equation}
 \label{eq:MISlatetime}
    \mathcal{A}(w)= \sum_{k=1}^\infty \f{a_k}{w^k}
\end{equation}
with
\begin{equation}
    \label{eq:MISgradevals}
    a_1 = 8 C_\eta, \quad 
    a_2 = \f{16}{3}\, C_\eta( C_{\tau\Pi}-C_{\lambda_1})~.
\end{equation}
Each term appearing here corresponds to a specific order of the gradient
expansion.  If this series is truncated, one obtains an
approximation which one would like to identify with the hydrodynamic
prediction for the asymptotic behaviour of $\pa(w)$. There is an important
subtlety however: the series appearing in \rf{eq:MISlatetime} has a vanishing
radius of convergence. This will be discussed at length below, but for the
moment we will adopt a pragmatic attitude and simply truncate the expansion,
keeping only a couple of the leading terms.

It is important to realise that there are corrections to \rf{eq:MISlatetime}
which are not of the form of a power of $1/w$ -- instead, they are damped exponentially
in the limit of large $w$. To see this, one can linearise this equation around
the truncated asymptotic solution 
\be
\pa(w) = \f{a_1}{w} + \f{a_2}{w^2} + \delta\pa(w)~,
\ee
by treating $\delta\pa$ as small.  This leads to the equation
\begin{equation}
\delta\mathcal{A}'(w)+
\left(
\frac{3}{2 C_{\tau\Pi}} +
\frac{2C_{\lambda_1} - C_\eta}{C_{\tau\Pi}}\f{1}{w} + O\left(\f{1}{w^2}\right)
\right)\delta\mathcal{A}=0~,    
\label{eq:deltaA}
\end{equation}
whose solution is
\begin{equation}
    \delta\mathcal{A}(w)=\sigma w^{\frac{C_\eta-2
    C_{\lambda_1}}{C_{\tau\Pi}}}e^{-\frac{3w}{2C_{\tau\Pi}}} \left(1
    +O\left(\f{1}{w}\right) \right)~,
\end{equation}
where $\sigma$ is an integration constant. 
A more systematic analysis along the lines sketched above reveals solutions of
the form of a {\em
transseries}~\cite{Heller:2015dha,Aniceto:2015mto,Basar:2015ava}:
\begin{equation}
    \mathcal{A}(w)=\sum_{m=0}^\infty\sigma^me^{-mAw}\Phi_m(w)~,
    \label{eq:MIStrans}
\end{equation}
where 
\begin{equation}
    \Phi_m(w)=w^{m\beta}\sum_{n=0}^\infty\frac{a^{(m)}_n}{w^n}~,
\end{equation}
with
\begin{equation}
    \label{eq:singulant}
    A=\frac{3}{2C_{\tau\Pi}}~, \qquad \beta=\frac{C_\eta- 2 C_{\lambda_1}}{C_{\tau\Pi}}~.
\end{equation}
and $\Phi_0(w)$ is just the series \rf{eq:MISlatetime}. Each transseries sector
provides a set of corrections weighted by a power of an exponential damping
factor. The damping rate is set by the relaxation time -- the constant factor of
$3/2$ is explained in Ref.~\cite{Janik:2006gp}.  Crucially, each transseries
sector is also weighted by a power of the undetermined transseries parameter --
the integration constant $\sigma$. This integration constant can in principle be
determined by setting an initial condition, but that information is
exponentially dissipated away in the course of evolution. The transient effects
of the nontrivial transseries sectors can actually be seen in numerical
experiments~\cite{Spalinski:2018mqg}.  It is important to note that the presence
of the transseries sectors is a consequence of the presence of nonhydrodynamic
modes in MIS theory. This connection is quite general and will manifest itself a
number of times in the following. 

The transseries structure is a beautiful metaphor of how information about the
initial state is dissipated in the course of evolution as the system approaches
equilibrium: this data is effectively lost due to the exponential damping,
leaving only a universal hydrodynamic tail: the hydrodynamic attractor.  The
early-time $1/w^4$, expansion-driven approach to the attractor is replaced at
later times by the exponential nonhydrodynamic mode decay whose rate is set by
the relaxation time.  

\subsection{Determining the attractor}

While there exist hydrodynamic models where the attractor can be found
exactly~\cite{Denicol:2019lio,Strickland:2019hff}, in general attractors can be
found be studying the behaviour of multiple solutions obtained by numerical
means. In the simple case of Bjorken flow in conformal MIS theory this can be
done by setting initial conditions at decreasing values of $w$, as illustrated
in \rff{fig:MISattractor}. 

Another approach to capturing the attractor is a variant of the slow-roll
approximation best known in the context of inflationary
cosmology~\cite{Liddle:1994dx,Spalinski:2007kt}. This method is approximate, but
can be pursued analytically. The idea is to treat the derivative term in
\rf{eq:MISAeom} as a perturbation, which ensures the regularity of the obtained
solution at $w=0$.  This can be implemented systematically by inserting a formal
gradient-counting parameter $\epsilon$ into \rf{eq:MISAeom} and seeking a
solution as a series in this quantity. The zeroth order solution is determined
by a quadratic equation.  The attractor solution corresponds to positive root,
and one finds~\cite{Heller:2015dha}
\begin{equation}
    \mathcal{A}_{\rm slow roll}(w)=
   \frac{6}{8C_{\tau\Pi}+\frac{3C_{\lambda_1}w}{C_\eta}}\sqrt{64C_\eta C_{\tau\Pi}+24 C_{\lambda_1}w+9w^2}~.
\end{equation}
This is just the nullcline of \rf{eq:MISAeom}. Corrections are easily calculated
and provide a very accurate representation of the attractor, but its analytic
form quickly becomes very complex.

Another way to obtain approximate attractors analytically in certain hydrodynamic models was
proposed in~\cite{Jaiswal:2019cju}, where the authors considered 
a family of relaxation equations of the form 
\begin{equation}
    \frac{d\pi}{d\tau} = -\frac{\pi}{\tau_\pi} +\frac{1}{\tau}\left[\frac{4}{3}\beta_\pi-\left(\lambda+\frac{4}{3}\right)\pi-\chi\frac{\pi^2}{\beta_\pi}\right]~,
    \label{eq:summary_pi}
\end{equation}
where $\pi\equiv\edens\pa$.  By suitable choices of the parameters $\beta_\pi$,
$\tau_\pi$, $\lambda$ and $\chi$ one can describe the original MIS
model~\cite{Israel:1979wp}, the DNMR model~\cite{Denicol:2012cn} or the
"third-order" model of Ref.~\cite{Jaiswal:2013vta}. All three models possess an
attractor solution, but it can only be found numerically. In a conformal theory,
the relaxation time is determined by the effective temperature, i.e.
$\tau_\pi\sim1/T(\tau)$. One can obtain an analytic approximation of the
attractor by treating this dependence in a sort of ``mean field'' spirit.
Instead of keeping the exact temperature dependence the authors of
Ref.~\cite{Jaiswal:2019cju} study three possible options which amount to taking
the temperature to be constant, or taking one or two terms in the expansion
given in  \rf{eq:MISTasym}.  In each of these cases one can obtain a general
analytic solution to \rf{eq:summary_pi}, which depends on an integration
constant. It is possible to choose this integration constant to obtain a
solution regular at $w=0$. This solution provides a rather good approximation to
the numerically calculated attractor, with the error not larger than
$3\%$~\cite{Jaiswal:2019cju}. This approximate attractor solution was used in
practice for the computations of thermal particle production~\cite{Naik:2021yph}.

Further analytic results for boost-invariant attractors can be found in
Refs.~\cite{Denicol:2017lxn,Denicol:2019lio,Blaizot:2020gql}.

We will also describe two systematic approaches to finding attractors in an
approximate way. One is 
based directly on the
gradient expansion, and leads to some very interesting developments which we
review in the following subsection. 
The other, perhaps the most general approach to
identifying attractors, albeit purely numerically, involves studying sets of solutions on time slices of phase space; it will be described in
\rfs{sec:PhaseSpace}. 

\subsection{The gradient expansion at large orders}
\label{sec:largeorders}

We now turn to an important point of both mathematical and physical
significance: the infinite series appearing in~\rf{eq:MIStrans} have a vanishing
radius of convergence.  At sufficiently late times, the asymptotic behaviour of
all solutions is given by the leading order of the gradient expansion, which
corresponds to Navier-Stokes theory. In many cases it has been possible to
calculate a large number of terms, which offers the possibility to extend the
late-time approximation of the attractor toward early times. This was studied in
the case of the large proper time expansion of \symm\ in
Ref.~\cite{Heller:2013fn} where the series was found to have a vanishing radius
of convergence. It was subsequently found that such expansions diverge in many
other cases, including models of
hydrodynamics~\cite{Heller:2015dha,Aniceto:2015mto} and kinetic
theory~\cite{Denicol:2016bjh,Heller:2016rtz,Florkowski:2016zsi}.  It has been
demonstrated that in the context of MIS theory the gradient expansion has a
vanishing radius of convergence also beyond the relatively simple setting of
Bjorken flow, and it can only be avoided by fine-tuning of the initial
conditions~\cite{Heller:2020uuy,Heller:2021oxl}. In fact, the only known example
of where the hydrodynamic gradient expansion is convergent for generic initial
conditions occurs for Bjorken flow in the model of an ultrarelativistic  gas of
hard spheres of Ref.~\cite{Denicol:2019lio}.  The implication of these findings
is that the gradient series does not define a unique solution. However, it
captures the asymptotic behaviour of all solutions in the late-time limit.  

The simplest approach to such divergent asymptotic series is truncation at low
order, as we have been tacitly assuming until now. It is known from countless
examples (such as the Stirling formula for the Gamma function $\Gamma(z)$ at
large values of $|z|$) that keeping only the leading terms of a divergent
asymptotic series often gives excellent results also quite far from the
asymptotic limit. This can be made quite precise using the notion of optimal
truncation~\cite{Bender78:AMM}. While this approach is very useful in practice,
from a conceptual point of view it is very interesting and rewarding to examine
the nature of the divergence in more detail, since it reveals the physics behind
it. 

The gradient expansion of the pressure anisotropy is of the form 
\begin{equation}
    \mathcal{A}(w)=\sum_{n=0}^\infty\frac{a_n}{w^n}~,
    \label{eq:Agradexp}
\end{equation}
where the leading terms can be read off from \rf{eq:MISlatetimeL}. When referring to
this series in the case of MIS, for definiteness we will assume numerical values for the
coefficients $a_n$ given in \rf{eq:MISlatetime}.  It is straightforward to compute hundreds of
these coefficients numerically. Simple convergence tests lead to the conclusion that the series is
divergent factorially (see \rff{fig:divergeMIS}): at large $n$, up to a constant factor, one has 
\be
\label{eq:dingle}
a_n\sim\Gamma(n+\beta) A^{-n}, 
\ee
where $A, \beta$ are constants which carry important information
about the physics. In particular, the quantity $A$ reflects the damping rate ot
transient, nonhydrodynamic effects. Since \rf{eq:dingle} arises in many
contexts, $A$ is referred to by various names. We will follow Dingle and refer
to is as the {\em singulant}~\cite{Dingle}. In the case of MIS theory
$A=3/2C_{\tpi}$, which shows that the divergence originates in the
nonhydrodynamic sector.


\begin{figure}[t]
\begin{center}
\includegraphics[width =.6\textwidth]{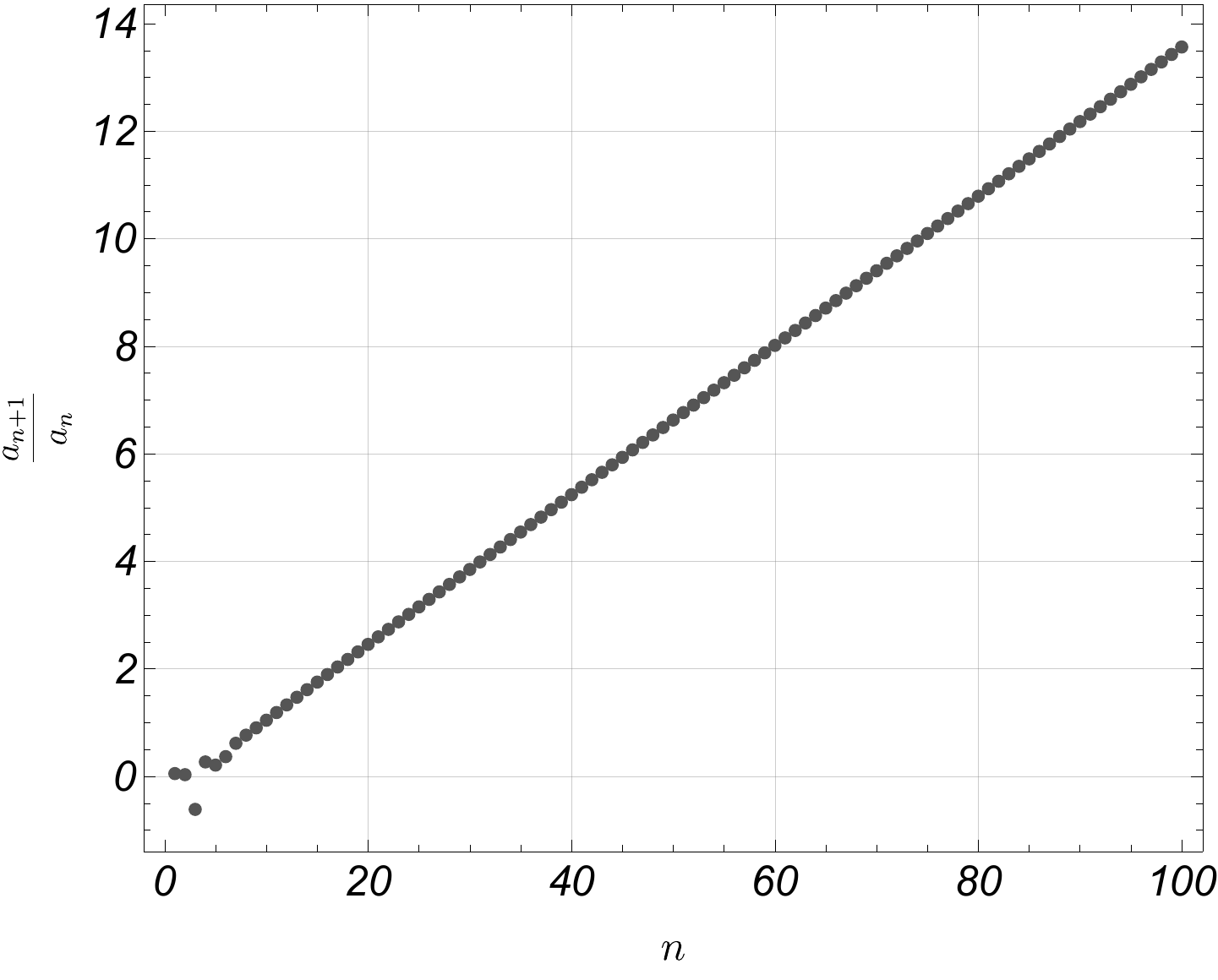}
\caption{The ratio of the coefficients of the gradient expansion \rf{eq:Agradexp}.}
\label{fig:divergeMIS}
\end{center}
\end{figure}


There is a large and growing body of work aimed at understating the role of
corrections to asymptotic series such as \rf{eq:Agradexp}, sometime referred to
as ``asymptotics beyond all orders''~\cite{Berry1991}.  An effective approach to
this problem is to consider ``resumming'' the series in Eq.~(\ref{eq:Agradexp}).
By this one means finding a function whose asymptotic expansion matches the
original series (see e.g.~\cite{Bender:2017fyz}). Given a factorially divergent sequence $\{c_n\}$ this can be
done by Borel summation, whose basic idea is captured by the formal manipulation 
\begin{equation}
\sum_{n=0}^{\infty} c_n
=
\sum_{n=0}^{\infty} c_n
\underbrace{
\left(
\frac{1}{n!} \int_0^\infty  t^n e^{-t} dt
\right)}_{1}
=
\int_0^\infty
\underbrace{\left(\sum_{n=0}^{\infty}\frac{c_n}{n!}\right)}_{\mathrm{Borel\ transform}}
t^n e^{-t} dt.
\end{equation}
To implement this idea in practice, one first defines the Borel
transform of the original factorially divergent series 
\rf{eq:Agradexp} by
\begin{equation}
    \mathcal{BA}(\xi)=\sum_{n=1}^\infty\frac{a_n}{n!}\xi^{n}~,
    \label{eq:ABorel}
\end{equation}
which defines an analytic function inside a disc of radius 
$|A|$ at the origin. 
The Borel sum of the original divergent series is defined by the inverse Borel
transform 
\begin{equation}
    \label{eq:Borelsum}
    \mathcal{A}_{\rm sum}(w)=w\int_\mathcal{C}d\xi\, e^{-w\xi}
    \,\widetilde{\mathcal{BA}}(\xi)~.
\end{equation}
The tilde over the Borel transform indicates that the domain where the series
\rf{eq:ABorel} is defined will need to be extended by means of analytic
continuation so that one can find a contour $\mathcal{C}$ which extends to
infinity. 

In most cases of interest one cannot carry out this prescription exactly. Typically, the
number of coefficients $a_n$ which are available in practice is finite, and the
coefficients which are available are often given numerically with some finite
precision. One also has to rely on approximate methods of analytic continuation.
The quality of this procedure is also critically important for the accuracy of
the result of the
resummation~\cite{Costin:2020hwg,Costin:2020pcj,Costin:2021bay,Costin:2022hgc}.

The most straightforward and widely-used way to carry out the required analytic continuation 
is to adopt the Pad\'e approximant
\begin{equation}
    \mathcal{BA}_{\rm Pade}(\xi)=\frac{P_m(\xi)}{Q_n(\xi)}~,
\end{equation}
where $P_m(\xi)$ and $Q_n(\xi)$ are polynomials of degree $m$, $n$ respectively,
with coefficients properly fitted to match the expansion (\ref{eq:ABorel}). Due
to the approximate nature of this procedure, the singularities of the
analytically continued Borel transform can only be poles.  However, given an
adequate number of terms in the series \rf{eq:Agradexp} and with polynomials of
high enough degree, the poles appear in dense sequences accumulating at the
actual branch points (``condensing'', as it were, along branch cuts). 
This procedure can thus provide a quantitative approximation to the true
singularities of the Borel transform. 

In the case of the MIS gradient expansion, the singularities of the analytically
continued Borel transform the are shown in \rff{fig:borelpadeMIS}.  This pattern
indicates the existence of a branch point at $\xi=A$ (given in
\rf{eq:singulant}) and this can be shown to be related to the large order
behaviour expressed by \rf{eq:dingle}. The fact that this branch point is found
on the real axis means that the integration contour in \rf{eq:Borelsum} must be
deformed to run either below or above the real axis. This leads to a complex
ambiguity of the Borel sum. This ambiguity is in fact cancelled once
contributions from nontrivial transseries sectors are included, and the
imaginary part of the transseries parameter is set correctly.  The consistency
of this procedure relies on the phenomenon of resurgence, which is an intricate
relationship between the expansion coefficients appearing in the different
transseries sectors. For details of these matters we refer the Reader to
Refs.~\cite{Heller:2015dha,Aniceto:2015mto,Basar:2015ava,Aniceto:2022dnm} and
for resurgence in general to Ref.~\cite{Aniceto:2018bis,Mitschi:2016fxp}. 


\begin{figure}[t]
\begin{center}
\includegraphics[width =.6\textwidth]{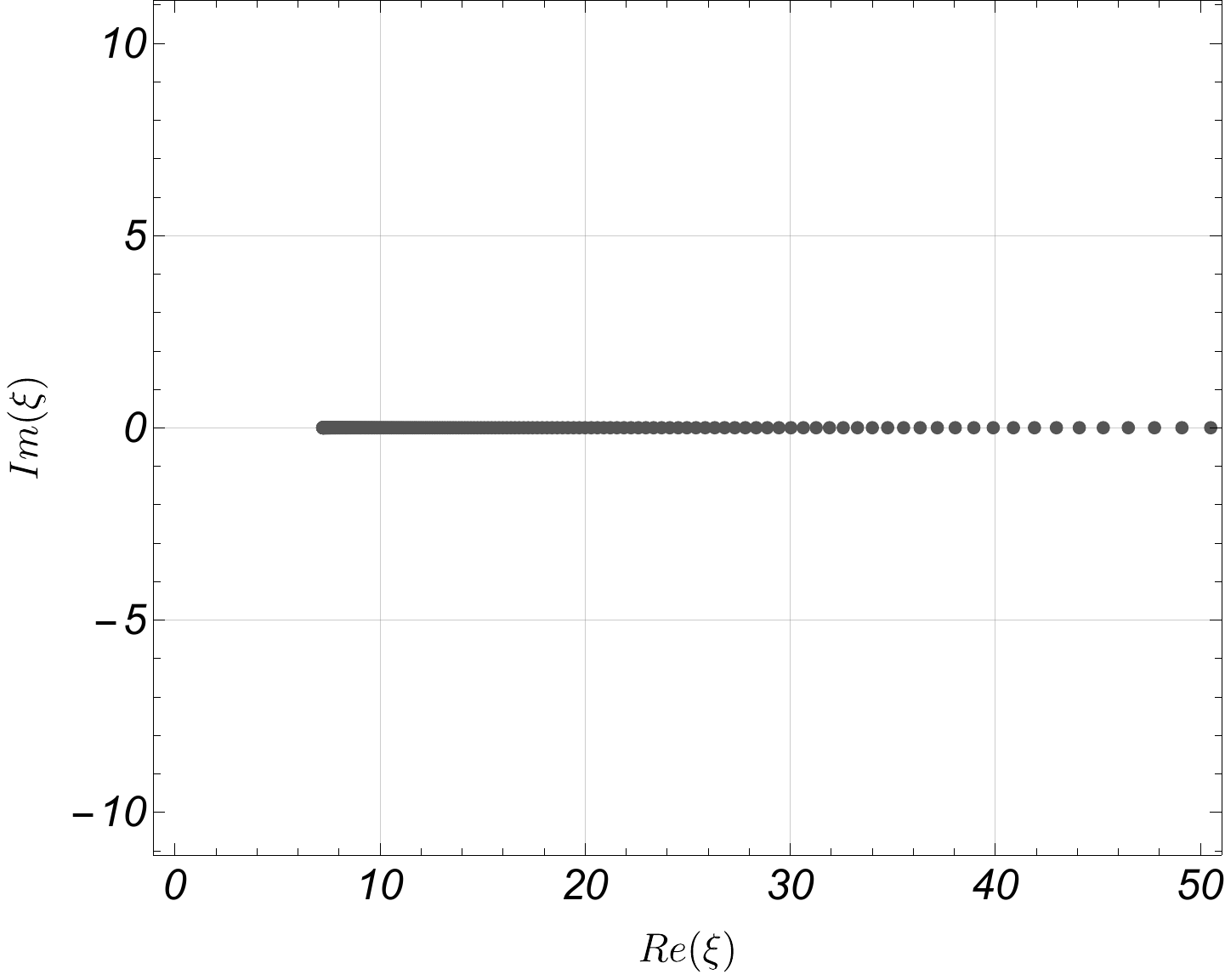}
\caption{The poles of the Borel transform \rf{eq:ABorel}.}
\label{fig:borelpadeMIS}
\end{center}
\end{figure}


\subsection{The attractor in HJSW hydrodynamics}


\begin{figure}[t]
\begin{center}
\includegraphics[width =.6\textwidth]{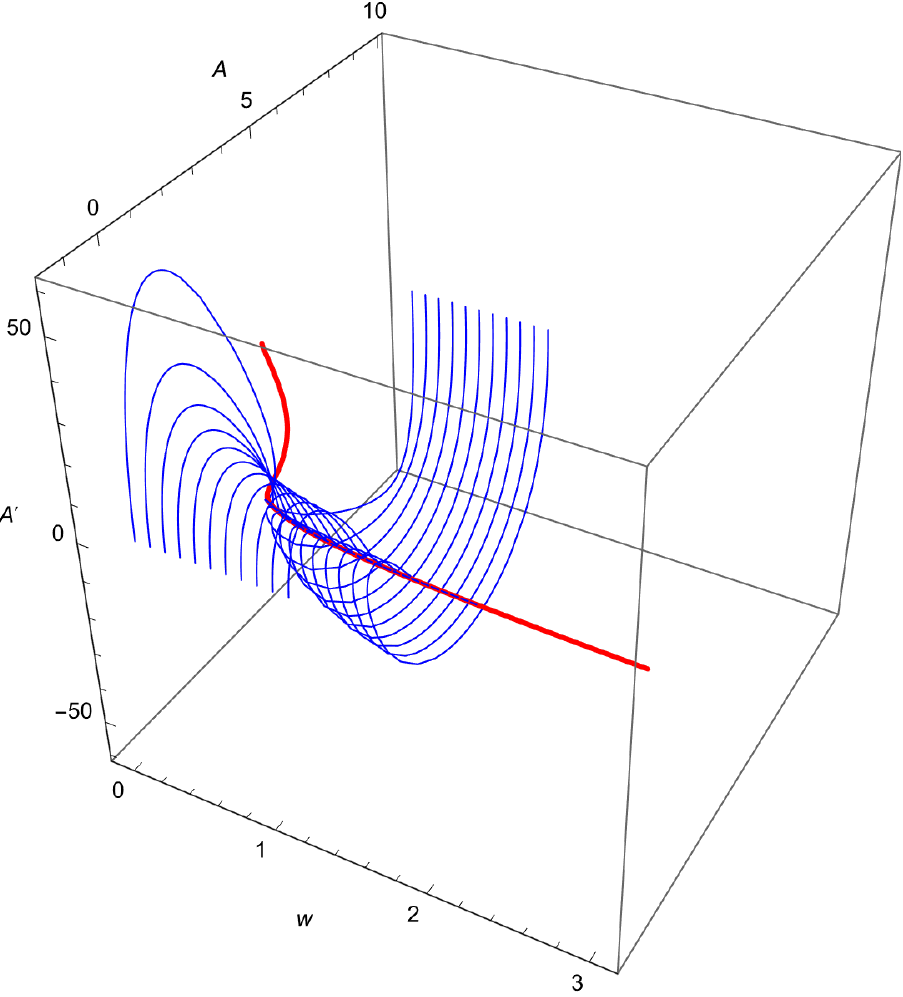}
\caption{The blue curves depict solutions whose initial
  conditions were set at several values of $w$ between $0.05$ and $0.3$. The red
  curve represents the attractor. The parameter values used when making the plot
were those for \symm\ and  $C_\sigma=0$.}
\label{fig:HJSWattractor}
\end{center}
\end{figure}


So far this Section has focused on the attractor of MIS theory, but the same ideas
can be applied to other hydrodynamic models discussed in \rfs{sec:MIS}. One
point of interest is that in such models one sometimes encounters
higher-dimensional phase spaces.  For example, this happens in the HJSW model
introduced in Ref.~\cite{Heller:2014wfa}, which leads to a second order equation
replacing \rf{eq:MISAeom}. In consequence, the full phase space is three dimensional.
Explicitly, this relaxation equation reads (see also Ref.~\cite{Florkowski:2017olj})
\bel{eq:vcp}
\alpha_1 \pa''+ \alpha_2 \, \pa'^2+\alpha_3 \, \pa'+12 \, \pa^3+\alpha_4 \, \pa^2+\alpha_5 \, \pa+\alpha_6 = 0,
\ee
where
\bel{vcp.coeffs}
\alpha_1 &=&  w^2 \, (\pa+12)^2,\nn\\
\alpha_2 &=& w^2 \, (\pa+12),\nn\\
\alpha_3 &=& 12 \,  w \,  (\pa+12) \, (\pa+3 \, w \,  \Omega_I),\nn\\
\alpha_4 &=& 48 \,  (3 \,  w\, \Omega_I - 1),\nn\\
\alpha_5 &=& 108 \, \left(- 4\,  C_\eta \, C_{\sigma} + 3 \, w^2 \, \Omega ^2\right),\nn\\
\alpha_6 &=& -864 \, C_\eta \, \left(- 2 \, C_{\sigma}+3\, w\, \Omega^2\right).
\ee
At early times
there is a unique power series solution regular at $w=0$:
\bel{eq:hjsw.smallw}
\pa(w) =  4 + \f{54 \, C_\eta \, |\Omega|^2- 48 \, \Omega_I}{20 - 9  \, C_\eta \,
  C_{\sigma}} \, w + \dots
\ee
This is the attractor, as seen in \rff{fig:HJSWattractor}, where this curve is
plotted in the full phase space. 

At large $w$, the gradient expansion takes the form
\bel{eq:hjsw.largew}
\pa(w) =  \f{8 C_\eta}{w} +
\f{16 C_\eta ( 2 \Omega_I- C_{\sigma} )}{3 |\Omega|^2 w^2} + \dots
\ee
As expected, the first term captures the shear viscosity, as in MIS theory. The
higher order terms differ from the corresponding expansion given in
Eq.~\eqref{eq:MISlatetime}, \eqref{eq:MISgradevals}.  Similarly to the case of
MIS theory, this series has vanishing radius of
convergence~\cite{Aniceto:2015mto}.  One can use this expansion in conjunction
with Borel summation to obtain a useful estimate of the attractor. We will
return to this point in \rfs{sec:attractorHolo}.

\subsection{Attractors in general frame models}


\begin{figure}[t]
\begin{center}
\includegraphics[width =.6\textwidth]{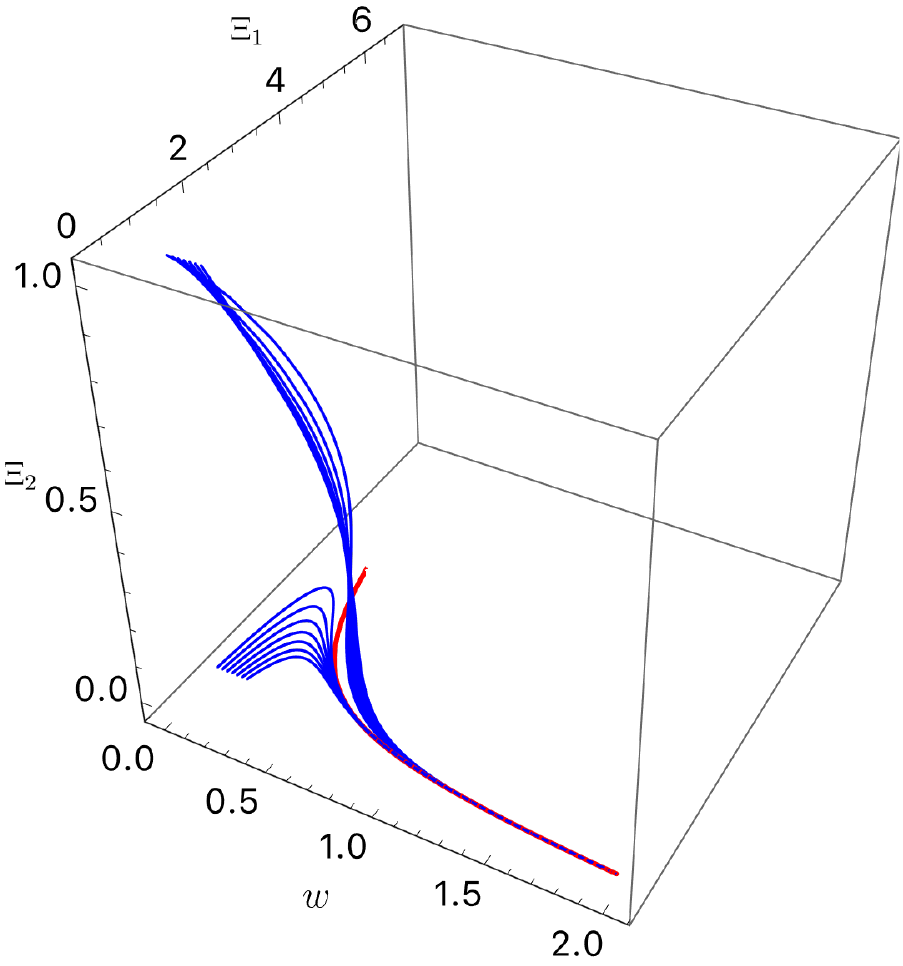}
\caption{The blue curves depict solutions whose initial conditions were set at
several values of $w$ between $0.05$ and $0.3$. The red curve represents the
attractor. The parameter values used when making the plot were $C_\eta = 0.08,
C_\tau=0.2, C_\varphi=0.01$.}
\label{fig:NSSattractor}
\end{center}
\end{figure}


Attractors have also been studied in hydrodynamic models where the Landau frame
condition has not been
imposed~\cite{Shokri:2020cxa,Noronha:2021syv,Pandya:2021ief,Rocha:2022ind}.
Here we wish to highlight an interesting example of an attractor within a
$3$-dimensional phase space which arises in the general-frame MIS theory of
Ref.~\cite{Noronha:2021syv} (see \rfs{sec:gf}). Imposing the symmetries of
Bjorken flow implies that the energy-momentum tensor contains three functions of
proper time (instead of two, as would be the case had the Landau frame condition
been imposed). This leads to a system of coupled equations for two functions of
$w$, denoted by $\xa,\xb$: 
\be
\label{eq:nssA}
\frac{1}{12}(C_\tau - C_\varphi) w (\xa  + 12) \xa'
-\frac{3}{8} w \xa (\xb  - 4)
+\frac{(C_\tau - C_\varphi)}{3} \xa^2
-\frac{9}{2} w \xb - 12 C_\eta &=& 0,\\
\label{eq:nssB}
\f{1}{12} C_\tau w (\xa  + 12) \xb'
+ \f{1}{3}\xa ( C_\tau \xb+  C_\varphi) +
\f{3}{2} w \xb  &=& 0,
\ee
where the prime denotes differentiation with respect to $w$, and $C_\varphi,
C_\tau$ are dimensionless constants. The functions $\xa, \xb$ replace the
pressure anisotropy in parametrising the dissipative part of the general-frame
energy-momentum tensor and are defined in Ref.~\cite{Noronha:2021syv}. In the
special case where $C_\varphi=0$ these equations admit a solution with
$\xb\equiv 0$ and then \rf{eq:nssA} reduces to the equation satisfied by the
pressure anisotropy in MIS theory~\rf{eq:MISAeom}. The late time asymptotics of
solutions are $\xa \sim 8 C_\eta/w$ and $\xb \sim -63 C_\eta C_\varphi/27 w^2$
for all initial conditions.  

The phase space of solutions in this model is
three-dimensional rather than two-dimensional as in MIS theory. As in the examples
discussed earlier, there is a unique
solution regular at $w=0$ which acts as an attractor, as seen in
\rff{fig:NSSattractor}.

%% file: kinetic.tex
\section{Attractors from Kinetic Theory}
\label{sec:attractorKT}

The discovery of attractors in hydrodynamic models can be viewed as a strong
indication that similar phenomena should occur also in more elaborate
microscopic theories. This supposition has by now been confirmed in numerous
studies discussed further in this review.  The simplest class of models, whose
complexity goes beyond what is discussed in the previous Section, are models of
kinetic theory, where attractors have been identified and studied in many interesting cases~\cite{Heller:2016rtz,Blaizot:2017ucy,Romatschke:2017vte,Behtash:2017wqg,Heller:2018qvh,Behtash:2018moe,Strickland:2018ayk,Behtash:2019txb,Denicol:2019lio,Strickland:2019hff,Behtash:2019qtk,Kurkela:2019set,Chattopadhyay:2019jqj,Almaalol:2020rnu,Heller:2020anv,McNelis:2020jrn,Kamata:2020mka,Blaizot:2020gql,Alalawi:2022pmg,Jaiswal:2022udf}. 

Kinetic theory is 
based on the classical notion of a single particle distribution
function $f(x, p)$ obeying the Boltzmann equation
\begin{equation}
    p^\mu\partial_\mu f(x,p) = \mathcal{C}[f]~.
    \label{eq:Boltzmann}
\end{equation}
The collision kernel appearing on the right-hand side of \rf{eq:Boltzmann} can
in general be very complicated, since in principle it should account for all
scattering processes which can occur in a given theory. In practice, only a
subset is accounted for, or some other form of approximation has to be adopted
to capture essential features of the underlying microscopic theory. Here we will
review kinetic theory attractors assuming one of two options: the relaxation
time approximation (RTA)~\cite{Anderson:1974abc} and the Effective Kinetic
Theory for Quantum Chromodynamics  (EKT)~\cite{Arnold:2002zm}.

\subsection{Boost invariant flow in RTA}

A significant simplification, which has been the subject of numerous studies 
is the relaxation time approximation, 
where the collision kernel in~\rf{eq:Boltzmann} is replaced by 
\begin{equation}
    \mathcal{C}[f]=p^\mu u_\mu \frac{f-f_{\rm eq}}{\tau_R}~.
    \label{eq:C_RTA}
\end{equation}
Here $\tau_R$ is a momentum-independent relaxation time and $f_{\rm eq} =
\exp\left(-\frac{p_\mu u^\mu}{T}\right)$ is the equilibrium distribution
function. The resulting equation is linear in $f(x,p)$ and is much easier to
work with.  Recently, this ansatz has been generalised in various
ways~\cite{Rocha:2021zcw,Rocha:2021lze,Dash:2021ibx,Rocha:2022ind,Denicol:2022bsq}. 

The Boltzmann equation in the RTA applied to Bjorken flow is a
quasi-analytically solvable problem \cite{Baym:1984np,Florkowski:2013lza} which
provides a very useful environment for testing ideas of nonequilibrium dynamics. 
Since the one particle distribution function $f(x,p)$ is a scalar, boost
invariance implies that it may depend only on variables invariant under
longitudinal boosts: 
$\tau$, $p_T$, and $W=t p_L- zE$, 
where $E\equiv p_0=\sqrt{p_T^2+p_L^2+m^2}$ is the particle's energy~\cite{Bialas:1984wv,Bialas:1987en}
\footnote{The boost-invariance of $W$ is a consequence of the transformation law 
$(E,p_L)\mapsto(E\cosh(y)-p_L\sinh(y),p_L\cosh(y)-E\sinh(y))$, and analogously for $(t,z)$.}.
With the help of
$W$ one can define $ v(p_T,W,\tau)=Et-p_L z =\sqrt{W^2+(p_T^2+m^2)\tau^2}~,$
which allows us to express energy and longitudinal 
momentum of particles of mass $m$ in terms of boost-invariant variables
\begin{equation}
    E=\frac{v t+W z}{\tau^2}~, \qquad 
    p_L=\frac{W t+ vz}{\tau^2}~.
\end{equation}
In terms of $\tau$, $W$ and $v$ we can write 
$p^\mu\partial_\mu f=\frac{v}{\tau}\partial_\tau f$,
$p_\mu u^\mu =\frac{v}{\tau}$,
and the boost-invariant Boltzmann equation in the RTA takes the form
\cite{Strickland:2018ayk,Baym:1984np,Florkowski:2013lza}
\begin{equation}
    \partial_\tau f(\tau,W,p_T)=\frac{f_{\rm eq}(\tau,W,p_T)-f(\tau,W,p_T)}{\tau_R}~.
    \label{eq:Boltzmann_RTA}
\end{equation}
The equilibrium distribution function is explicitly given by
\begin{equation}
  \label{biedf}
    f_{\rm eq}(\tau,W,p_T)=\exp\left(-\beta u_\mu p^\mu\right)
    =\exp\left(-\frac{\sqrt{W^2+p_T^2 \tau^2}}{T(\tau) \tau}\right)~,
\end{equation}
where we have set $m=0$, since for the time being we will concentrate on models
respecting conformal symmetry.

In order to obtain a closed system of equations one needs a way to determine the
effective temperature $T(\tau)$ appearing in \rf{biedf}. This can be achieved by
imposing the Landau matching condition, which states that local energy density
determined by the function $f(\tau,W,p_T)$ should be equal to the equilibrium
configuration with temperature $T(\tau)$.
In order to do that in a Lorentz invariant way one uses the measure  
\begin{equation}
    dP = \frac{d^4p}{(2\pi)^4}2\pi\delta(p^2)2\theta(p^0)=\frac{dp_L}{(2\pi)^3p^0}d^2p_T=\frac{dW d^2p_T}{(2\pi)^3v}~,
\end{equation}
and the desired matching condition is expressed as
\begin{equation}
    \edens(\tau)= \int dP(p_\mu u^\mu)^2 f(\tau,W,p_T) = \frac{3T(\tau)^4}{\pi^2}~.
    \label{eq:LandauMatching}
\end{equation}
A beautiful fact of life is that Eq.~(\ref{eq:Boltzmann_RTA}) admits the general
solution~\cite{Baym:1984np,Florkowski:2013lza} 
\begin{equation}
    f(\tau,W,p_T)=D(\tau,\tau_0)f_0(W,p_T)+\int_{\tau_0}^\tau
    \frac{d\tau'}{\tau_R(\tau')} D(\tau,\tau')f_{\rm eq}(\tau',W,p_T)~,
    \label{eq:RTA_solution}
\end{equation}
where $f_0(W,p_T)$ is the initial distribution function at $\tau=\tau_0$, and $D(\tau_2,\tau_1)$ is 
given by 
\begin{equation}
    D(\tau_2,\tau_1)=\exp\left[-\int_{\tau_1}^{\tau_2}\frac{dt}{\tau_R(t)}\right]~.
\end{equation}
The first term in Eq.~(\ref{eq:RTA_solution}) expresses free streaming, which
dominates at early times, while the second term is captures 
relaxation toward local equilibrium, which is controlled by $\tau_R$.

\subsection{The gradient expansion}

The Boltzmann equation in the RTA \rf{eq:Boltzmann_RTA} can be used to calculate
the distribution function in the gradient expansion. The most direct way to
proceed is to solve it iteratively starting with the equilibrium distribution,
thus implementing the Chapman-Enskog expansion (see
e.g.~\cite{DeGroot:1980dk,Jaiswal:2013npa}). One can then calculate the late
proper time expansion of the effective temperature using \rf{eq:LandauMatching}
and translate it into a series for the pressure anisotropy, which can be
written in the form of \rf{eq:MISlatetime}, with the leading coefficients given
by~\cite{Heller:2016rtz} 
\begin{equation}
    a_1=\f{8}{5}\ \gamma,\quad
    a_2=\f{32}{105}\ \gamma^2, \quad
    a_3 = - \f{416}{525} \, \gamma^3~.
\end{equation}
This can be matched to the gradient expansion of any hydrodynamic
model~\cite{Florkowski:2016zsi}. Depending on the choice of model, one or more terms may be
matched. In the case of MIS theory, a comparison with \rf{eq:MISgradevals} shows
that to match RTA kinetic theory one needs $C_\eta = \gamma/5$. 

The large order behaviour of the gradient expansion reveals a nonhydrodynamic
mode with the expected relaxation time, but the results are actually much more
complex, because of the wealth of possible initial conditions in kinetic theory, where
the initial state is specified by the distribution function at some initial
time. This will not be discussed further here, but some details can be found
in Refs.~\cite{Heller:2016rtz,Heller:2018qvh}.  Note also that the spectrum of
nonhydrodynamic modes in RTA kinetic theory is very different from that of
hydrodynamic models~\cite{Romatschke:2015gic}.

\subsection{The initial value problem}

The additional input needed to evaluate \rf{eq:RTA_solution} is an initial
condition. An important example, used below, is the  
Romatschke-Strickland parametrisation \cite{Romatschke:2003ms}
\begin{equation}
    f_0(W,p_T)=
    \exp\left[-\frac{\sqrt{(p\cdot u)^2+\xi_0(z\cdot p)^2}}{\Lambda_0}\right]
    =
    \exp\left[-\frac{\sqrt{(1+\xi_0)W^2+p_T^2 \tau^2_0}}{\Lambda_0\tau_0}\right]~,
    \label{eq:f0RTA}
\end{equation}
where $-1<\xi_0<\infty$ measures initial momentum space anisotropy,
$z_\mu=(\frac{z}{\tau},0,0,\frac{t}{\tau}),$ and $\Lambda_0$ determines the
characteristic energy scale. Using this form, one can explicitly evaluate the
initial energy density
\begin{equation}
    \edens(\tau_0)=\int dP (p\cdot u)^2 f_0(\tau_0,W,p_T)=
    \frac{3T_0^4}{\pi^2}\frac{\mathcal{H}(\frac{\alpha_0\tau_0}{\tau_0})}{\mathcal{H}(\alpha_0)}~,
\end{equation}
where $\alpha_0=(1+\xi_0)^{-\frac{1}{2}}$.  Using the Landau matching condition,
along with the general solution presented in Eq. (\ref{eq:RTA_solution}) one
then 
obtains an integral equation for the $T(\tau)$, i.e., the effective temperature
as a function of proper time $\tau$~\cite{Baym:1984np,Florkowski:2013lza}
\begin{equation}
    T(\tau)^4=D(\tau,\tau_0)T_0^4\frac{\mathcal{H}\left(\frac{\alpha_0\tau_0}{\tau}\right)}{\mathcal{H}(\alpha_0)}+\int_{\tau_0}^\tau \frac{d\tau'}{2\tau_{\rm eq}(\tau')}D(\tau,\tau')T(\tau')^4\mathcal{H}\left(\frac{\tau'}{\tau}\right)~,
    \label{eq:Tint}
\end{equation}
where 
\begin{equation}
    \mathcal{H}(y)=y\int_0^\pi \sin(\phi)\sqrt{y^2\cos^2(\phi)+\sin^2(\phi)} d\phi~.
\end{equation}
Equation (\ref{eq:Tint}) can be solved in an iterative manner, with some initial
temperature profile $T(\tau)$ \cite{Strickland:2018ayk}. Knowing the temperature
as a function of $\tau$ one can carry out the integral in \rf{eq:RTA_solution}
to obtain the full distribution function $f(\tau,W,p_T)$.

\subsection{Attracting behaviour of the distribution function}
\label{subsec:RTAattr}

To establish the existence of an attractor in kinetic theory one may adopt one of two
approaches. The first is to look at the moments of the distribution function
\cite{Blaizot:2017ucy,Kurkela:2019set,Strickland:2019hff,Almaalol:2020rnu} while the second looks for attractor behaviour of the distribution function itself
\cite{Strickland:2018ayk}.  In this subsection we will follow the latter
approach, while the former will be described in Sec.~\ref{subsec:ETK} in the
context of more realistic approximation to the collisional kernel~\cite{Almaalol:2020rnu}. 


\begin{figure}
\begin{center}
\includegraphics[height =.4\textheight]{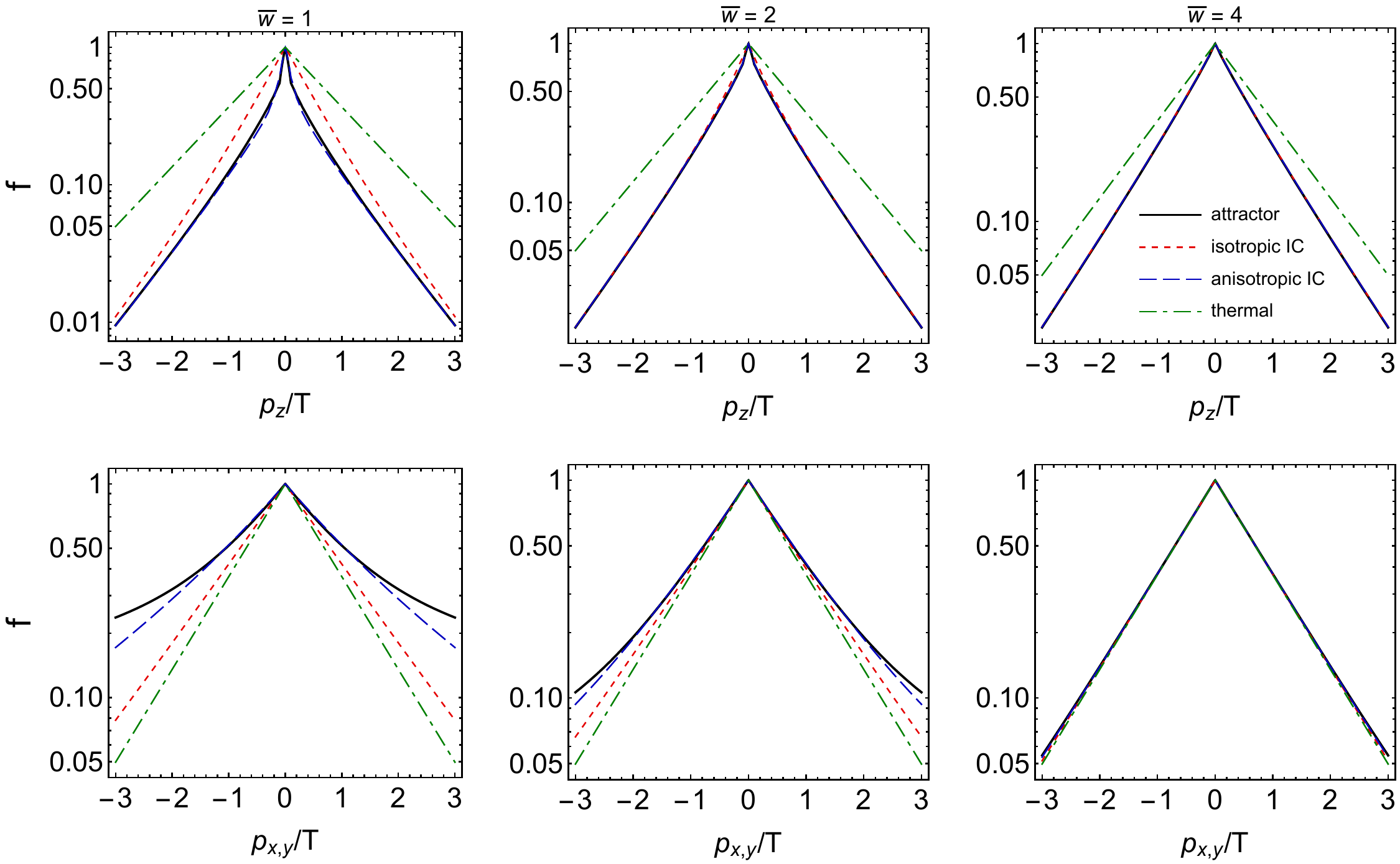}
\includegraphics[height =.4\textheight]{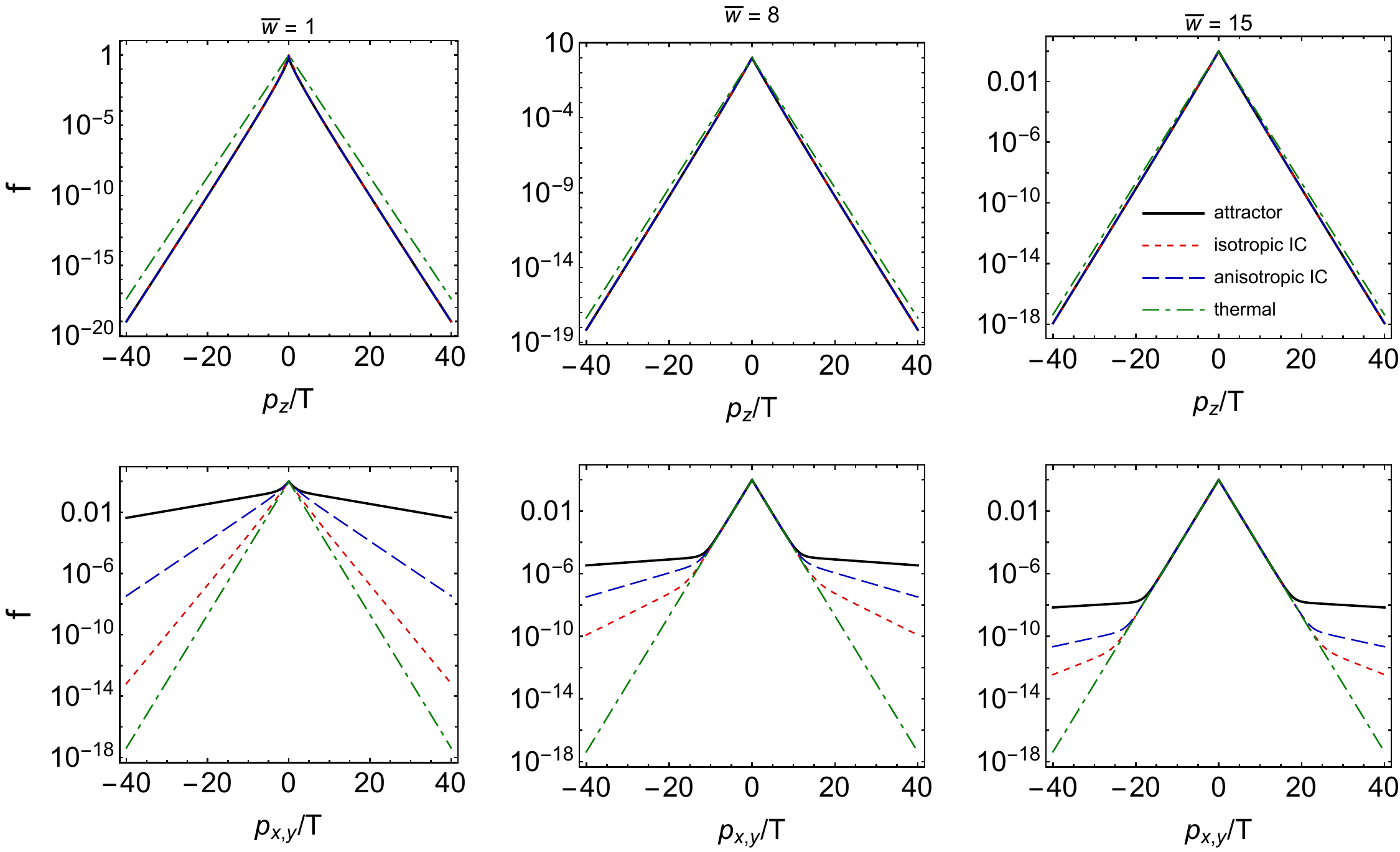}
\caption{Quantitative approach towards a hydrodynamic attractor
    $\alpha_0\simeq0.0025$ (solid black line) in the distribution function.
    Initial conditions are $T_0=1$~GeV at $\tau_0=0.1$~fm/c and
    $0.1\leq\alpha_0\leq1.5$ for dashed/dotted colour lines.   
The first and third rows show  $f(p_T=0,p_L)$,
while second and
fourth show $f(p_T,p_L=0)$.
First two columns are for lower
momenta $p_i/T\leq3$ while second two columns are for higher momenta
$p_i/T\leq40$ ($i=x,y,z$). Different columns represent different time
$\overline{w}$ instances marked in the top. The 
    scaled variable $\overline{w}=\tau/\tau_R = \frac{\tau
T(\tau)}{5\bar{\eta}}$, which differs by a constant factor from the variable
$w=\tau T(\tau)$ introduced earlier. 
Plots from Ref. \cite{Strickland:2018ayk}.}
\label{fig:fAttractor}
\end{center}
\end{figure}


In order to demonstrate that the full distribution function $f(\tau,W,p_T)$ has
an attractor one numerically solves the RTA Boltzmann equation
(\ref{eq:Boltzmann_RTA}) for the class of initial conditions parametrised by
Eq.~(\ref{eq:f0RTA}) and identifies the attractor by a "slow roll" approximation
\cite{Romatschke:2017vte,Strickland:2018ayk}:
\begin{equation}
    \left.\mathcal{A}'(\tau T)\right|_{\tau=\tau_0}\propto\left.\frac{\edens\partial_\tau\edens+\tau\edens\partial^2_\tau\edens-\tau(\partial_\tau\edens)^2}{\edens^2}\right|_{\tau=\tau_0}~,
\end{equation}
with $T(\tau_0)=1$~GeV and $\tau_0=0.1$~fm/c \cite{Strickland:2018ayk}.  Solving
$\mathcal{A}'|_{\tau=\tau_0}=0$ for $\alpha_0$ singles out the value of the initial anisotropy
parameter $\alpha_0\approx0.0025$, which determines the attractor solution.

The approach to the attractor for different anisotropic initial configurations
is shown in Fig.~\ref{fig:fAttractor}.  It is apparent that the infrared part of
the distribution function (the region close to $p=0$) approaches the attractor
earlier than the ultraviolet part, which is a manifestation of the ``bottom-up''
scenario characteristic of weakly coupled systems \cite{Baier:2000sb}.  The
approach to the attractor is also slower in the transverse direction ($p_z=0$)
than in the longitudinal direction ($p_T=0$).  Note also that in some momentum
regions the attractor is approached from below, while in others it is approached
from above.

\subsection{Weakly coupled QCD }
\label{subsec:ETK}

The discussion of previous section relied on the RTA collisional kernel.
An important question is whether similar results can be established within more
realistic models. Recently, this issue was addressed in the context of Effective
Kinetic Theory (ETK) of QCD~\cite{Arnold:2002zm}. The EKT Boltzmann equation
for a pure gluon system reads
\begin{equation}
    -\partial_\tau f +\frac{p_z}{\tau}\partial_{p_z} f =
    \mathcal{C}_{1\leftrightarrow2}[f]+\mathcal{C}_{2\leftrightarrow2}[f]~,
    \label{eq:BoltzmannEKT}
\end{equation}
where the inelastic $ \mathcal{C}_{1\leftrightarrow2}$ and elastic $
\mathcal{C}_{2\leftrightarrow2}$ collisional terms include physics of dynamical
screening and Landau-Pomaranchuk-Migdal damping. 
Although EKT is does not
account for the full complexity of QCD, for isotropic systems it incorporates
the leading $\alpha_s$-order description and has been extensively used to
address off-equilibrium perturbative QGP dynamics~\cite{Kurkela:2015qoa,Du:2020dvp,Du:2020zqg}.

To study the process of equilibration, Ref.~\cite{Almaalol:2020rnu} considers
the set of moments of the distribution functions defined by
\begin{equation}
    \mathcal{M}^{nm}(\tau):=\int\frac{d^3p}{(2\pi)^3}p^{n-1}p_z^{2m}f(\tau,\bf p)~,
\end{equation}
where $p=|{\bf p}|$. In terms of these, the energy density of a massless
particle gas is $\edens=\mathcal{M}^{20}$, particle density is
$n=\mathcal{M}^{10}$, while the longitudinal pressure 
reads $P_L=\mathcal{M}^{01}$.
The pressure anisotropy can be expressed in terms of these 
moments as
\begin{equation}
    \mathcal{A}=3\frac{\pT(\tau)-\pL(\tau)}{\edens(\tau)}=
    \frac{3}{2}-\frac{9}{2}\frac{\mathcal{M}^{01}(\tau)}{\mathcal{M}^{20}(\tau)}~.
\end{equation}
The distribution function can be obtained numerically by solving the Boltzmann
equation~(\ref{eq:BoltzmannEKT}) utilising the algorithm described in
Ref.~\cite{AbraaoYork:2014hbk,Kurkela:2015qoa}.

Two classes of initial conditions were 
considered in Ref.~\cite{Almaalol:2020rnu}.  The first one is given by a
spheroidally deformed thermal distribution function given by
\begin{equation}
  f_{0,\rm RS}(p)=\frac{1}{\exp\left(\frac{\sqrt{p^2+\xi_0^2p_z^2}}{\Lambda_0}\right)-1},
  \label{eq:f0RS}
\end{equation}
where $-1<\xi_0<\infty$, as in the RTA case, parametrises the initial momentum
anisotropy, while $\Lambda_0$ sets the initial energy scale.
The second group of initial conditions is given by the non-thermal CGC-motivated
distribution function, explicitly written as 
\begin{equation}
    f_{0, \rm CGC}(p)=\frac{2A}{\lambda_{\rm YM}}\frac{Q_0}{\sqrt{p^2+\xi_0^2p_z^2}}\exp\left(-\frac{2}{3}\frac{p^2+\xi_0^2 p_z^2}{Q_0^2}\right)~,
    \label{eq:f0CGC}
\end{equation}
where the scale $Q_0$ is related to the QCD saturation scale $Q_0=\langle
p_T\rangle_0\approx1.8Q_s$ \cite{Lappi:2011ju}.  Furthermore, $\lambda_{\rm
YM}=g_{\rm YM}^2N_c$  is the 't Hooft coupling.  The normalisation constant $A$
is fixed by matching the initial energy density with the predictions of
classical Yang-Mills theory $\tau_0\edens_0=0.358\nu_{\rm
eff}\frac{Q_s^3}{\lambda_{\rm YM}}$ \cite{Lappi:2006hq}. For both sets of
initial conditions, the scales $\Lambda_0$ and $Q_0$ play a  role similar to the
temperature in a thermal distribution, i.e., they determine which portion of
momentum space is occupied. The fact that distribution in \rf{eq:f0CGC} is
inversely proportional to $\lambda_{\rm YM}$ reflects the overpopulation of
gluons determined by multiple low energy scatterings at initial times. 

\begin{figure}
\begin{center}
\includegraphics[height =.17\textheight]{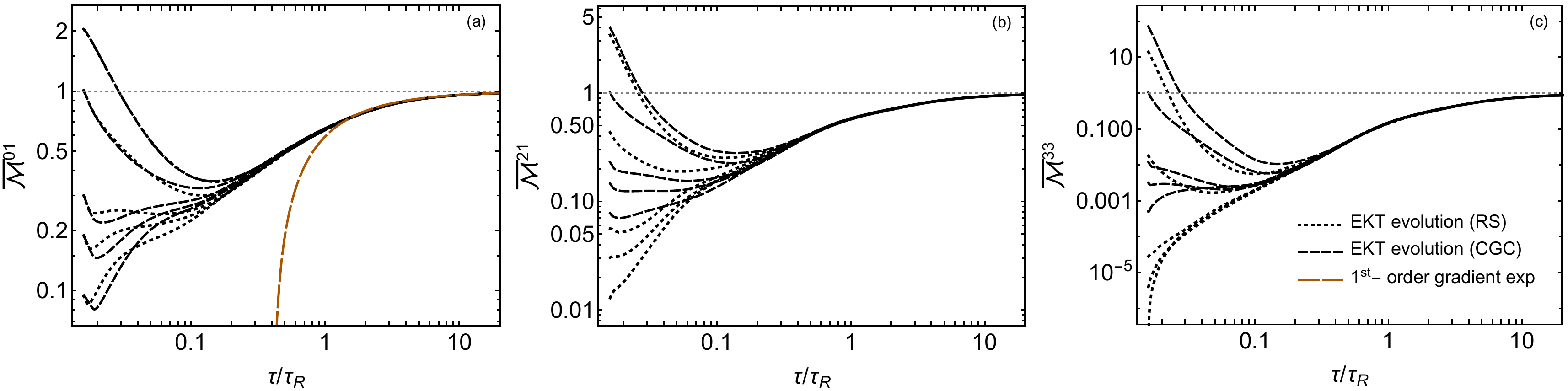}
\includegraphics[height =.17\textheight]{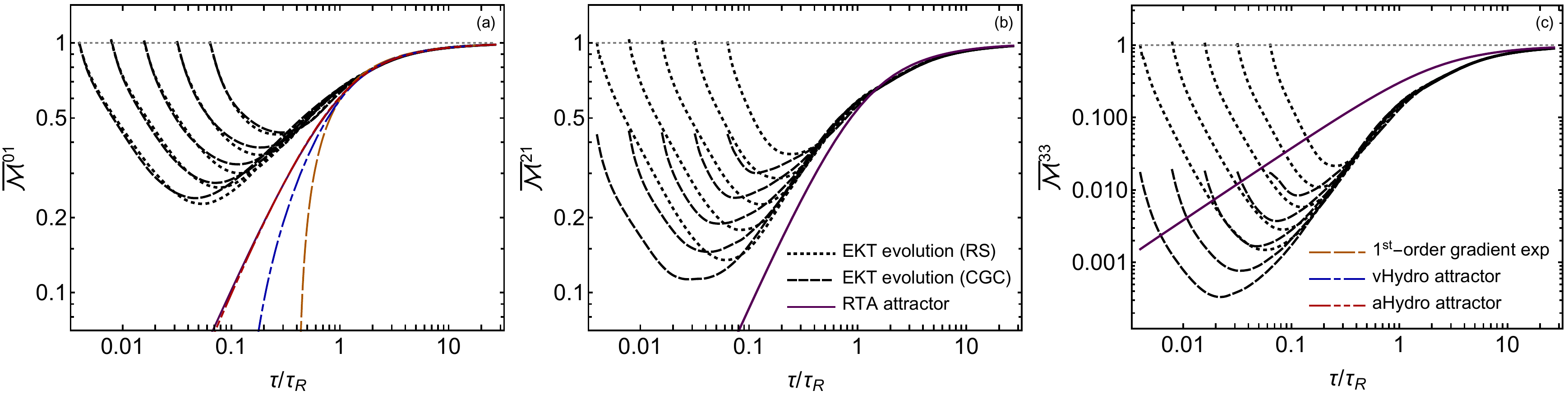}
\caption{Evolution of the scaled moments 
    $\overline{\mathcal{M}}^{nm}(\tau)=
    \mathcal{M}^{nm}(\tau)/\mathcal{M}^{nm}_{\rm eq}(\tau)$
computed as functions of rescaled time 
    $\overline{w}=\tau/\tau_R = \frac{\tau T(\tau)}{5\bar{\eta}}$ 
    for various initial conditions. Upper panel: fixed
    initial time different initial momentum space anisotropy. Lower panel:
    different initial times with fixed initial momentum anisotropy. The value
    the 't Hooft coupling used here is $\lambda_{\rm YM}=10$, which corresponds
to shear viscosity $\eta/s\approx0.63$ \cite{Arnold:2003zc,Keegan:2015avk}.  
The plots are taken from Ref.~\cite{Almaalol:2020rnu}.}
\label{fig:EKTattractor}
\end{center}
\end{figure}

The plots in Fig.~\ref{fig:EKTattractor} show the evolution of three sample moments
with different initial conditions,
parametrised by Eq.~(\ref{eq:f0RS}) and \rf{eq:f0CGC}.  
As 
seen in the upper panel of Fig. \ref{fig:EKTattractor}, all sampled initial
conditions merge into one universal line, the hydrodynamic attractor, before
they are well approximated by the viscous hydrodynamics. This happens on
the time scale $\tau/\tau_R\sim0.5$ common for all three moments of the
distribution function. Since this result holds also for higher moments, it is a
strong indication that, similarly to the RTA case, the attractor is present in the full
one particle distribution function \cite{Alalawi:2020zbx}.


The plots in the lower panel of Fig.~\ref{fig:EKTattractor} show evolution of
moments initialised at successively smaller initial times $\tau_0$.  As apparent
from the figure, earlier initialisation leads to faster decay to the attractor,
suggesting a scaling dependence on $\tau_0$ at early times.  At late times, both
RS and CGC initial conditions follow the same attractor, showing that details of
hydrodynamic evolution are insensitive not only to the initial pressure
anisotropy but also to microscopic features such as momentum distribution and
initial occupancy. While true at late times, this need not be the case
at very early times where different models predict different attractors, as also
indicated in Fig.~\ref{fig:EKTattractor}. This will be discussed further in
\rfs{sec:prehydro}.

%% file: sym.tex
\section{Attractor behaviour through AdS/CFT} 
\label{sec:attractorHolo}

In this Section we review studies of equilibration of Bjorken flow in \symm\
theory, which are possible to carry out in the strong coupling limit by virtue
of the AdS/CFT correspondence \cite{Maldacena:1997re,Witten:1998qj}. Excellent
reviews of the applications of holographic methods to heavy ion collisions
include
Refs.~\cite{DeWolfe:2013cua,Gubser:2010nc,Casalderrey-Solana:2011dxg,vanderSchee:2014qwa};
applications to Bjorken flow are reviewed in
Refs.~\cite{Janik:2010we,Florkowski:2017olj,Berges:2020fwq,Soloviev:2021lhs}. In
this Section we will therefore refrain from discussing the techniques involved,
our focus being on the results of such calculations and their interpretation in
terms of hydrodynamic attractors. 

\subsection{Thermal states in AdS/CFT}

The basic fact which lies at the heart of applying AdS/CFT to nonequilibrium
physics is that the equilibrium state of \sym\ supersymmetric Yang-Mills plasma
in $4$-dimensional Minkowski space corresponds to a black hole in
asymptotically-$AdS_5$ space. This object is often referred to as a black brane
due to the fact that the horizon of the gravitational solution is planar rather
than spherical.  The black brane temperature $T$ is equal to the temperature of
the plasma.  The duality map (sometimes referred to as the holographic
dictionary) leads to a formula for the energy density of the plasma: 
\begin{equation}
    \edens = \frac{3\pi^2}{8}N_c^2 T^4.
    \label{eq:EHEOS}
\end{equation}
Up to a factor of $3/4$, which is interpreted as
the effect of strong coupling, this coincides with the result for the energy
density of quanta comprising the plasma in the absence of interactions.

The interpretation of the equilibrium state in terms of a black hole immediately
suggests that its perturbations should correspond to perturbations of the black
hole. Their spectrum can therefore be computed by standard methods developed in
studies of general relativity, adapted to the specific challenges of
asymptotically AdS spaces. At the linearised level such perturbations are known
as quasinormal modes (QNM) of the black hole, and they describe damped
oscillations of the black hole horizon. Their  complex frequencies naturally
fall into one of two categories: a finite number of hydrodynamic modes whose
damping rate is proportional to the wave vector, and an infinite series of
nonhydrodynamic modes which are damped even for arbitrarily long-wavelength
perturbations.  This matches directly the picture of perturbations of
equilibrium expected on the basis of linear response theory, as reviewed in
\rfs{sec:linresponse}. 

The process of equilibration can also be described analytically at the nonlinear
level.  This was pioneered in
Refs.~\cite{Janik:2005zt,Janik:2006ft,Heller:2007qt,Heller:2008mb} by studies of
the asymptotic late-time expansion of Bjorken flow (reviewed in
\rfs{sec:HoloAsym} below).  These results were subsequently generalised to
generic near-equilibrium states~\cite{Bhattacharyya:2007vjd}, which are mapped
to perturbed black objects in asymptotically $AdS$ spaces described in a
gradient expansion analogous to what is done in hydrodynamics.  This connection
is sometimes referred to as the fluid-gravity correspondence. It has also led to
an interpretation of off-equilibrium entropy in field theory in terms of
slowly-evolving horizons in the dual gravitational
representation~\cite{Bhattacharyya:2008xc,Booth:2010kr,Booth:2011qy}.  These
analytic studies have provided a number of insights which were later applied to
numerical simulations and have greatly aided the interpretation of their
results.

\subsection{Numerical solutions and early time behaviour}

Numerical approaches to solving the initial value
problem in the gravitational representation of Bjorken flow and translating the
result into field theory
language have also been critically important~\cite{Chesler:2008hg,Chesler:2009cy,Heller:2011ju,Heller:2012je,Chesler:2013lia,Jankowski:2014lna}.
One of the first steps in such calculations is the selection of consistent
initial geometries.  This is a nontrivial task, as it requires satisfying the
constraints following from Einstein equations.  A basic result is that the
early-time behaviour of the energy density on the field theory side has the form
of a Taylor series with only even powers of the proper time~\cite{Beuf:2009cx}: 
\begin{equation}
    \label{eq:epsearly}
    \edens = \edens_0 + \edens_2\tau^2 + O(\tau^4)~.
\end{equation}
In Refs.~\cite{Heller:2011ju,Heller:2012je,Jankowski:2014lna} the initial conditions were set in
such a way that the leading coefficient $\edens_0\neq 0$. This corresponds to
the initial value of the pressure anisotropy $\pa(0)=6$. A number of such
solutions are plotted in \rff{fig:symnumerics}. It is apparent that the pressure
anisotropy reaches the hydrodynamic attractor while the system is still very
anisotropic. It is not clear however whether an expansion dominated regime
exists at early times. This is partly due to the oscillatory behaviour, which is
interpreted as a consequence of the rich spectrum of nonhydrodynamic modes whose
frequencies are not purely imaginary, in contrast to models of MIS hydrodynamics
or kinetic theory.  A possibly more significant issue is that the effective
phase space of the theory is multidimensional. While two real numbers suffice to
specify the initial data for the equations for Bjorken flow in MIS theory, in
the AdS/CFT calculation the initial data is specified by a function of the
radial (holographic) coordinate in the asymptotically-AdS space, which
defines the initial geometry.  If this were coarse-grained in some way, one
could represent the phase space as a having effectively a finite number of
dimensions, but there is no reason to believe that a two-dimensional truncation
would provide a reasonable
account. The plot of $\pa(w)$ should therefore be viewed as a projection from a
high-dimensional phase space and may obscure the picture at early times. 
An indication of this can be seen in \rff{fig:HJSWattractor} and
\ref{fig:NSSattractor}. 


\begin{figure}[t]
\begin{center}
    \includegraphics[width=.75\textwidth]{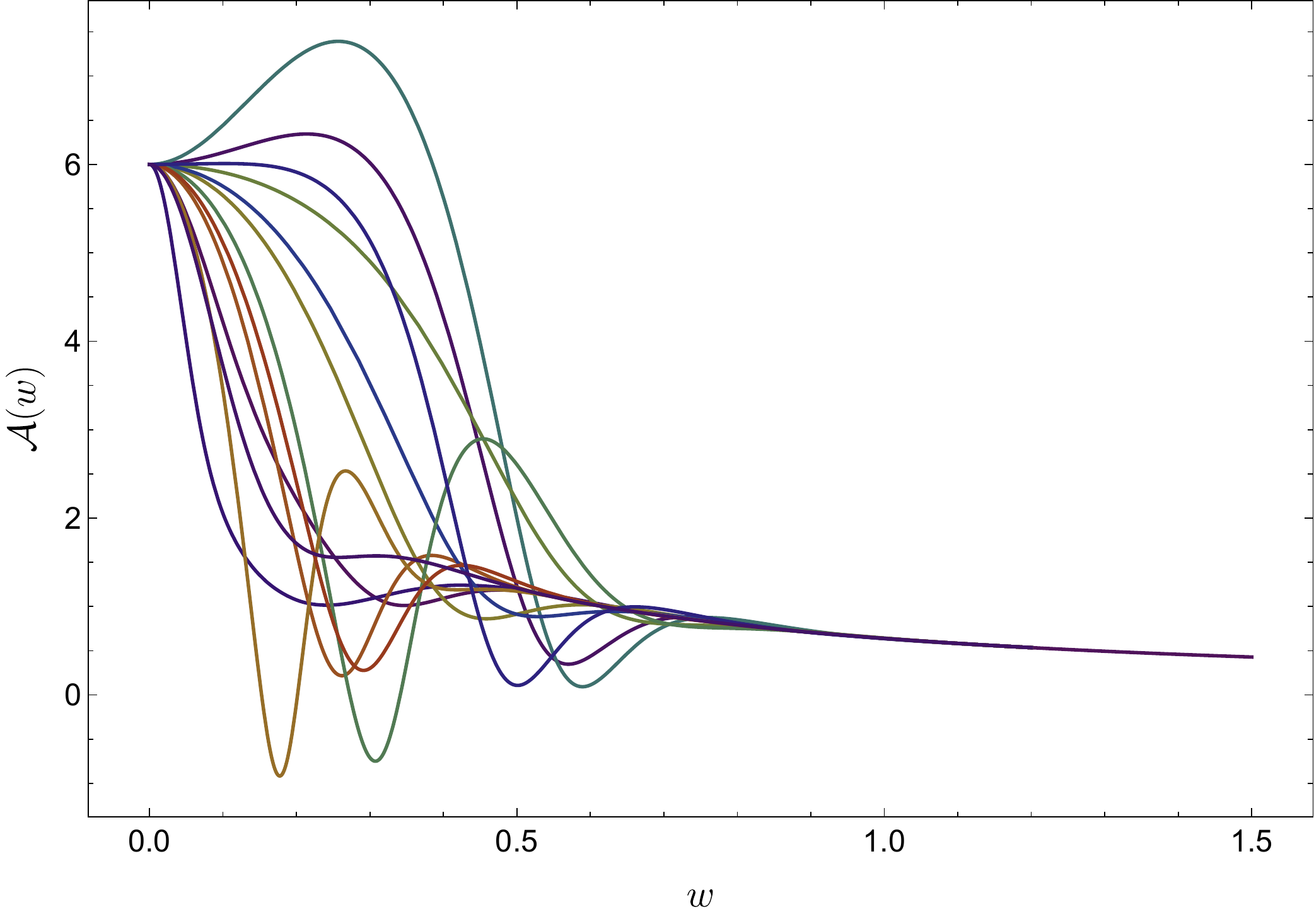}
\caption{
    Pressure anisotropy as a function of $w$ for
$\mathcal{N}=4$ SYM at strong coupling \cite{Jankowski:2015uva}.
}
\label{fig:symnumerics}
\end{center}
\end{figure}


In contrast with hydrodynamic models, where a regular solution at $w=0$ was a
natural candidate for an attractor, there is no such natural candidate here. In
Ref.~\cite{Romatschke:2017ejr} an attempt was made to find a physical argument
which would single out a special initial condition close to $w=0$. An
interesting feature of this proposed attractor is that it appears to be close to
free-streaming at early times, just as what is found in kinetic theory.  
A somewhat complementary approach to identifying the early-time attractor based
on Borel summation of the gradient expansion is also not conclusive, since it
looses predictivity at very early times, as  
reviewed in \rfs{sec:BorelAtr} below.

\begin{figure}
\centering
\includegraphics[width =.75\textwidth]{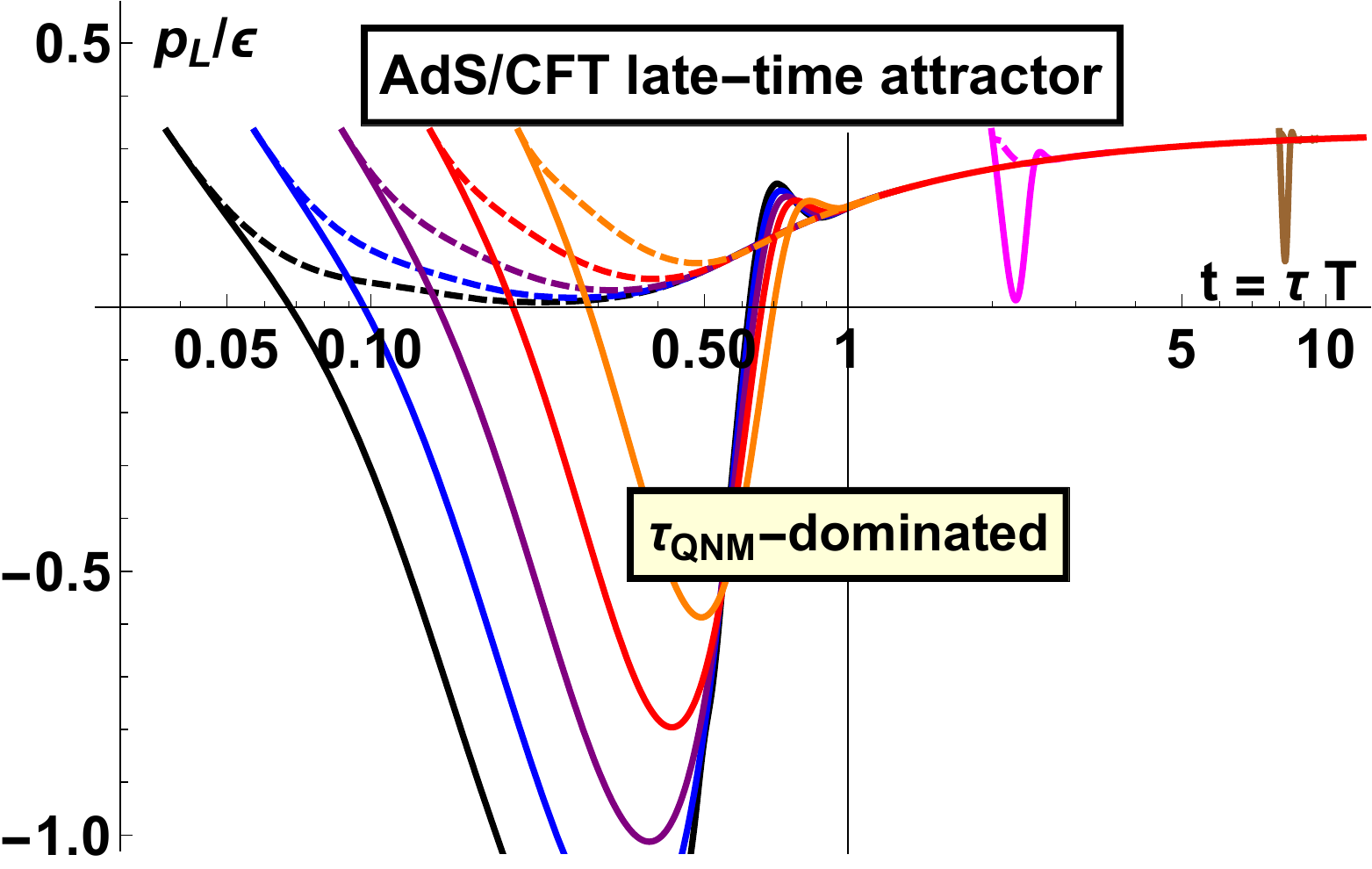}
\caption{Evolution of various initial configurations in holography. 
Note that the pressure anisotropy is connected to the quantity in the plot through
the relation 
$\pa=-\frac{3}{2}+\frac{9P_L}{2\epsilon}$.
Plot form Ref.~\cite{Kurkela:2019set}.
}
\label{fig:LateHolo}       
\end{figure}

The existence of an early-time attractor in this theory is of great interest,
and this question was revisited recently in Ref.~\cite{Kurkela:2019set}.  In
this paper the authors looked for the early, expansion-dominated phase and did
not find evidence for it.  The
evolution of the pressure anisotropy for a number of solutions is shown in
Fig.~\ref{fig:LateHolo}, where it is apparent that the approach to the attractor
is determined by the $1/T$ scale irrespective of the initialization time.

In summary, the status of the early-time attractor in the case of \symm\ is not entirely
clear at this time. It is expected to exist on the basis of general, kinematic arguments,
but it could also be that due to the strong coupling limit the expansion
dominated region is artificially smeared out and effectively only a late time
attractor exists.


\subsection{The large proper time expansion}
\label{sec:HoloAsym}

The emergence of fluid behaviour in boost-invariant \symm\ was first 
demonstrated in Refs.~\cite{Janik:2005zt} by studying the behaviour
of the energy density at large values of the proper time.  
The asymptotic behaviour of the energy density is given by
\begin{equation}
    \edens_{\rm hydro}(\tau)\sim \frac{\Lambda^4}{(\Lambda\tau)^{4/3}}{\sum_{n=0}^\infty \edens_n^{(0)}(\Lambda\tau)^{-2n/3}}~,
    \label{eq:epsHydro}
\end{equation}
where the energy scale $\Lambda$ reflects the initial conditions, as in other
cases of Bjorken flow, and $\edens_0^{(0)}=1$.  The expansion coefficients
$\edens_n^{(0)}, n>0$ can be determined using the AdS/CFT correspondence. The first
three subleading orders were calculated
analytically~\cite{Janik:2006ft,Heller:2008mb,Booth:2009ct}, and higher orders
numerically~\cite{Heller:2013fn,Aniceto:2018uik}. For $n\gg1$ these coefficients
diverge factorially, i.e.
\begin{equation}
    \edens_n^{(0)}\sim\frac{\Gamma(n+\beta_1)}{A_1^{n+\beta}}+{\rm h.c.}~
    \label{eq:hydro_diverge}
\end{equation}
where $A_1,\beta_1$ are complex constants; the singulant $A_1$ turns out
to be related to the lowest
nonhydrodynamic quasinormal mode frequency of the dual black hole
$\omega_1=3.1195-{\rm i}2.7467$ \cite{Kovtun:2005ev} by the relation 
$A_1={\rm i}\frac{3}{2}\omega_1$. This connection of large
order behaviour of the gradient expansion to the nonhydrodynamic modes is therefore fully analogous to
what was discussed in \rfs{sec:largeorders} in the case of MIS theory. 


\begin{figure}[t]
\centering
\includegraphics[height =.35\textheight]{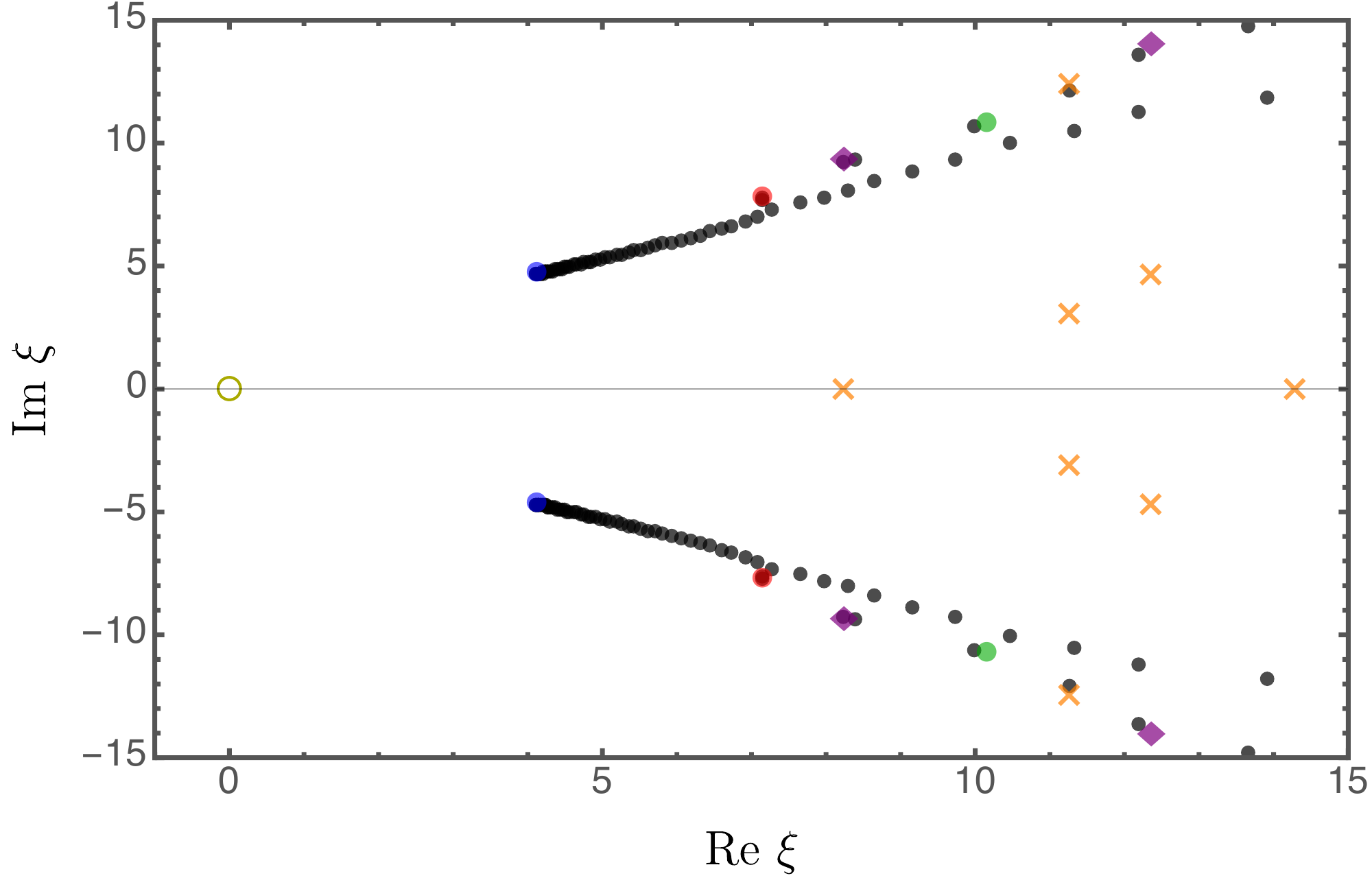}
\caption{Poles of the Borel-Pad\'e approximant $\mathrm{BP}_{189}\left[\edens_{\rm hydro}\right]$,
           in the complex $\xi-$plane.  We can see the appearance of the
           different fundamental sectors (shown as filled circles) as well as
           mixed sectors (shown as filled purple diamonds).  The predicted
           branch points for each sector are marked by colours:  $A_{1}$ and
           $\overline{A_{1}}$ (blue), $A_{2}$ and $\overline{A_{2}}$ (red),
           $A_{3}$ and $\overline {A_{3}}$ (green). This plot is taken from Ref.~\cite{Aniceto:2018uik}.
}
\label{fig:AdSBorel}       
\end{figure}


Equation~(\ref{eq:epsHydro}) cannot be the whole story, since initial states in
$AdS$ contain much more information than just the scale $\Lambda$. In fact, each
nonhydrodynamic mode of \symm\ plasma introduces an exponentially damped
transseries sector, 
with an independent transseries parameter which corresponds to a
piece of initial data.  Thus, the full expansion of $\edens(\tau)$ takes form of
multi-parameter transseries with infinitely many transseries
parameters~\cite{Aniceto:2018uik}: 
\begin{equation}
    \edens(\tau)\sim
    \underbrace{ \frac{\Lambda^4}{(\Lambda\tau)^{4/3}}{\sum_{n=0}^\infty
    \edens_n^{(0)}(\Lambda\tau)^{-2n/3}}}_{\edens_{\rm hydro}(\tau)} +
    \frac{\Lambda^4{\sigma_1}}{(\Lambda\tau)^{4/3}}{\sum_{n=0}^\infty
    \edens_n^{(1)}(\Lambda\tau)^{-2n/3}e^{-A_1(\Lambda\tau)^{2/3}}} +{\rm h.c.} +\cdots
    \label{eq:asymEps}
\end{equation}
For simplicity, only one nontrivial transseries sector is written explicitly in 
\rf{eq:asymEps}; it is the contribution of the least-damped, transient mode determined
by the complex quasinormal mode frequency $A_1$. The imaginary part of $A_1$
controls the exponential damping, while the real part determines the oscillation
frequency.  The full solution also includes supplementary sectors representing mutual
couplings between distinct modes.  This intricate structure is reflected in
\rff{fig:AdSBorel}, where the branch points correspond to quasinormal modes (as
well as their products)~\cite{Aniceto:2018uik}.

All the coefficients $\edens_n^{(k)}$ appearing in the transseries expansion in
\rf{eq:asymEps} can be determined using the AdS/CFT correspondence; many of them
have been calculated numerically for the most relevant sectors in
Ref.~\cite{Aniceto:2018uik}.  The original hydrodynamics series $\edens_n^{(0)}$
diverges factorially, as in \rf{eq:hydro_diverge}, and so do the series appearing
in each transseries sector. For instance, in the first sector one finds that for
$n\gg1$
\begin{equation}
    \edens_n^{(1)}\sim\f{\Gamma(n+\beta_2)}{A_2^{n+\beta_2}} + \mathrm{h.c.}
\end{equation}
where $A_2,\beta_2$ are complex constants; the singulant $A_2={\rm
i}\frac{3}{2}\omega_2$, where $\omega_2=5.1695-{\rm i}4.7636$ is second least
damped QNM in the sense that $|{\rm Im} \ A_1| < |{\rm Im} \ A_2|$.  This type of
relation between series appearing in different transseries sectors is a
manifestation of resurgence~\cite{Aniceto:2018bis,Mitschi:2016fxp}. The resurgence property of
the transseries implies that the complete structure of the nonhydrodynamic
sectors can be recovered from the original hydrodynamic gradient expansion.

\subsection{The attractor by Borel summation} 
\label{sec:BorelAtr}
 
As reviewed in \rfs{sec:attractors}, one can estimate the location
of the attractor by Borel summation of the hydrodynamic gradient expansion. To
do this in the present case one has to calculate the expansion of the pressure
anisotropy in powers of the $w$ variable using \rf{eq:epsHydro}.  This series is
also factorially divergent and the analytic continuation of the Borel transform
has the same pattern of singularities as seen in \rff{fig:AdSBorel}.  In
particular, there are no singularities on the real axis.  This means that the
Borel sum of this series does not suffer from the complex ambiguity encountered
in the case of MIS theory.  Of course the transseries contributions are still
present, and will become significant for values of $w$ sufficiently far from the
asymptotic region. Thus, at small values of $w$ this approach
looses predictability, since the no longer negligible exponentially suppressed
contributions eventually bring in dependence on the transseries coefficients and
it is not known which values correspond to the attractor. 


\begin{figure}[t]
\begin{center}
\includegraphics[width=.48\textwidth]{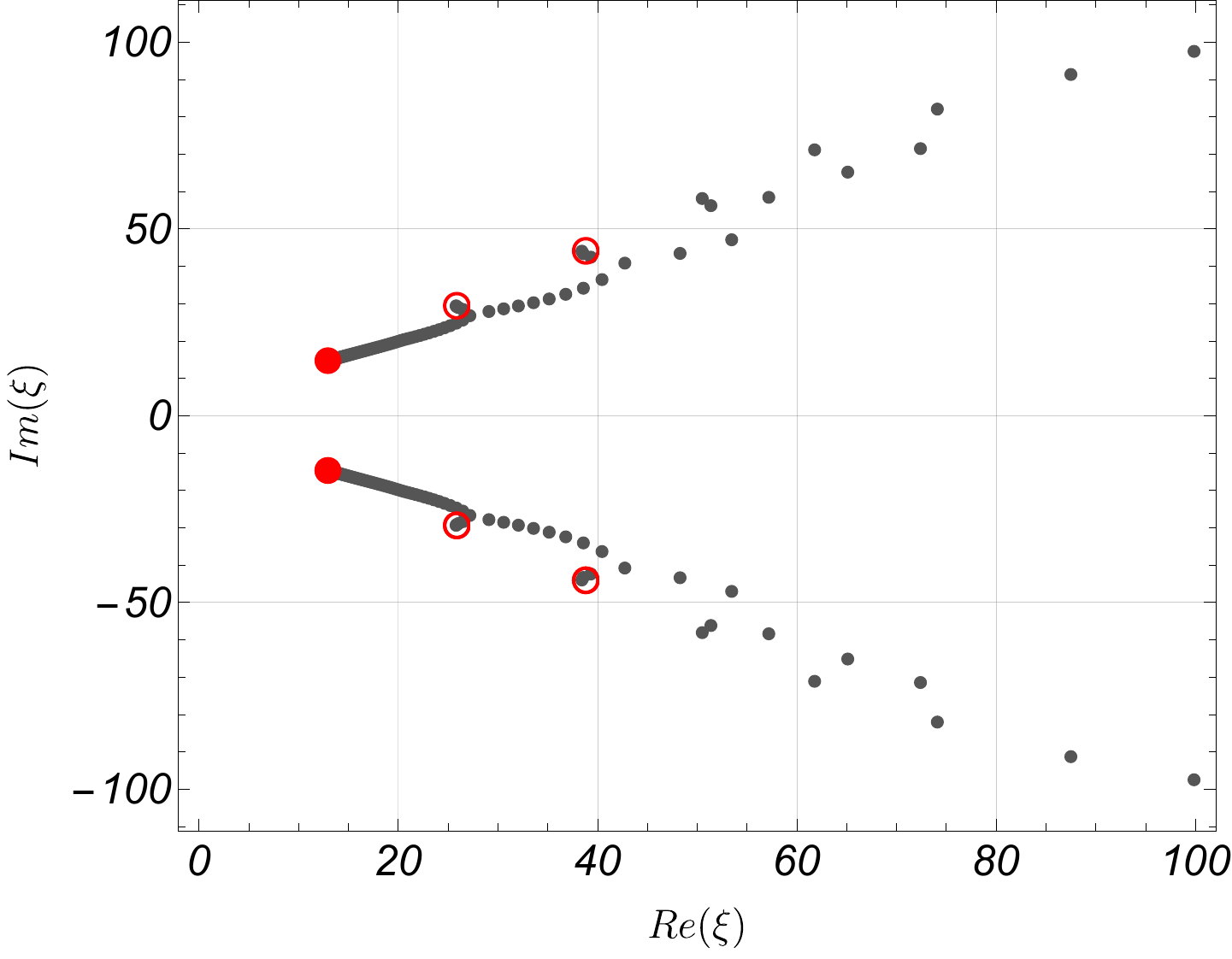}
\includegraphics[width=.48\textwidth]{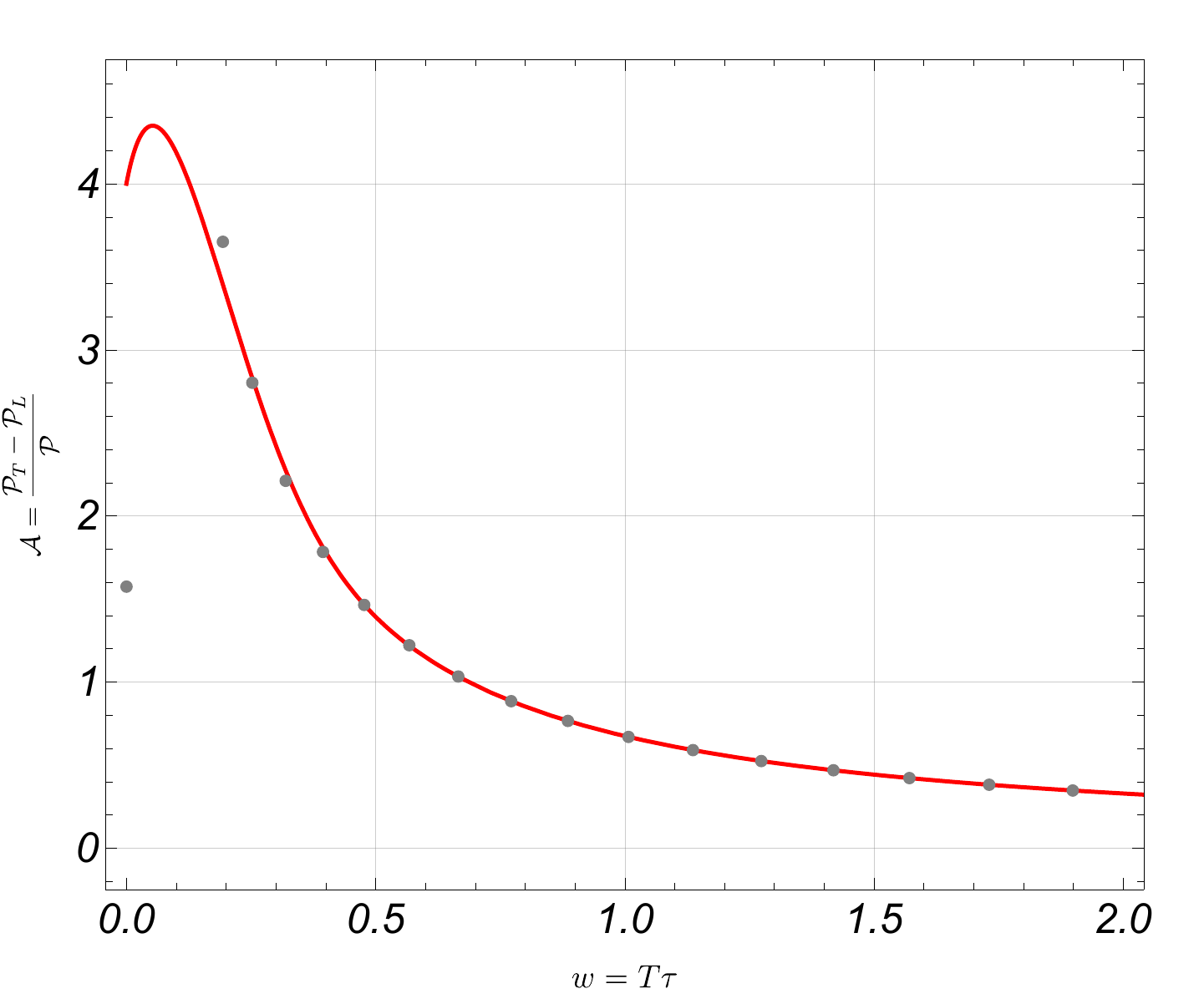}
\caption{
        Left panel: poles of the Borel transform of the gradient series of HJSW
    hydrodynamics; the red dots represent the known complex frequencies
    $\Omega$ of the
    nonhydrodynamic modes; the circles show the locations of their multiples. 
Right panel: the numerical attractor in HJSW hydrodynamics projected on the $(w,
\pa)$ plane (red curve) compared with the result of Borel summation of the
gradient expansion (black dots). }
\label{fig:borelpadeHJSW}
\end{center}
\end{figure}


This procedure has been in tested in the case of the HJSW
model~\cite{Heller:2014wfa} which has a similar singularity structure  and where
the true attractor is easily found numerically.  The relevant equation which is
satisfied by $\pa(w)$ is \rf{eq:vcp}, which can easily be solved in a power
series for large $w$; the leading terms appear in \rf{eq:hjsw.largew}.  This
series is factorially divergent and the singularities of the analytic
continuation of its Borel transform resemble those of \symm, as seen in the
right panel of \rff{fig:borelpadeHJSW}.  The series can then be Borel-resummed
at various values of $w$ following the procedure outlined in
\rfs{sec:largeorders}. We can interpret the outcome as an approximation to the
attractor. As seen from \rff{fig:borelpadeHJSW}, the result is in excellent
agreement with the numerical attractor down to $w\approx 0.4$. At earlier times
the transient transseries contributions are no longer negligible. Moreover,
these effects depend on the transseries parameters, and it is not known how to
determine their values so as to reproduce the attractor.  This would require
connecting the late-time transseries with the convergent series describing the
attractor \rf{eq:hjsw.smallw}. The analogous issue was recently discussed in MIS
theory using transasymptotic summation~\cite{Aniceto:2022dnm}, so perhaps this
point could be addressed in the future. 


\begin{figure}
\begin{center}
\includegraphics[width=.82\textwidth]{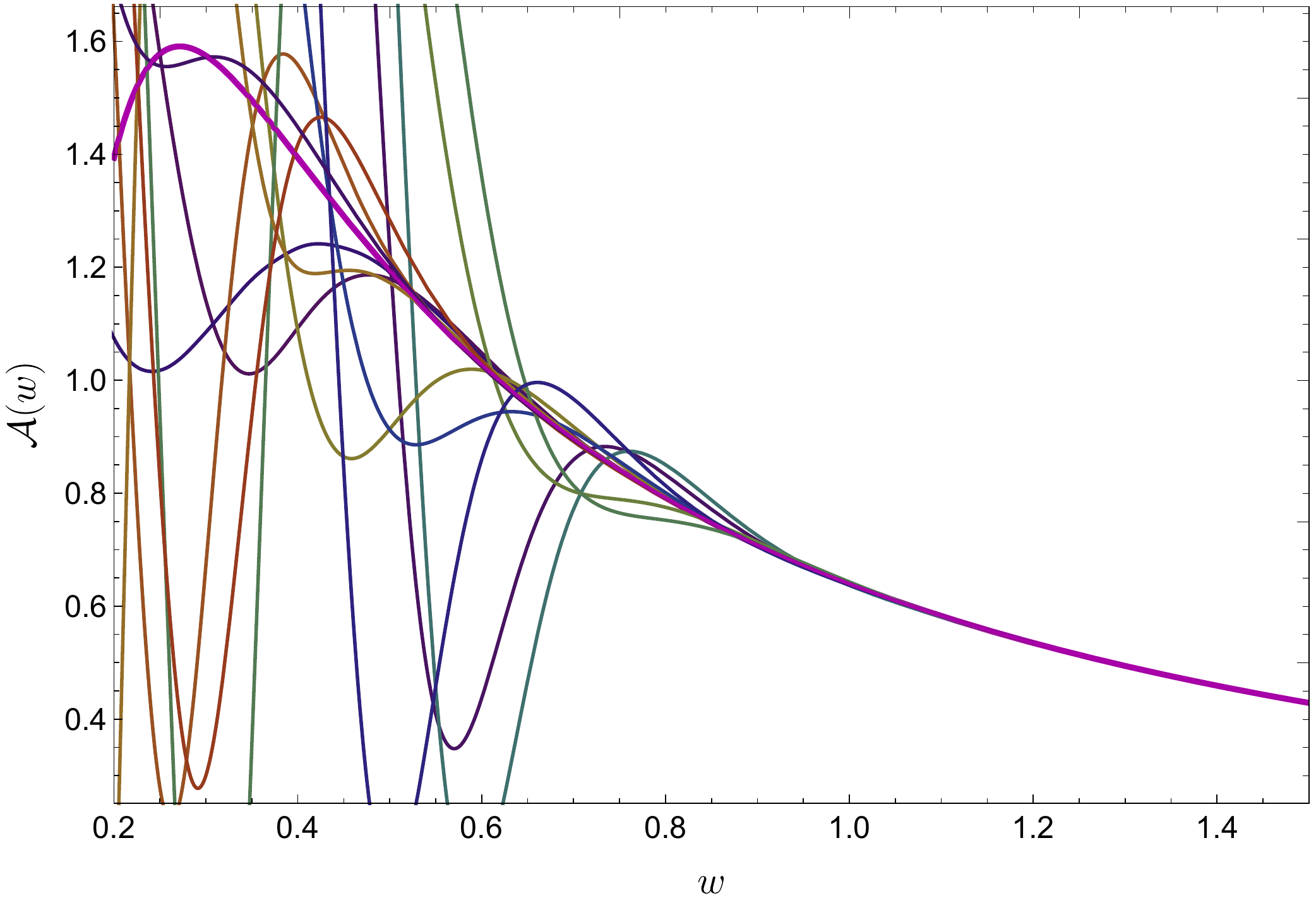}
\caption{
The pressure anisotropy as a function of $w$ for $\mathcal{N}=4$ SYM at strong
coupling, along with the parametrized attractor (thick magenta line), from
Eq.~(\ref{eq:Aparam}). 
}
\label{fig:AttractorParam}
\end{center}
\end{figure}
%


In the case of \symm\ theory one can apply Borel summation to the gradient
expansion in exactly the same way~\cite{Spalinski:2017mel}. The resulting
estimate of the attractor can be represented by fitting the numerical result to
a rational function:
\begin{equation} 
    A_{\rm attr}(w)=\frac{-276w+2530}{3975 w^2 - 570 w + 120}~.
\label{eq:Aparam} 
\end{equation}
It is apparent from \rff{fig:AttractorParam}
that this resummation breaks down below $w\approx0.4$ and therefore sheds little
light on the early-time behaviour. 

To conclude this section, let us mention some results concerning finite coupling
corrections~\cite{Casalderrey-Solana:2017zyh}. Such corrections can be included
by modifying the dual gravitational representation of the leading approximation
discussed so far. The ensuing Bjorken flow gradient expansion can also be
calculated to high order. The analytic continuation of its Borel transform
reveals an interesting structure of singularities which interpolates between the
one found in at infinite coupling \cite{Aniceto:2018uik} and the one familiar
from kinetic theory \cite{Heller:2016rtz}.

%% file: phase.tex
\section{The phase space perspective}
\label{sec:PhaseSpace}

We have seen that some part of the information about the initial state becomes 
suppressed during dissipative evolution. In the case of Bjorken flow this is
especially clearly visible in the behaviour of $\pa(w)$, where at late times the
approach to equilibrium is independent of initial conditions up to exponentially
suppressed corrections.
Furthermore, in this case the attractor locus is one dimensional, and is in fact
a solution of a differential equation.  These features are a consequence of
expressing the dynamics through a set of convenient variables. To explore
hydrodynamic attractors beyond the simplest settings one needs to understand how
to analyse the problem without such special variables, because in more
complicated situations such variables may not be known, or even exist. In this
Section we review a very general approach, which does not rely on any symmetry assumptions such as
those of Bjorken flow~\cite{Heller:2020anv}.  It addresses the
emergence of attractors by considering the behaviour multiple solutions in a
suitably defined phase space.  It is worth emphasising that this perspective can be
applied to any dynamical model of equilibration, including  models formulated in
the language of kinetic theory and the AdS/CFT correspondence. 

\subsection{Dimensionality reduction}


\begin{figure}[t!]
  \centering
  \includegraphics[width=0.45\columnwidth]{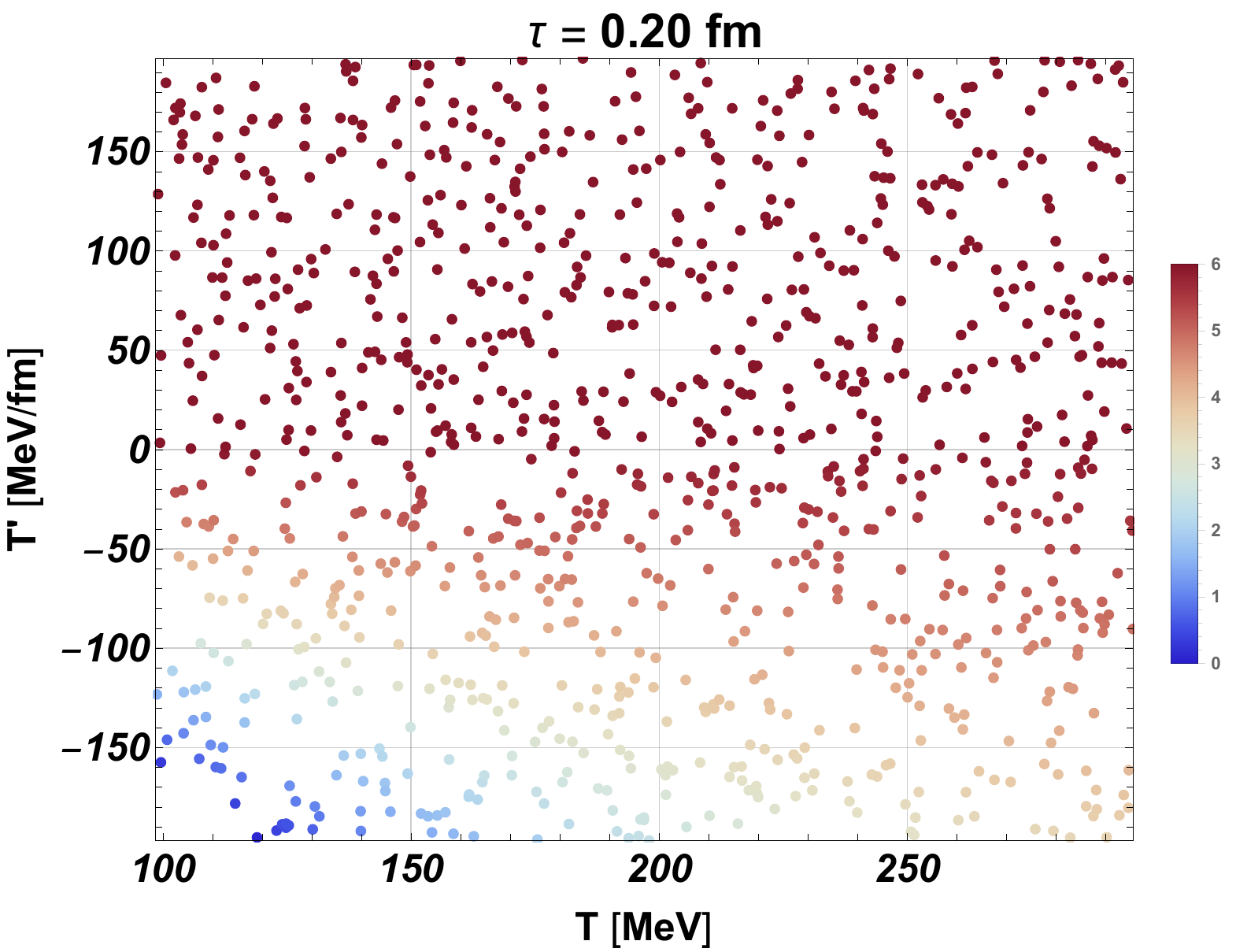}
\hspace{4mm}
  \includegraphics[width=0.45\columnwidth]{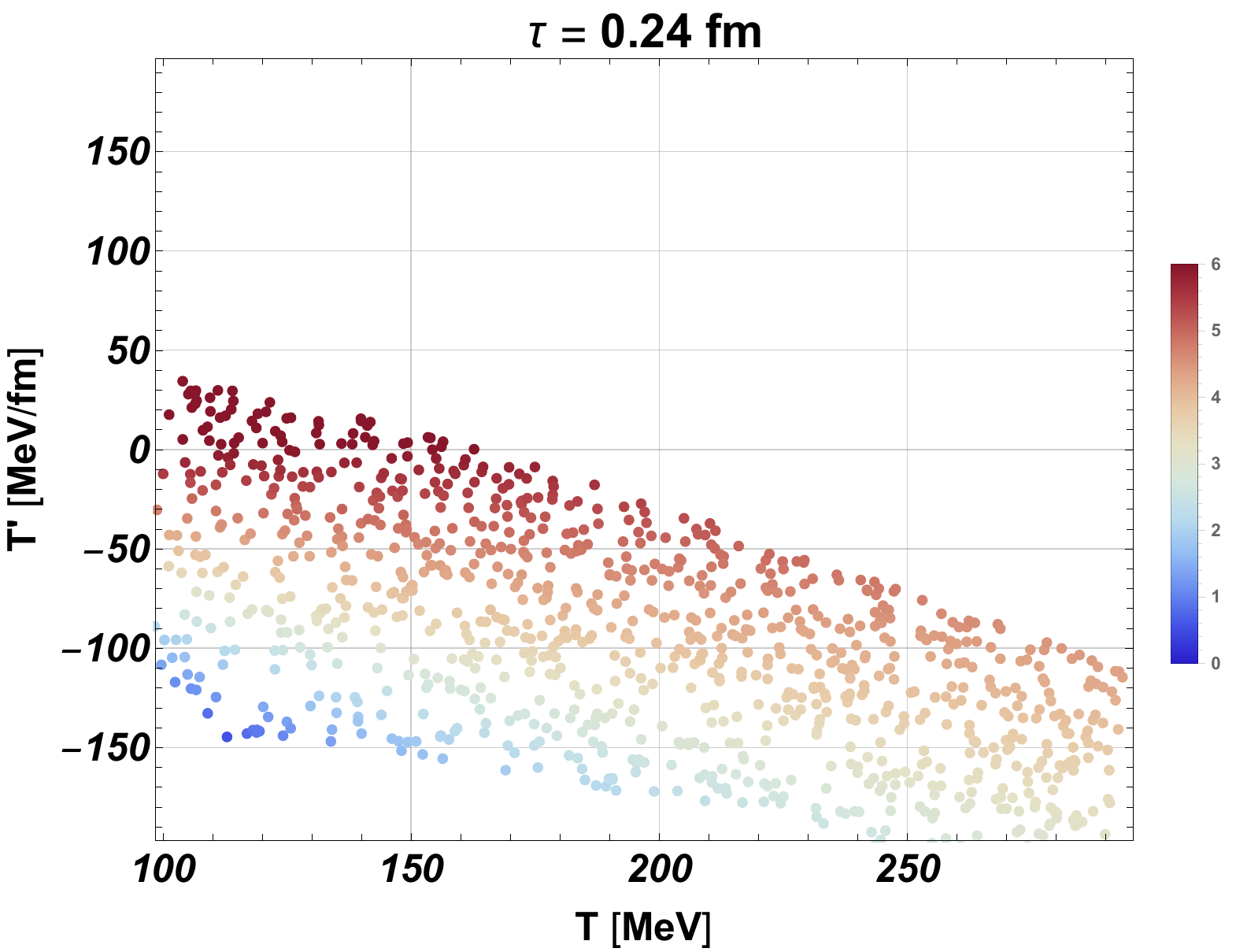}
\vspace{1mm}\vfill
  \includegraphics[width=0.45\columnwidth]{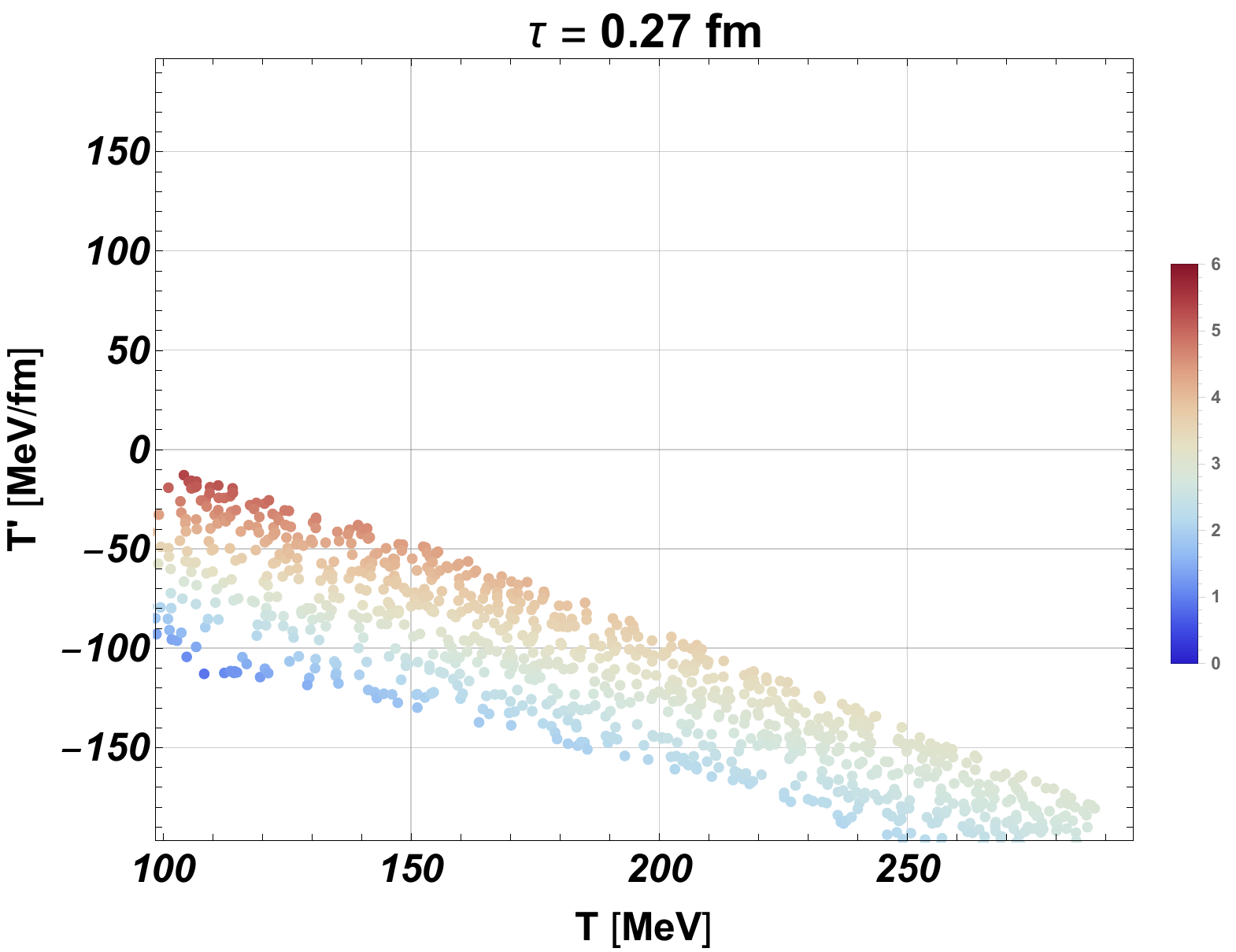}
\hspace{4mm}
  \includegraphics[width=0.45\columnwidth]{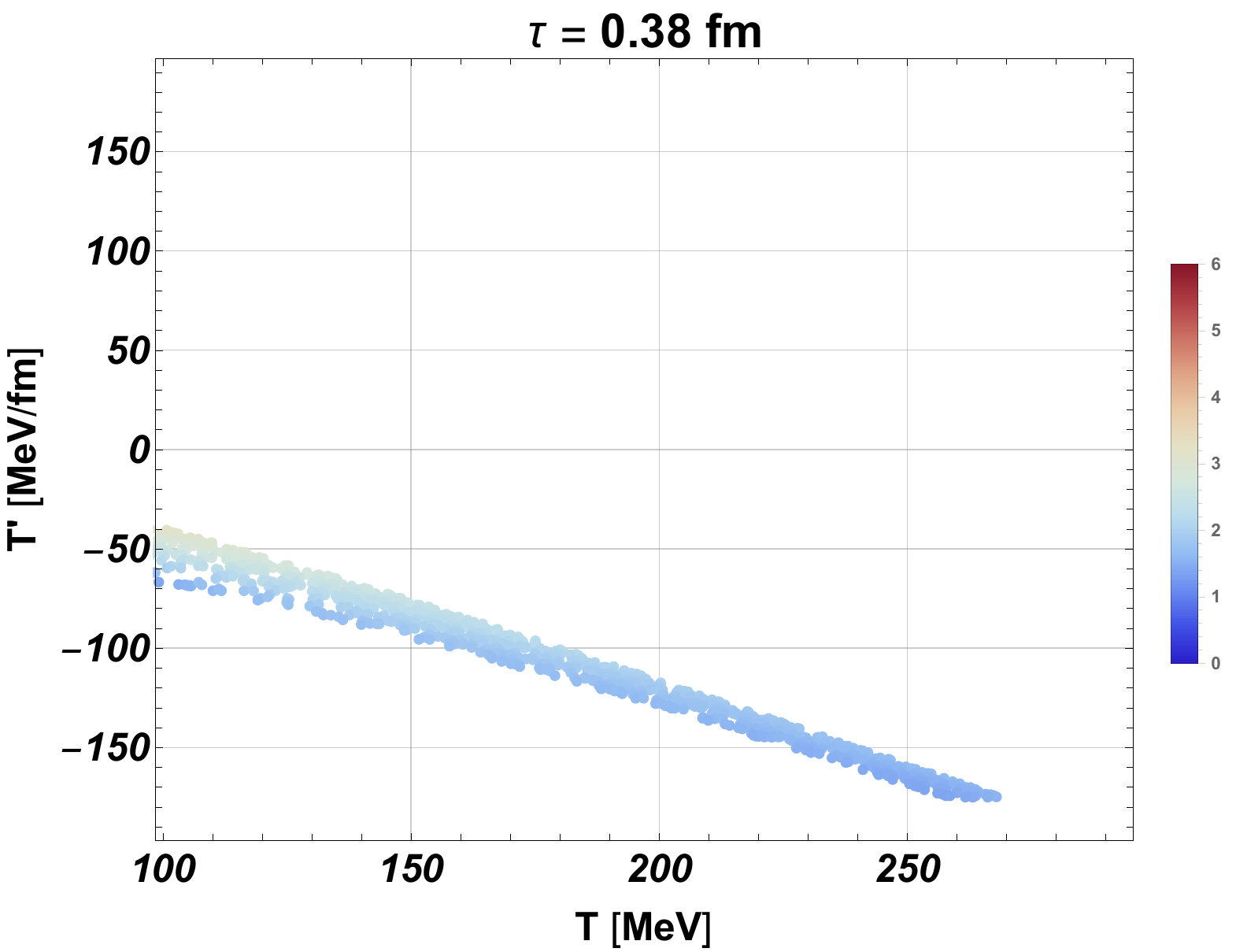}
\vspace{1mm}\vfill
  \includegraphics[width=0.45\columnwidth]{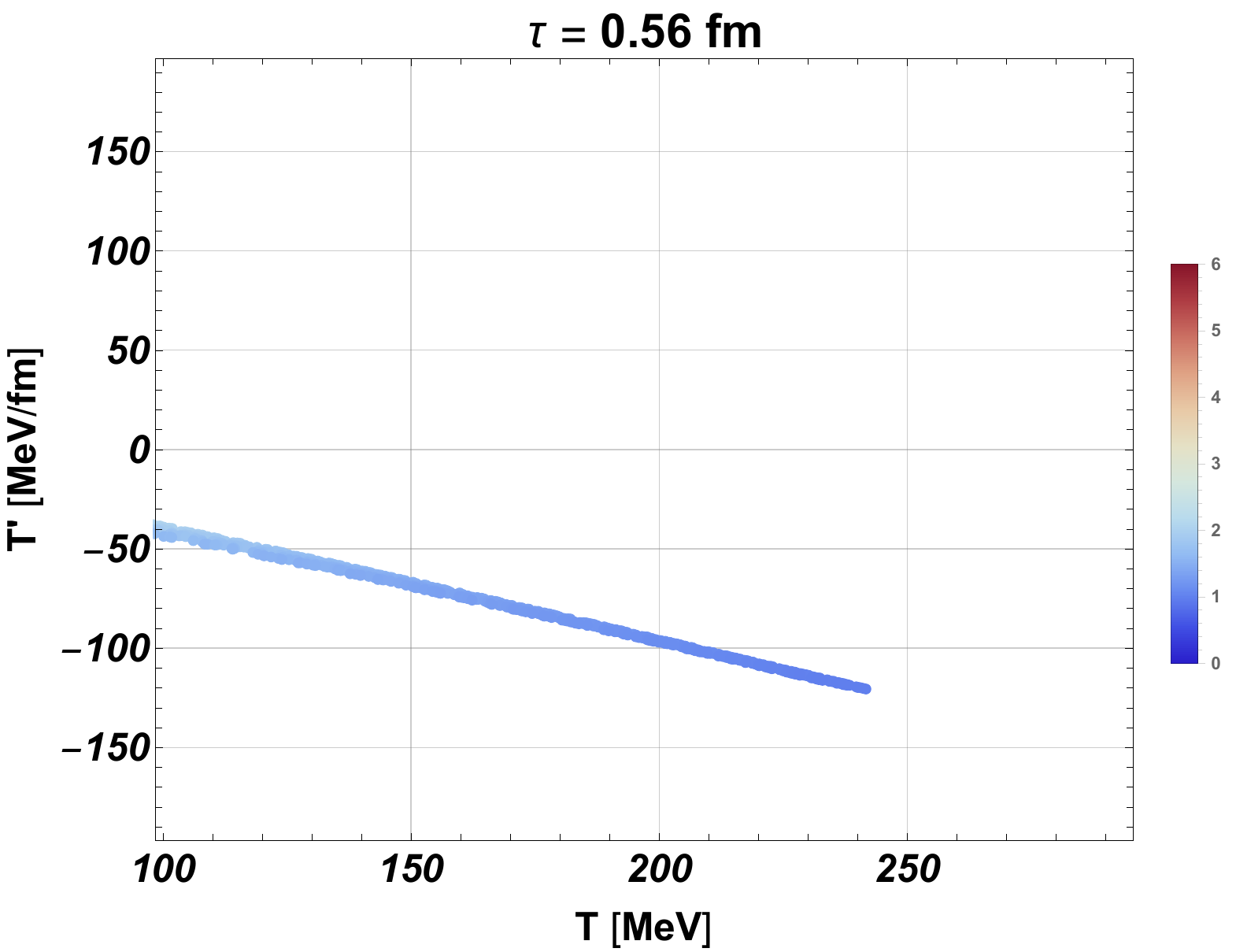}
\hspace{4mm}
  \includegraphics[width=0.45\columnwidth]{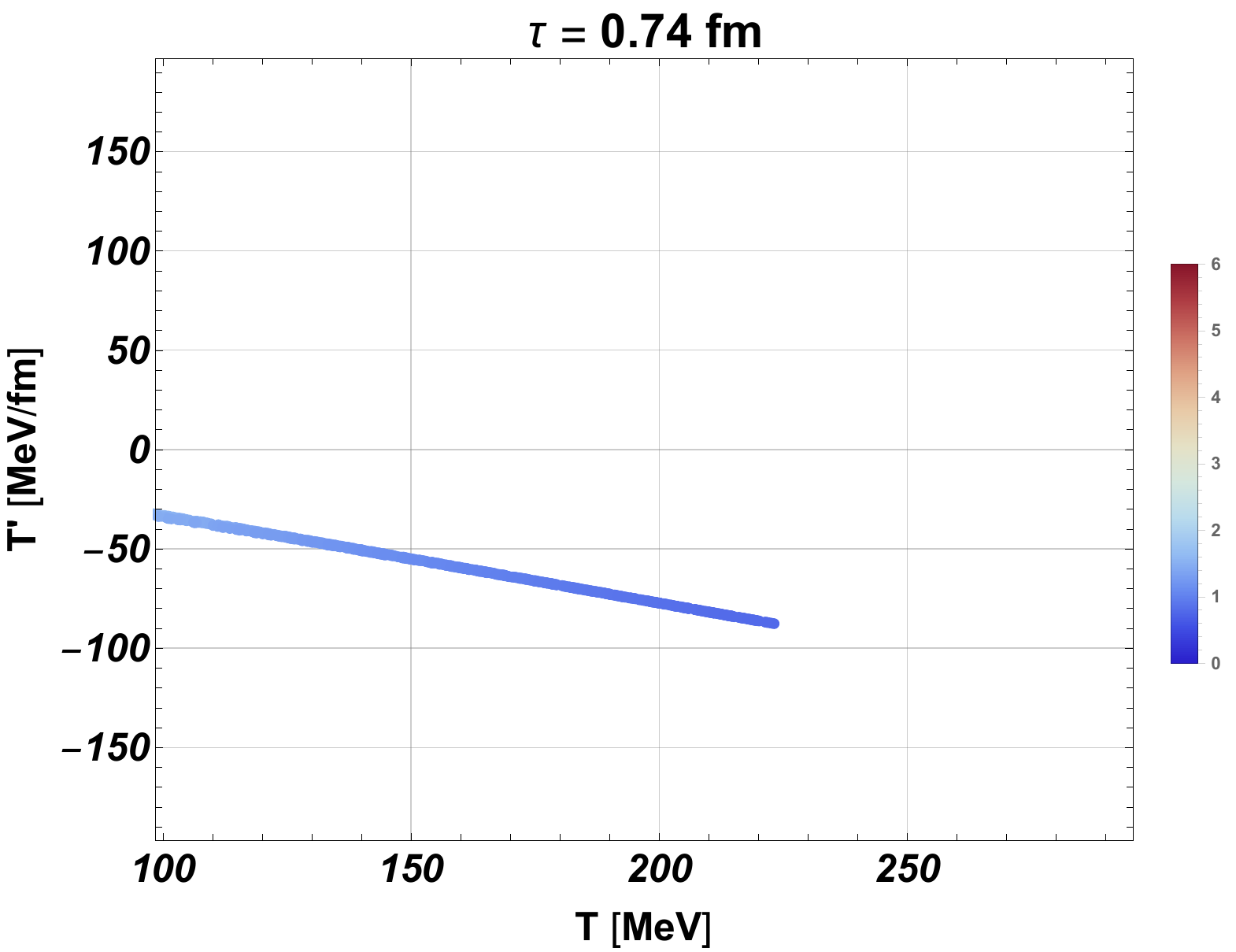}
\caption{A sequence of snapshots expressing the evolution of a point-cloud of
solutions plotted on a proper-time slice in boost-invariant MIS theory. Initially the depicted region is
uniformly filled, but in subsequent plots we see the dimensionality reduced from
$2$ to $1$. The colour of a dot encodes the effective
temperature~\cite{Spalinski:2022cgj}.}
\label{fig:dimred}
\end{figure}


To introduce the basic idea we return to the simplest models of equilibration,
formulated in the language of hydrodynamics. Given an equation such as
\rf{eq:MISTeom}, the most generic parametrisation of phase space would be to
use $T(\tau), \dot{T}(\tau)$.  The late time behaviour of the temperature
\rf{eq:MISTasym} shows that any set of solutions whose initial conditions are set on some
proper-time slice $\tau=\tau_0$, in the course of evolution collapses
approximately onto a one-dimensional locus: a curve parametrised by the value of
$\Lambda$, which is the only remnant of the initial state.   In the simplest MIS
model this means that only one combination of two integration constants is still
accessible at late times, but in more complex models the initial state could
carry much more information which is effectively dissipated by the time the
asymptotic form \rf{eq:MISTasym} is reached.  This suggests that even in more
general settings one may view the attractor as a locus of low dimensionality
embedded in a potentially high-dimensional phase space. One can thus say that
evolution toward the hydrodynamic attractor is tantamount to {\em dimensionality
reduction} of sets of solutions viewed on phase space slices. 

We illustrate this perspective by considering the full phase space for Bjorken
flow in  MIS theory, parametrised by $(\tau, T, \dot{T})$ -- the proper time
is included as one of the phase space variables because equations of motion
\rf{eq:MISTeom} depend explicitly on~$\tau$.  The plots in \rff{fig:dimred}
allow us to follow a collection of solutions starting with a
uniformly-distributed set of points on an initial proper-time slice. In the
course of dissipative evolution one sees them all approaching the attractor
locus, which in this case is a straight line, whose alignment depends on 
$\tau$. 

\subsection{Machine learning}

The notion of dimensionality reduction in phase space is a promising perspective
on hydrodynamic attractors, one which is not tied to the simplicity of Bjorken
flow. A key element of this approach is some convenient, but essentially
arbitrary parametrisation of phase space. In situations more generic than
Bjorken flow this will certainly involve some coarse-graining, but one will
still need to consider phase spaces of large dimensionality. This suggests
applying machine learning techniques to this problem. This has not yet been
fully explored in the published literature, but some encouraging pilot studies exist.
Here we will comment briefly on the approach suggested in
Ref.~\cite{Heller:2020anv}, which made use of one of the simplest
dimensionality-reduction methods -- Principal Component Analysis (PCA).  PCA
analyses the variations in a data set in different directions and associates an
\textit{explained variance} with each of them. In this way the number of
principal components of a set of solutions on a proper time slice reflects the
effective dimensionality of this \textit{point cloud}. In the case of Bjorken flow, when
the system is close to equilibrium this cloud will be one-dimensional,
reflecting the single integration constant of the asymptotic Bjorken solution
\rf{eq:bjorken}. 

\begin{figure}[t!]
    \centering
\includegraphics[width=.47\columnwidth]{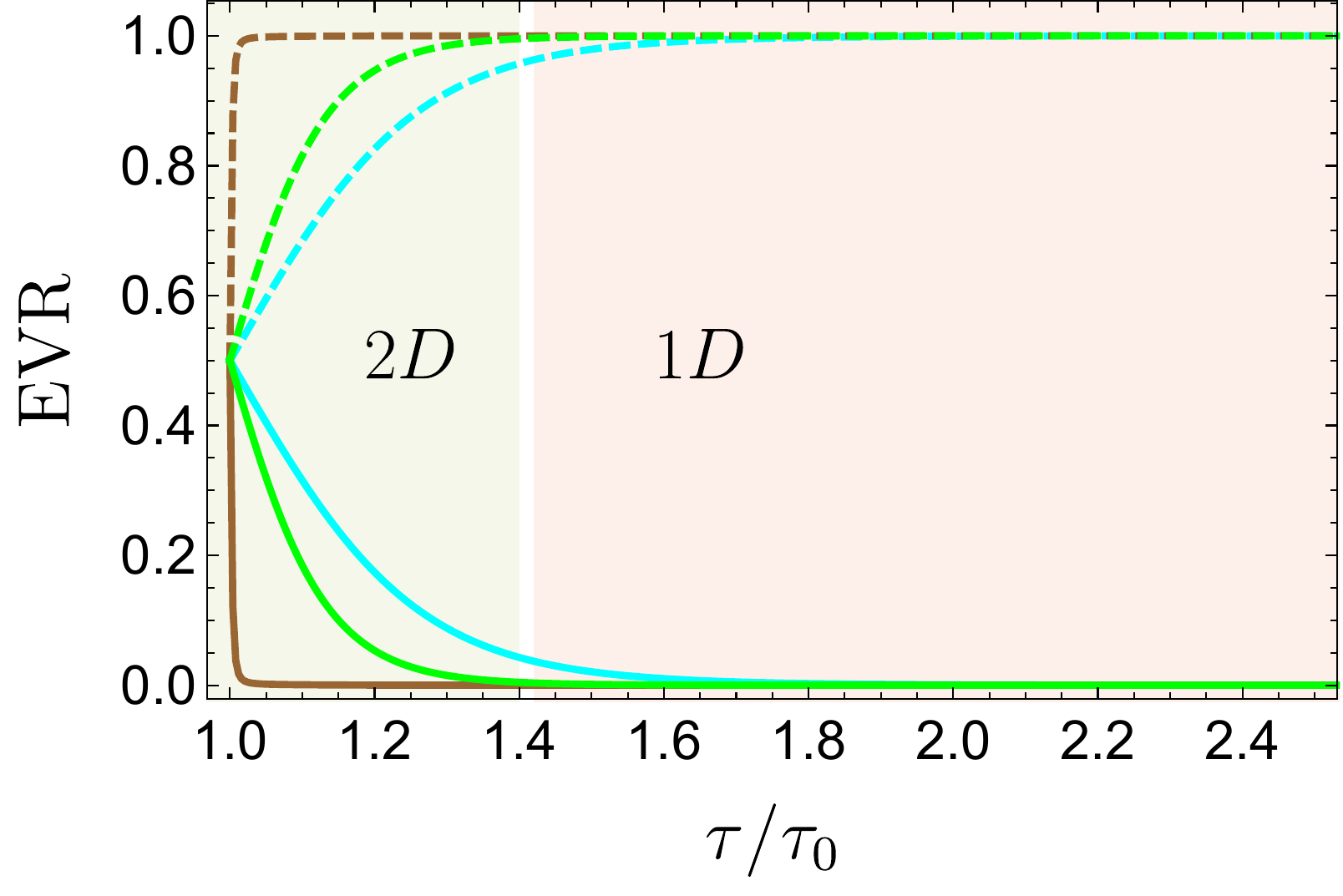} 
\includegraphics[width=.47\columnwidth]{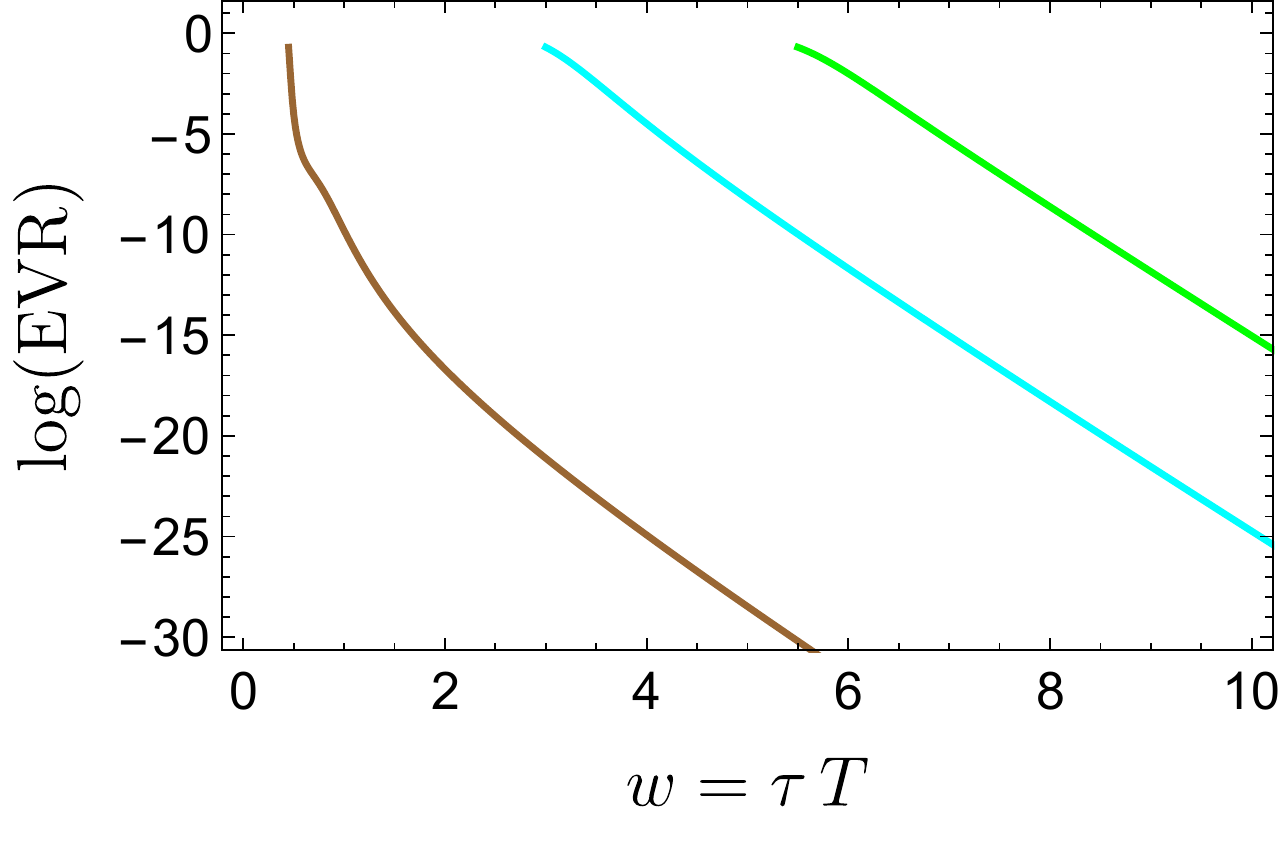}
\caption{Left: evolution of explained variance ratio of each principal component
  in MIS/BRSSS for circles (radius: $10^{-4}$) of initial conditions with
  centres lying in the middle of initial dots of corresponding colour in
  Fig.~\ref{plot:descent}. Right: logarithm of decaying principal components
  plotted as a function of $w$. For large enough values of~$w$ one clearly sees
  persistent exponential decay. Plot from Ref.~\cite{Heller:2020anv}. }
    \label{fig:brsss_pca}
\end{figure}

As a simple illustration, we begin by applying PCA to the two-dimensional phase
space of MIS theory, so as to quantify the pattern of behaviour seen in
\rff{fig:dimred}. On the initial time slice we pick a state $(T,\dot{T})$ and
consider a random set of points within a disc around it. For this set of points,
the two principal components are approximately equal in magnitude. At each time
step we recompute the principal components of this evolving point cloud; their
evolution is shown in Fig.~\ref{fig:brsss_pca}.  Dimensionality reduction is
signalled when one of the components is much smaller than the other one.  It is
significant that the dimensionality reduction splits into two stages: the first
one can interpret as the effect of the longitudinal expansion, and the second as
nonhydrodynamic mode decay. 


\begin{figure}[t!]
    \centering
\includegraphics[width=.95\columnwidth]{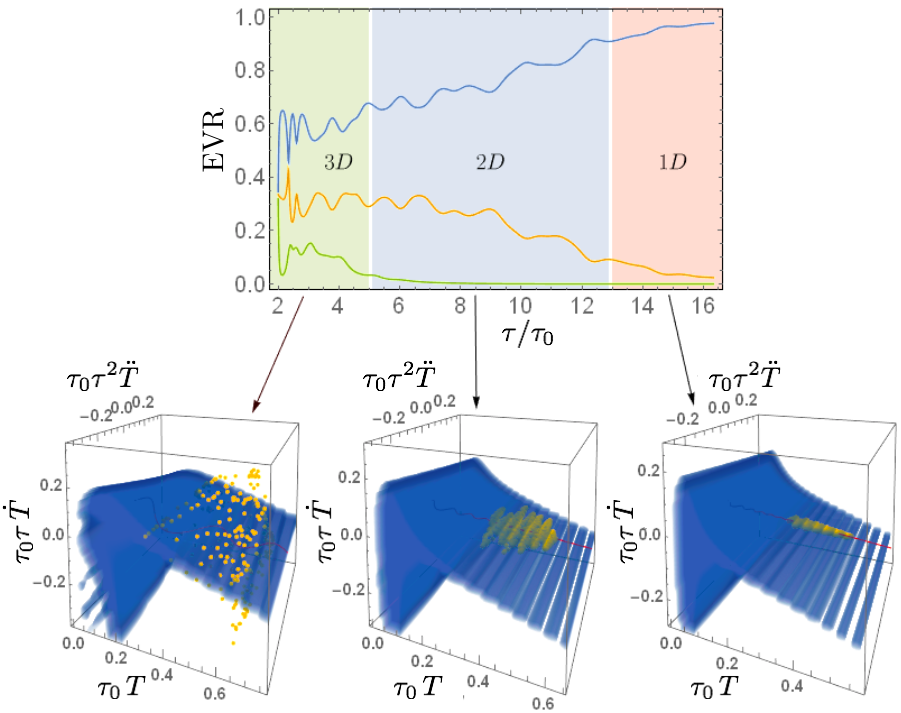}
\caption{In the HJSW model, the evolution of a cloud in phase space can
be split into three stages, corresponding to the dimensionality
of the cloud. The reduction from three to two dimensions
corresponds to a collapse onto the slow region (blue region in plots). Figure performed with
$C_\eta=0.75$, $C_{\tau_\pi}=1.19$, and $C_\omega=9.8$, taken from Ref.~\cite{Heller:2020anv}. }
    \label{dimred:hjsw}
\end{figure}

A similar picture emerges in the case of the HJSW model discussed in
\rfs{sec:HJSW}, where the phase space s three-dimensional.  In this case the
dimensionality reduction is even more striking, as seen in \rff{dimred:hjsw},
whose bottom panel depicts three stages in the evolution of a uniform cloud of
solutions at three instances of proper time. Initially, the boost-invariant
expansion leads to a rapid, parameter-independent collapse of the
three-dimensional region to a two-dimensional locus -- this is the early-time,
expansion-dominated phase. Subsequently the reduced two-dimensional cloud
evolves until it shrinks to a line, as it must for conformal Bjorken flow. The
dynamics of the second and third stages depend on the parameter values, as
expected on the basis of the interpretation in terms of nonhydrodynamic mode
decay.  The evolution of principal components is shown in the upper part of
Fig.~\ref{dimred:hjsw}, where one can clearly discern the three different stages
with different dimensionality. 


This type of analysis can be applied to phase spaces of arbitrary
dimensionality, in any dynamical model. This will in general involve some coarse
graining and truncation of the phase space, which in itself need to be
finite-dimensional. An enlightening example which illustrates these issues is
provided by a model of Bjorken flow in kinetic theory in the RTA which can be
found in the Supplemental Material of Ref.~\cite{Heller:2020anv}. A more
realistic study of the phase space approach to Bjorken flow in the effective
kinetic theory of QCD can be found in Ref.~\cite{Du:2022bel}.  These studies
rely on PCA, but a number of other machine learning techniques exist which may
provide a more refined picture, such as Topological Data Analysis (see e.g.
\cite{chazal}) which has recently been explored in a somewhat related physical
context~\cite{Spitz:2023wmn}.

\subsection{The attractor as a slow region}


\begin{figure}[t!]
\centering
  \includegraphics[width=.47\columnwidth]{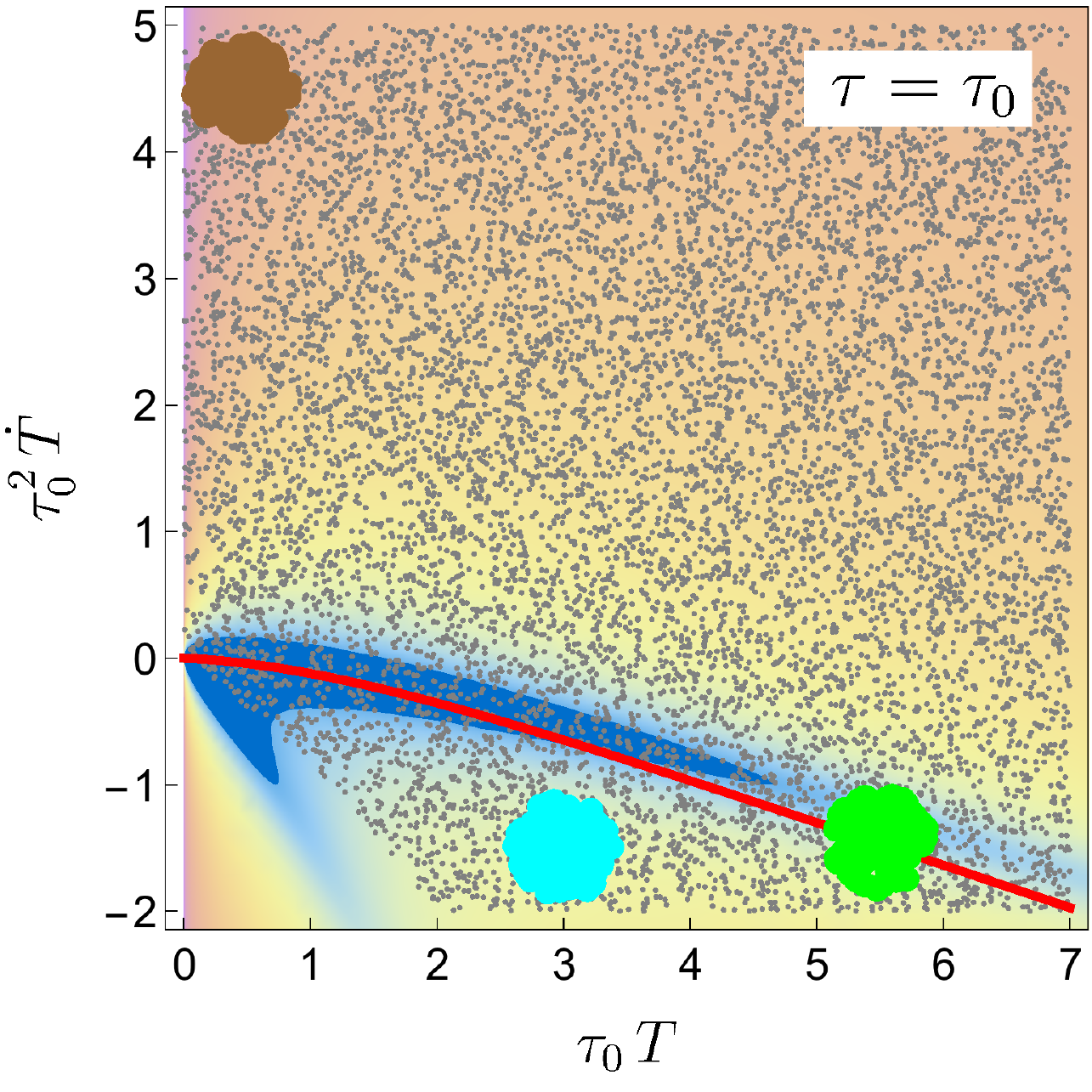}
\includegraphics[width=.47\columnwidth]{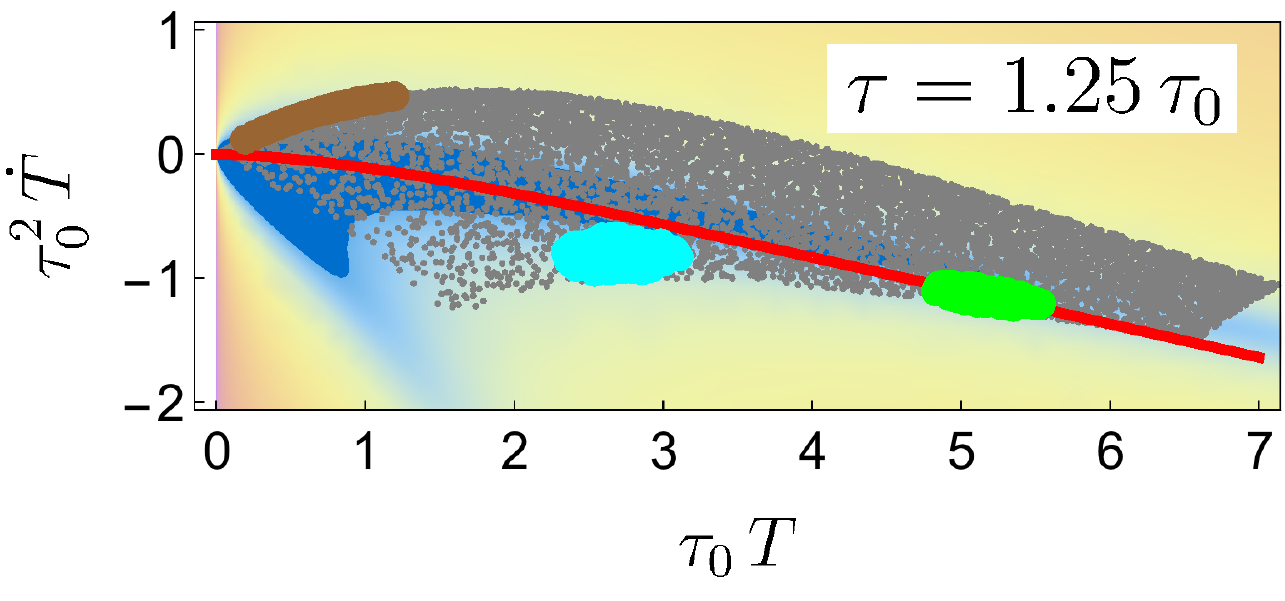}
\vspace{3mm}
\includegraphics[width=.47\columnwidth]{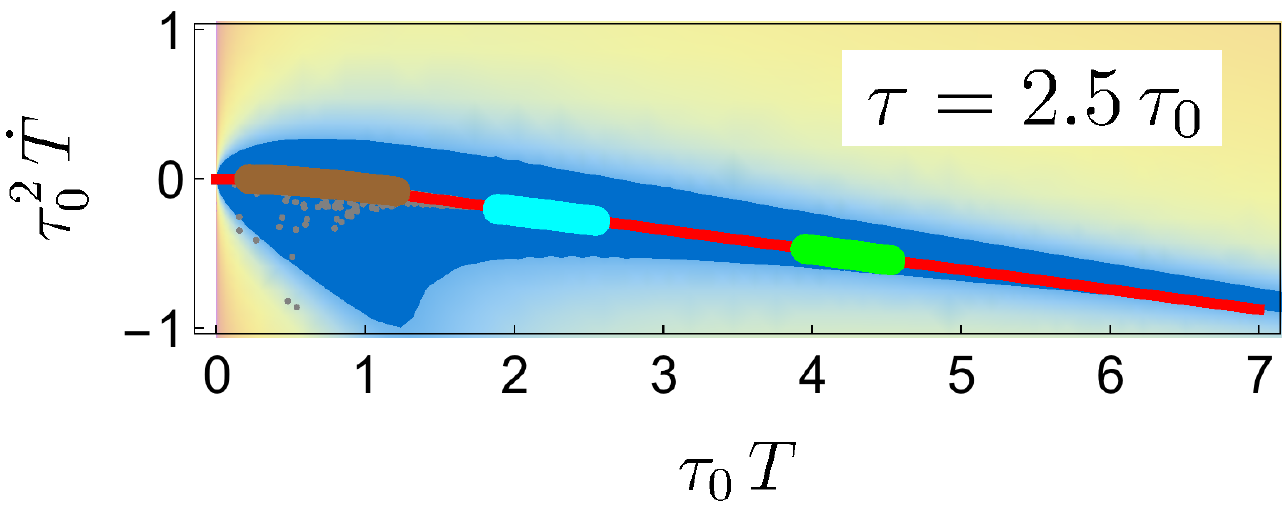}
\caption{Three proper time slices of phase space tracking a cloud of
  about 10,000 solutions of MIS theory. The red
  curve denotes the attractor locus
${\cal A}_\star$.  The background colour encodes the speed at which
the points move in phase space, with magenta faster than blue according to
the norm of velocity vector~\eqref{eq.Vdef}; the dark blue denotes the
slow region. The plots were made for $C_{\eta} = 0.75$ and
$C_{\tau_{\pi}} = 1$; 
$\tau_{0}$ denotes the initialisation time.
 Plot from Ref.~\cite{Heller:2020anv}.}
\label{plot:descent}
\end{figure}


In models with phase spaces whose dimension is greater than two, the attractor
is not a single solution but rather a region of phase space onto which actual
solutions condense. In conformal examples of Bjorken flow it was possible to
project this region onto a single solution in the $\pa, w$ variables, but this
is not expected to be possible in more general situations. Therefore, a more
general characterisation of the attractor is needed. Intuitively one would
expect that the attractor locus should correspond to a ``slow region'' where the
flow in phase space is slowest, because it takes a long time to escape it, while
the fast regions can be quickly traversed.

In the case of MIS this idea correctly identifies the
attractor on any given proper time slice. A point on such a slice of phase space can be
described by the vector 
$\vec{X}(\tau) = \left(\tau_{0} \,
T(\tau),\tau_0^2 \, \dot{T}(\tau) \right)$; its velocity is given by 
\be
\label{eq.Vdef}
  \vec{V}=\tau_{0} \frac{\partial \vec{X}}{\partial \tau} ~,
\ee
where the factors of $\tau_{0}$ have been introduced for dimensional reasons. The slow
region can be defined by the Euclidean norm
$V$ of this vector, which has a minimum at
asymptotically late times when the system approaches local thermal equilibrium. 
The resulting picture can be seen in Fig.~\ref{plot:descent}; the
background colour is determined by~$V$, where  bluer colour implies lower speed.
There is a slow region stretching out from local thermal equilibrium, and the
attractor $\mathcal{A}_{*}(w)$ lies along it. 
The identification of the attractor as a slow region at least in principle
generalises directly to phase spaces of any dimension.

%% file: prehydro.tex
\section{Attractors and prehydrodynamic flow}
\label{sec:prehydro}

As discussed in Sec.~\ref{sec:HIC}, the standard picture of heavy-ion collisions
involves formation of QGP followed by a stage of nonequilibrium evolution until
a time when conventional fluid dynamics can be applied. The developments
reviewed in the last four Sections suggest that approximate Bjorken symmetry of
the early time dynamics could be responsible for a far-from-equilibrium
attractor capable of providing a bridge between these two stages. This attractor
could then be modelled using some much simpler effective description such as MIS
theory used outside its naive domain of applicability.  In this Section we will
review some efforts aiming to apply this idea to QGP
dynamics~\cite{Giacalone:2019ldn,Jankowski:2020itt}. Other work which makes
practical use of attractors in the context of heavy-ion collisions includes
Refs.~\cite{Du:2020dvp,Du:2020zqg,Du:2022bel,Dore:2020fiq,Dore:2020jye,Dore:2021xqq,Dore:2022qyz,Ambrus:2022qya,Ambrus:2022koq,Ambrus:2023oyk,Coquet:2021cuv,Coquet:2021gms}.

As currently understood (see \rfs{sec:attractors}), the existence of a
nonequilibrium hydrodynamic attractor in Bjorken flow is contingent upon there
being a definite, finite, physically distinguished value $\paz(0)$ of the
pressure anisotropy at $w=0$~\cite{Heller:2015dha,Aniceto:2015mto}. One can
translate this into a statement about the early time behaviour of the energy
density. Indeed, under the above assumptions, the conservation of
energy-momentum~\rf{eq:Adef3} implies that for asymptotically small proper-time
$\tau$
\be
\label{eq:mu}
  \edens \sim \f{\mu^4}{(\mu\tau)^{\beta}}~,
\ee
where the scale $\mu$ is an integration constant which reflects the initial
conditions, and the exponent $\beta$ is related to the attractor by the relation
\be
\label{eq.abeta}
\paz(0) = 6\left(1-\f{3}{4}\beta\right)~.
\ee
We will consider $0\leq\beta < 4$ (where $\beta=1$ corresponds to free
streaming).  While different initial conditions will correspond to different
values of $\mu$, the parameter $\beta$ characterises the attractor
itself and is therefore a feature of the particular microscopic theory under
consideration.  For instance, in MIS theory the attractor is
the unique stable solution which is regular at $w=0$, where 
\be \label{eq.mis} \paz(0) = 6\sqrt{\frac{C_\eta}{C_{\tau\Pi}}} \iff \beta =
\f{4}{3} \left(1 - \sqrt{\f{C_\eta}{C_{\tau\Pi}}}\right)~.
\ee
%

\subsection{Entropy and particle production}
\label{subsect:Entropy}

The hydrodynamic attractor connects early and late-time behaviour of the system,
and this fact makes it possible to relate final state entropy to characteristics
of the initial state.  We will denote the value of $w$ at very early proper time
$\tau_0$  by $w_0$, and its value at late times $\tau_\infty$ by $w_\infty$.
Since at late times the system is approaching thermal equilibrium, one may use
standard thermodynamic relations to write the entropy density as\footnote{Note
that the arguments of this Section do not require invoking any concepts of
entropy far from equilibrium.}
\be
s(\tau_\infty) = \f{4}{3} \f{\edens(\tau_\infty)}{T(\tau_\infty)}
\ee
and then apply \rf{eq.approx} to express the right hand side in terms of
quantities evaluated at proper time $\tau_0$. This leads to the key relation
between the entropy density per unit rapidity at late time and the initial
energy density
\be
\label{eq:entrofinal}
s(\tau_\infty) \tau_\infty= h(\beta) \left(\edens(\tau_0) \tau_0^{\beta}\right)^{\frac{2}{4-\beta}}~,
\ee
where
\be
\label{eq:hbeta}
h(\beta) = \f{4}{3} w_\infty w_0^{\f{2\beta}{\beta-4}} \Phi_\paz(w_\infty, w_0)^2~.
\ee
The reason for writing \rf{eq:entrofinal} in this particular way is that the
left hand side as well as both factors on the right hand side are well defined
as $\tau_0\rightarrow 0$ and $\tau_\infty\rightarrow \infty$. Given \rf{eq:mu}
this is obvious for the second (parenthesised) factor, but one can also check
that the function in \rf{eq:hbeta} is finite in this limit because $\Phi_\paz$
in \rf{eq.phidef} diverges for small $w_0$ and vanishes for large $w_\infty$
precisely in such a way that the dependence on the initial and final values of
$w$ drops out, leaving a finite and nonzero result.  This can be shown in
general based on the asymptotic behaviours of the pressure anisotropy. 

The importance of \rf{eq:entrofinal} rests on the fact that it is an explicit
relation between the initial energy density and final entropy density of
expanding plasma, which accounts for entropy production as the
system evolves along the attractor.  Furthermore, it can be translated into a
statement about centrality dependence of observed particle multiplicities by
utilising the following relation~\cite{Yagi:2005yb}:
\begin{equation}
\label{eq:dndy}
    \frac{d N_{\rm ch}}{d \eta}\approx A(s\tau)_{\rm hydro}~,
\end{equation}
where $A$ is a constant whose value will not be relevant in the subsequent
analysis. In the context of hydrodynamic attractors such a calculation was
first described in Ref.~\cite{Giacalone:2019ldn} for the special case of
free-streaming attractors, and generalised to any Bjorken attractor in
Ref.~\cite{Jankowski:2020itt}. We will now review these developments. 

Given the entropy density in \rf{eq:entrofinal}, and using \rf{eq:dndy}, the charged particle
multiplicity of a specific event can be
expressed as 
\be
\label{eq:dndyEps}
\dndy = A \tau_0^{\frac{2\beta}{4-\beta}} h(\beta) \int d^2\mathbf{x}_\perp
\edens(\tau_0, \mathbf{x}_\perp)^{\frac{2}{4-\beta}}~.
\ee
The new element here is allowing for a non-trivial dependence of the
initial energy density on the location in the plane transverse to the collision
axis.  This brings in dependence on the impact parameter of a given event. The
underlying assumptions and applicability of this procedure are discussed in
Ref.~\cite{Giacalone:2019ldn}, where it was introduced. Formula
(\ref{eq:dndyEps}) can be used to estimate the expected multiplicity by
averaging over Monte Carlo generated events.  

For a given event, the calculation of the initial energy density requires a model of the
initial state.  Ref.~\cite{Giacalone:2019ldn} considered free-streaming
attractors and showed that the results are consistent with experiment if one
chooses the dilute-dense
model~\cite{Dumitru:2001ux,Blaizot:2004wu,Gelis:2005pt,Blaizot:2010kh,Schlichting:2019bvy}
to describe the initial energy density profile. We will first review that study,
and then turn to the analysis of Ref.~\cite{Jankowski:2020itt}, where this
approach was generalised by dropping the assumption of free streaming at early
times. 

\subsection{The dilute-dense model and free-streaming attractors}

A standard approach to quantify fluctuations of nucleon positions is the Glauber
model~\cite{Miller:2007ri}. A basic object used to formulate a description of
the initial state in this approach is the thickness function
$T(\mathbf{x}_\perp)$ \cite{Miller:2007ri,Yagi:2005yb}, which is determined by
the integral of the average nuclear matter density along the longitudinal
direction~\cite{Yagi:2005yb}
\begin{equation}
    T({\bf x}_\perp)=\int_{-\infty}^\infty dz~ \rho({\bf x}_\perp,z)~.
\end{equation}
The nuclear density is usually parametrised by the Woods-Saxon distribution function
\begin{equation}
    \label{eq:saxonw}
    \rho(r)=\rho_0\left[1+\exp\left(\frac{r-R}{a}\right)\right]^{-1}~,
\end{equation}
where $\rho_0$ is chosen such that $\rho(r)$ is normalised to the number of
nucleons. For the two systems considered in
Refs.~\cite{Giacalone:2019ldn,Jankowski:2020itt} we have $a_{\rm Pb}=0.55$~fm,
$R_{\rm Pb}=6.62$~fm for $^{208}\rm Pb$ and $a_{\rm
Au}=0.53$~fm and $R_{\rm Au}=6.40$~fm for $^{197}\rm Au$~\cite{deVries:1987atn}.  

To quantify off-central collisions one introduces the impact parameter, that is
a vector ${\bf b}=(b_x,b_y)$ in the transverse plane which connects the centres of
the projectiles. Then, given the thickness functions $T^{A/B}({\bf
x}_\perp)\equiv T({\bf x}_\perp\pm{\bf b}/2)$ of two nuclei A and B colliding at a
given impact parameter $\bf b$, one defines centrality as~\cite{Teaney:2009qa}
\begin{equation}
    centrality = \frac{\pi |{\bf b}|^2}{\sigma_{\rm tot}}~,
    \label{eq:centrality}
\end{equation}
where $\sigma_{\rm tot}$ is a total inelastic nucleus-nucleus cross section. For
the data considered here we have $\sigma_{\rm tot}=767$
fm$^2$ for Pb-Pb collisions, and $\sigma_{\rm tot}=685$~ fm$^2$ for Au-Au
collisions. In a fluctuating Glauber model, the positions ${\bf x}_i$ of nucleons in
each of the nuclei are sampled with the distribution $\rho(r)$ given in
\rf{eq:saxonw} and their collisions are
determined by the mutual distance not larger than $\sqrt{\sigma_{\rm nn}/\pi}$,
where $\sigma_{\rm nn}$ is a nucleon-nucleon cross section equal to $\sigma_{\rm
nn}\left(\sqrt{s}=200~\rm GeV\right)=4.2$~fm$^2$ and $\sigma_{\rm
nn}\left(\sqrt{s}=2.76~\rm TeV\right)=6.4$~fm$^2$. The thickness function is
then determined on an event-by-event basis by summing up all density profiles of
all participating nucleons
\begin{equation}
    T_{A/B}({\bf x_\perp})=\frac{1}{n_{A/B}}\sum_{i=1}^{n_{A/B}}\rho_c\left({\bf x}-{\bf x}_i\pm\frac{\bf b}{2}\right)~,
\end{equation}
where $n_{A/B}$ is number of nucleons in the $A/B$ nuclei, and each nucleon is modelled by a Gaussian,
\begin{equation}
    \rho_c({\bf x_\perp})=\frac{1}{(2\pi v^2)^{\frac{3}{2}}}\exp\left(-\frac{\bf x^2_\perp}{2v^2}\right)~,
\end{equation}
with a fixed width $v=0.5$~fm determining its transverse size.

In the dilute-dense
model~\cite{Giacalone:2019ldn,Dumitru:2001ux,Blaizot:2004wu,Gelis:2005pt,Blaizot:2010kh,Schlichting:2019bvy}
of the initial energy deposition which is used in Ref.~\cite{Giacalone:2019ldn} the
initial energy density is taken as
\begin{equation}
    \epsilon_0^{\rm dilute-dense}({\bf x}_\perp) =C T^<({\bf x}_\perp)\sqrt{T^>({\bf x}_\perp)}~,
    \label{eq:eps_diluted}
\end{equation}
where $C$ is a constant. 
Centrality dependence enters through the relations
\begin{equation}
    T^<({\bf x}_\perp)={\rm min}\left(T({\bf x}_\perp-{\bf b}/2),T({\bf x}_\perp+{\bf b}/2)\right)~,
\end{equation}
\begin{equation}
     T^>({\bf x}_\perp)={\rm max}\left(T({\bf x}_\perp-{\bf b}/2),T({\bf x}_\perp+{\bf b}/2)\right)~.
\end{equation}
When the resulting energy density is used in \rf{eq:dndyEps}, the results are
consistent with the assumption of a free-streaming
attractor~\cite{Giacalone:2019ldn}. 

\subsection{Beyond free-streaming}

The dilute-dense model is one of a number of initial state models currently
being explored, and it is interesting to ask to what extent are other models
compatible with a free-streaming attractor. The study of
Ref.~\cite{Jankowski:2020itt} considered two options. The first is defined
by~\cite{Lappi:2006hq,Romatschke:2017ejr}
\begin{equation} \epsilon_0^{\rm dense-dense}({\bf x}_\perp)=C T({\bf
x}_\perp-{\bf b}/2)T({\bf x}_\perp+{\bf b}/2)~.  \label{eq:eps_dd}
\end{equation}
The second model assumes 
\begin{equation} 
    \label{eq:trento1}
    \epsilon_0^{p=-1}({\bf x}_\perp)=C\frac{T({\bf x}_\perp-{\bf
b}/2)T({\bf x}_\perp+{\bf b}/2)}{T({\bf x}_\perp-{\bf b}/2)+T({\bf x}_\perp+{\bf
b}/2)}~, \end{equation}
which is a special case of Trento parametrisation~\cite{Moreland:2014oya}.  The
normalisation constant $C$ appearing in these equations is independent of the
impact parameter $\bf{b}$.

\begin{figure}
\begin{center}
\includegraphics[width=0.7\textwidth]{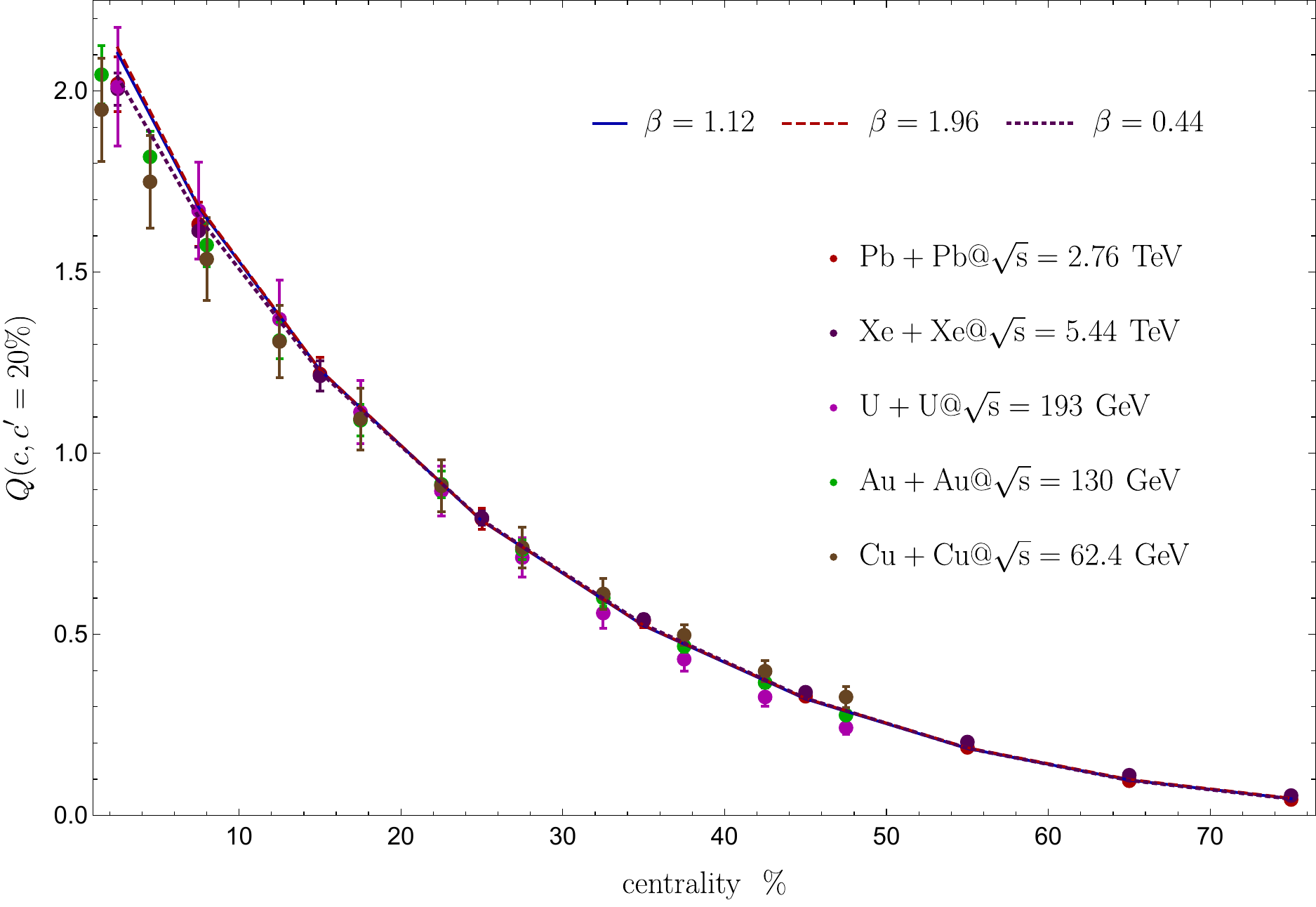}
\caption{
  Universal centrality dependence of $Q(c,c'{=}20))$, i.e. the number of
  produced charged particles normalised to $20$ centrality for each of the
  three models we consider. Experimental data
  shown for different collision systems: Xe+Xe \cite{ALICE:2018cpu}, Pb+Pb
  \cite{ALICE:2010mlf}, Au+Au \cite{PHOBOS:2010eyu},U+U \cite{PHENIX:2015tbb},
  Cu+Cu \cite{PHOBOS:2010eyu}.
  The plot is taken from Ref.~\cite{Jankowski:2020itt}.
  }
   \label{fig:wideExp} 
  \end{center}
\end{figure}

The impact of changing the initial state model can be assessed by fitting the
parameter $\beta$ using the observed centrality dependence of the measured
multiplicities. 
The formula \rf{eq:dndyEps} leads to a prediction in each centrality class.  We
then define the following ratios of multiplicities at different centralities 
\be Q(c,c') \equiv \f{\langle\dndy\rangle_{c^{\ }}}{\langle\dndy\rangle_{c'}}~,
\label{eq:Qij} \ee
where the angle-brackets denote the mean value over events in the specified
centrality class. These quantities are independent of the normalisation factors
$C$ entering Eqns.~(\ref{eq:eps_diluted}), (\ref{eq:eps_dd}),
(\ref{eq:trento1}); they are also independent of the factor $h(\beta)$, which
depends on the shape of the presumptive attractor, not just its behaviour at
early time.  However, they retain dependence on the parameter $\beta$ itself,
which is related to the attractor by \rf{eq.abeta}. In this way, for any value
of $\beta$, we obtain a set of numbers $Q(c,c')$ which can be directly compared
with  published experimental results (see \rff{fig:wideExp}). The best fit for
each of the three models is found to be 
\be 
\beta^{\rm dilute-dense} = 1.12, \quad \beta^{\rm IP} = 1.96, \quad
\beta^{\rm dense-dense} = 0.44  
\ee
with statistical errors not exceeding $0.02$. 
These values differ by a factor of almost $4.5$, which shows that if indeed an
attractor determines early time behaviour, it is strongly connected to the
initial state model. The implication here is that if we believe that QGP is
free streaming at early times, then this places a constraint on the initial state
model. Conversely, if we have reasons to favour a particular initial state
model, then this may require accounting for corrections to free streaming at
early times.  These remarks should be relevant for Bayesian
studies~\cite{Nijs:2020ors,Nijs:2020roc,JETSCAPE:2020mzn,Bernhard:2016tnd},
which scan over families of initial state models, but assume free-streaming
prehydrodynamic evolution for all of them.

%% file: beyond.tex
\section{Beyond conformal Bjorken flow}
\label{sec:beyond}

\subsection{The origin of attractor behaviour}


\begin{figure}
\centering
\includegraphics[height =.35\textheight]{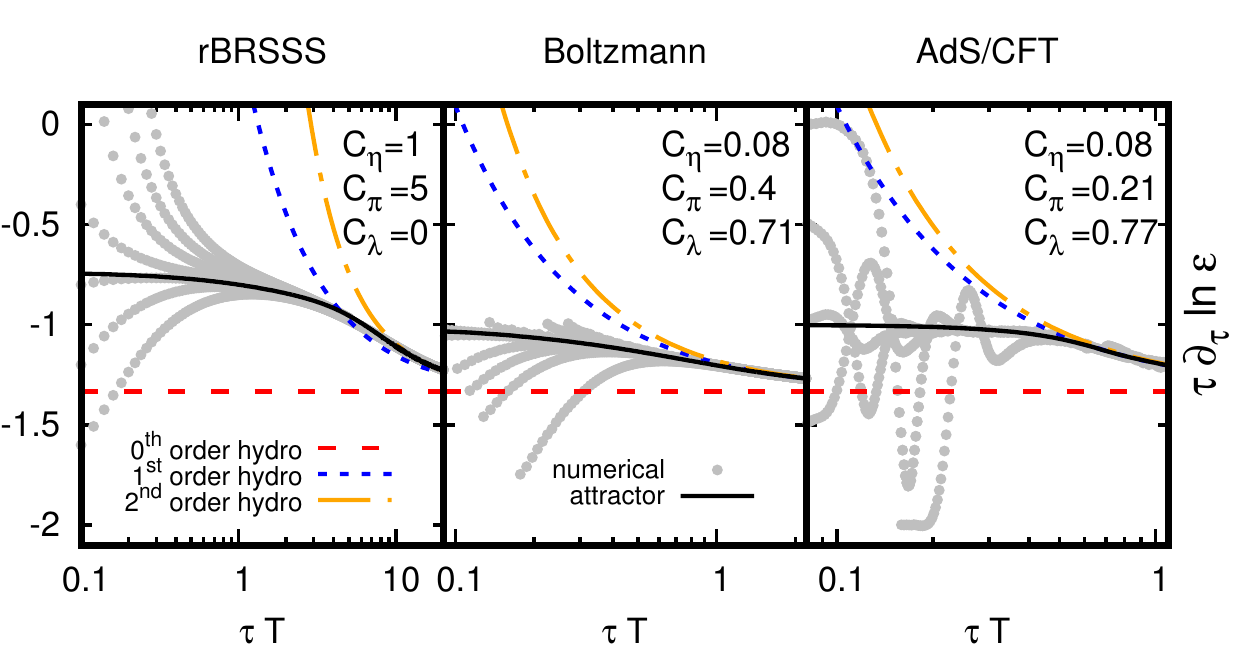}
\caption{Hydrodynamic attractors in different theories.
Plot form Ref.~\cite{Romatschke:2017vte}.
}
\label{fig:attractors}       
\end{figure}


We have seen that in the case of Bjorken flow in conformal models of
equilibration one can identify hydrodynamic attractors which extend into the
far-from-equilibrium region. This is illustrated in Fig.~\ref{fig:attractors}
which presents the approximate attractors obtained for three different
microscopic models. The appearance of attractors in these very different systems
suggests that they are a generic feature of conformal theories undergoing
Bjorken flow and are ultimately due to the specific kinematics of the earliest
stages of ultrarelativistic heavy-ion collisions.  These kinematic circumstances
lead to simplifying symmetry assumptions such as boost invariance, conformal
symmetry and the suppression of transverse dynamics.  The resolution of the
early thermalisation puzzle based on the notion of attractors relies on these
assumptions remaining approximately valid for a sufficiently long time.

The physics of a heavy ion collision can be pictured as a competition between
the longitudinal expansion resulting from the initial conditions and the
interactions which drive the system toward equilibrium~\cite{Blaizot:2017ucy}.
While the kinetic theory perspective makes this very explicit, the presence of
an early-time, expansion dominated phase followed by a late-time regime
interpreted in terms of nonhydrodynamic mode decay is also apparent in
hydrodynamic models, as discussed in \rfs{sec:MISearly}. Recently this point was
amplified in Refs.~\cite{Kurkela:2019set,Heller:2020anv}, where these two
regimes were clearly distinguishable. As shown in \rff{fig:timescales}, at late
times generic solutions approach the attractor exponentially, with the rate set
by the nonhydrodynamic mode frequency at vanishing wave vector.  This
interpretation is consistent with linear response, and has been verified in many
examples.  However, the behaviour at very early times, at least in the case of
kinetic theory and hydrodynamic models, is not exponential, but follows a power
law which is independent of the transport coefficients. The compelling
explanation of this behaviour is that it is a consequence of the longitudinal
expansion dominating over any transverse dynamics.  

To assess the relevance of far-from-equilibrium hydrodynamic attractors to the
physics of QGP at early times one needs to understand primarily the effects of
conformal symmetry breaking and the onset of transverse dynamics. While a full
exploration of these important issues remains a task for the future, a few notable
results concerning these effects already exist in the
literature~\cite{Romatschke:2017acs,Chattopadhyay:2021ive,Jaiswal:2021uvv,Chen:2021wwh,Kamata:2022jrc,Jaiswal:2022udf,Chattopadhyay:2022sxk,Alalawi:2022pmg,Alqahtani:2022dfm} and some of them will be
reviewed in this Section. 


\begin{figure}
\begin{center}
\includegraphics[height = .4\textheight]{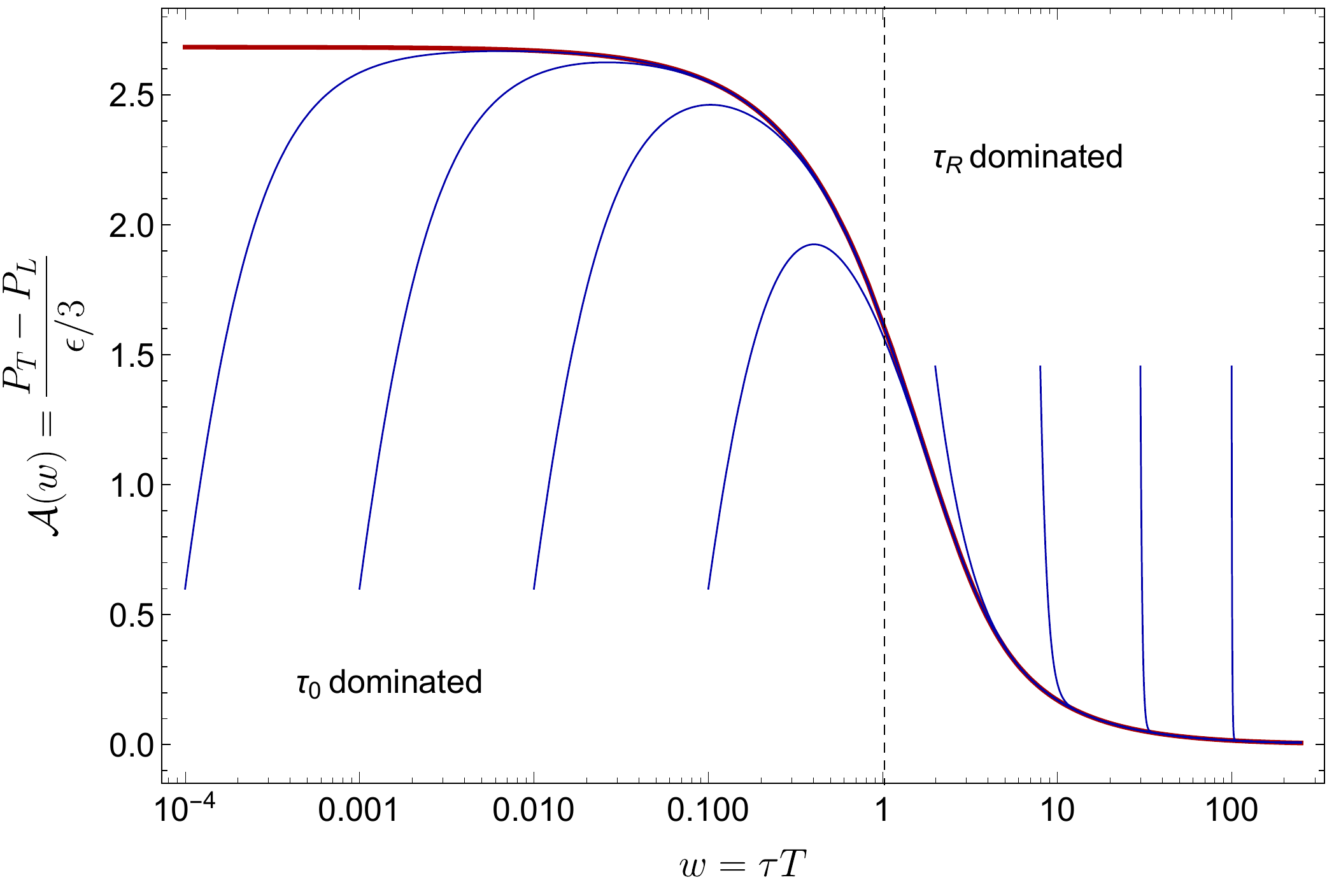} 
\caption{Log-linear plot showing different attraction mechanisms: expansion domination at early times, non-hydro mode decay at late times \cite{Kurkela:2019set,Blaizot:2017ucy}.}
\label{fig:timescales}
\end{center}
\end{figure}


\subsection{Breaking conformal symmetry}

The assumption of approximate conformal symmetry is valid in QCD at
sufficiently high energies. Systematic theoretical studies and detailed
comparison to available experimental data have provided constraints on the
parameters appearing in the hydrodynamic description of QGP evolution, notably
the bulk viscosity which is a sign of departures from conformality. The results
of a multiparameter Bayesian fit of Ref.~\cite{Nijs:2020ors,Nijs:2020roc} show
that the effects of bulk viscosity are subdominant relative to shear, making the
assumption of conformal invariance at early times look plausible. To quantify
this effect in the pre-hydrodynamic stage the easiest thing to do is to assume
that the free-streaming particles move with some effective velocity $v_{\rm
fs}<1$, where $v_{\rm fs}=1$ is conformal.  Experimental data suggest that
$v_{\rm fs}\approx0.82$~\cite{Nijs:2020ors,Nijs:2020roc}, indicating some
departure from conformality (see \rff{fig:BayesianTransp}). 
Other recent studies which provide evidence for the effects of conformal
symmetry breaking at the prehydrodynamic stage are reported in
Refs.~\cite{daSilva:2022xwu,NunesdaSilva:2020bfs}.  


\begin{figure}
\begin{center}
\includegraphics[height = .21\textheight]{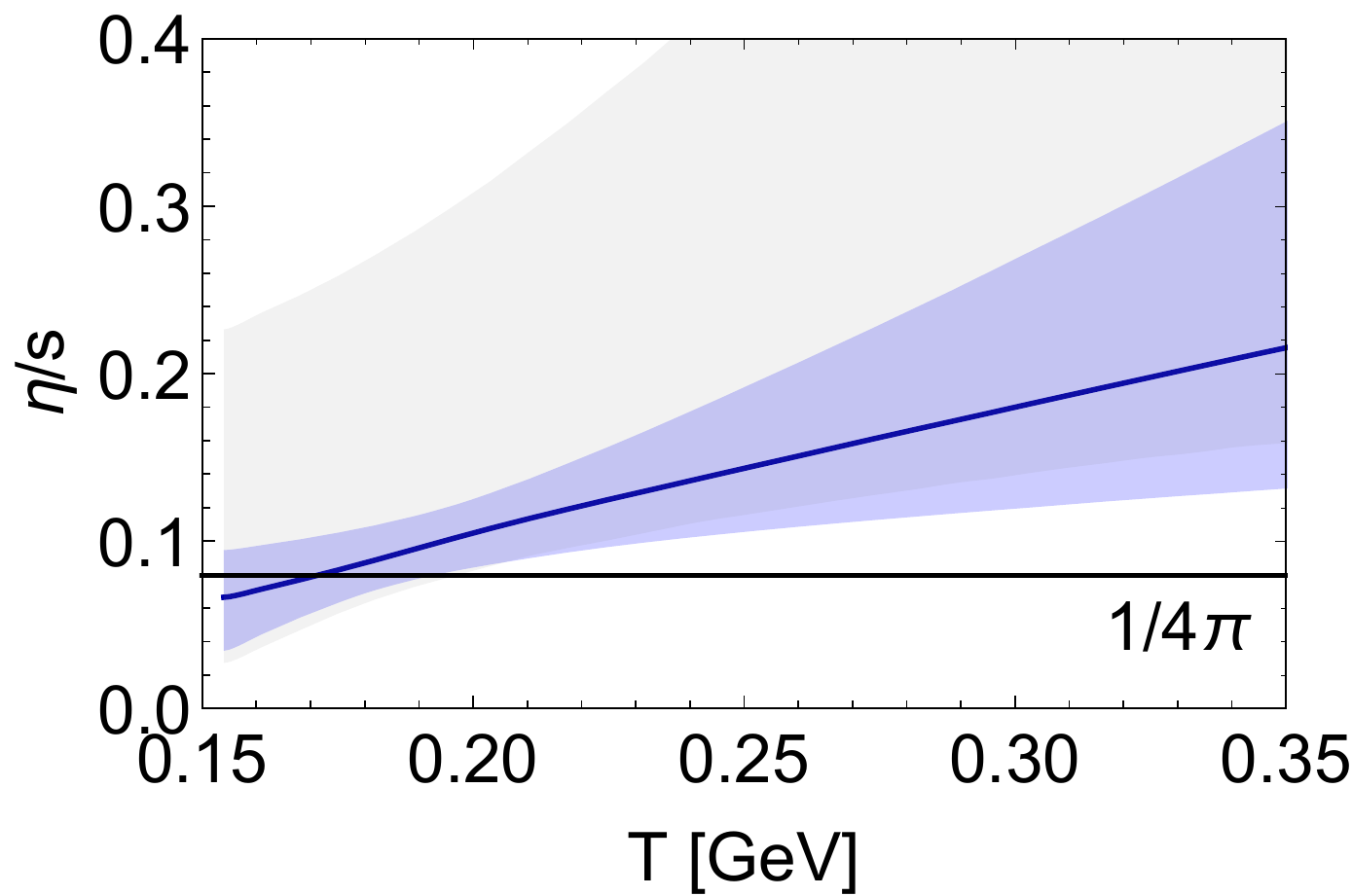} 
\includegraphics[height = .21\textheight]{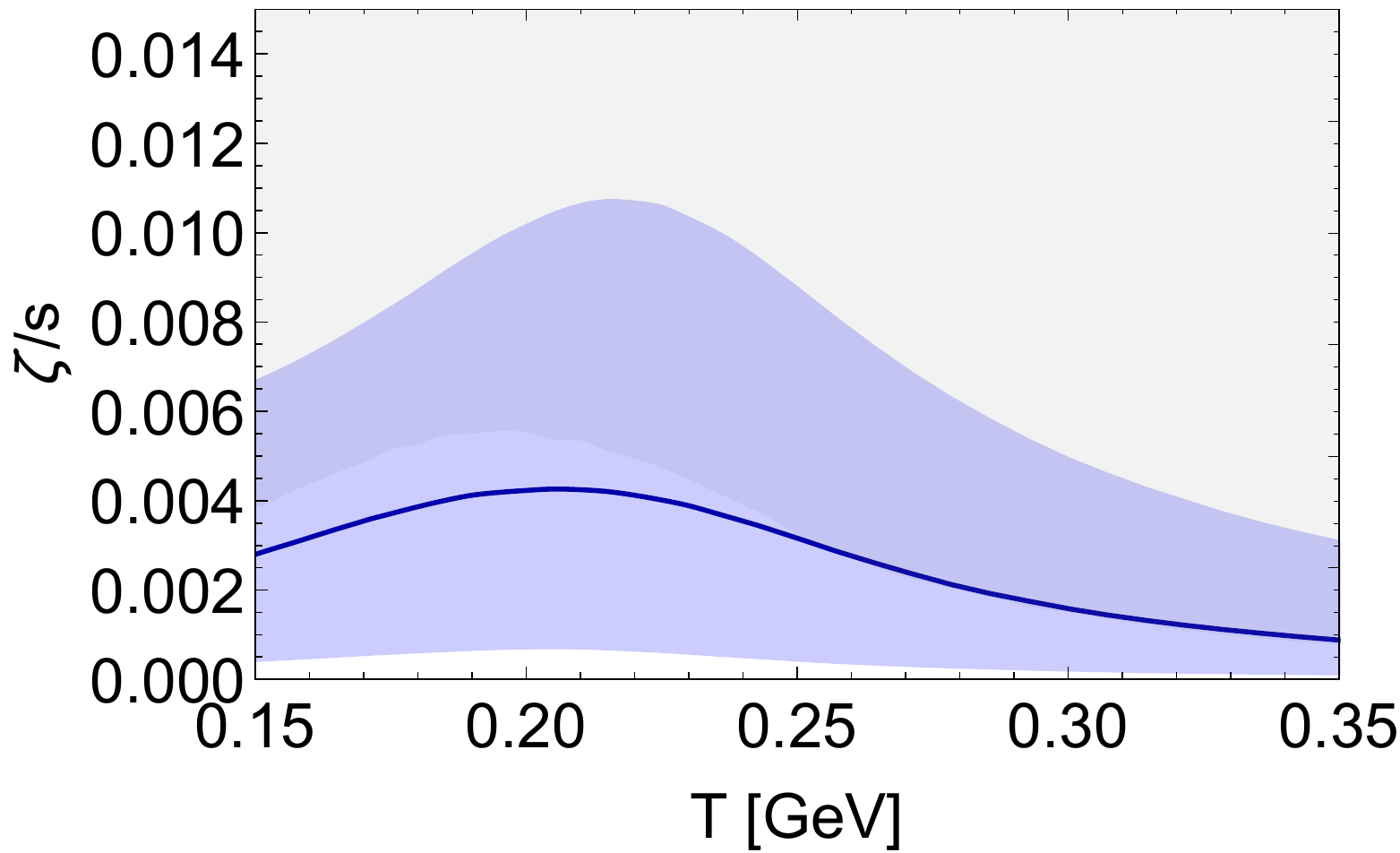}
\caption{Values of the shear (left) and the bulk (right) viscosity extracted from a Bayesian  analysis of a $20$ parameter model of a HIC from Ref.~\cite{Nijs:2020ors,Nijs:2020roc}.
The horizontal black line on the left panel is the holographic prediction~\cite{Kovtun:2004de}.
}
\label{fig:BayesianTransp}
\end{center}
\end{figure}


An important difference between conformal and non-conformal Bjorken flow is that
in the latter case the energy momentum tensor \rf{eq:Tmn} contains an additional
degree of freedom, which (as discussed in \rfs{sec:confsym}) is eliminated by
the tracelessness condition for conformal systems.  Attractors in Bjorken flow
without conformal symmetry were recently investigated in
Refs.~\cite{Chattopadhyay:2021ive,Jaiswal:2021uvv,Chen:2021wwh,Kamata:2022jrc,Alalawi:2022pmg,Chattopadhyay:2022sxk}.
Here we would like to mention one particular model of kinetic theory in the RTA,
where conformal symmetry is broken due to quasiparticles of nonvanishing mass
$m$~\cite{Jaiswal:2014isa,Denicol:2014vaa}; this model also assumes a constant
relaxation time $\tau_R$.
Refs.~\cite{Chattopadhyay:2021ive,Jaiswal:2021uvv,Chen:2021wwh} have identified
a free-streaming far-from-equilibrium attractor in this model, however the
hydrodynamic description employed there showed only a late-time attractor (in
the near-equilibrium,  hydrodynamic region).  Here we will review the analysis
of Ref.~\cite{Jaiswal:2022udf} which also studies this model, generalising the
very suggestive approach to conformal Bjorken flow in RTA kinetic theory
developed in
Refs.~\cite{Blaizot:2017ucy,Blaizot:2018rft,Blaizot:2019scw,Blaizot:2021cdv}.
This analysis leads to a different hydrodynamic description of the system, which
was found to reproduce the free-streaming attractor seen at the level of kinetic
theory.  

The basic idea of
Refs.~\cite{Blaizot:2017ucy,Blaizot:2018rft,Blaizot:2019scw,Blaizot:2021cdv} is
to convert the Boltzmann kinetic equation to an infinite hierarchy of coupled
ordinary differential equations describing a set of moments of the distribution
function \footnote{In this section we use the notation $\int_{\bf
    p}\equiv\int\frac{d^3p}{(2\pi)^3}$.
}
\be
\LL_n &\equiv&  \int_{\bf p} p_0 \ P_{2n}(\cos \psi) \ f(\tau, p)~, 
\qquad \forall n \geq 0
\ee
where $\cos\psi=p_z/p_0=v_z$ is constituent's velocity along the $z-$direction,
and $P_{2n}$ are Legendre polynomials of degree $2n$. 
These moments satisfy the following infinite hierarchy of equations
which can be derived from the RTA Boltzmann equation \rf{eq:Boltzmann_RTA} 
\begin{subequations}
\label{eq:byh}
\begin{align}
\pd{\LL_0}{\tau} =& -\frac{1}{\tau} \left( a_0 \LL_0 + c_0 \LL_1 \right) \,,\\
\pd{\LL_n}{\tau} =& -\frac{1}{\tau} \left( a_n \LL_n + b_n \LL_{n-1} + c_n \LL_{n+1} \right) 
	- \frac{\left( \LL_n - \LL_n^{\rm eq} \right)}{\taur}  \, ,
\quad \forall n \geq 1 \qquad
\end{align}
\label{eq:L}
\end{subequations}
where the coefficients $a_n,b_n,c_n$ are known explicitly~\cite{Blaizot:2017ucy,Jaiswal:2022udf}
\begin{equation}
    a_n=\frac{2(14n^2+7n-2)}{(4n-1)(4n+3)}, \quad
    b_n=\frac{2n(2n-1)(3n+3)}{(4n-1)(4n+1)}, \quad
    c_n=\frac{(1-2n)(2n+1)(2n+2)}{(4n+1)(4n+3)}~,
\end{equation}
and are determined entirely by the free-streaming part of the kinetic equation. The
moments $\mathcal{L}_n^{\rm eq}$ are computed with the Boltzmann equilibrium
distribution function $f_{\rm eq}(p_0/T)=\exp(-p_0/T)$, so 
for example $\mathcal{L}_1^{\rm eq}=\frac{1}{2}(\edens-3P)$. 

While in the conformal case the moments $\LL_n$ are sufficient to provide a
representation of the dynamics, in the nonconformal case the
energy-momentum tensor is no longer traceless and this requires
introducing a second set of moments.  Indeed, from the general form of the
energy-momentum tensor \rf{eq:Tmn} for Bjorken flow it follows that 
\be
\edens = \LL_0 ,  
\qquad \pL = \frac{1}{3} \left( \LL_0 + 2 \LL_1 \right) , 
\qquad \pT = \frac{1}{3} \left( \LL_0 -\LL_1 -\frac{3}{2} T_\mu^\mu  \right)~.
\ee
This motivates the introduction of another set of moments
\cite{Jaiswal:2022udf} in the following way:
\be
\MM_n &\equiv& m^2 \int_{\bf p} \frac{1}{p_0} P_{2n}(\cos \psi) \ f(\tau, p), 
	 \qquad \forall n \geq 0~.
\ee
For example, the moment $\MM_0$ is equal to the trace of the energy-momentum tensor:
\be
\MM_0 = m^2\int_{\bf p} \frac{1}{p_0}f(\tau, p)  = T^\mu_\mu = \edens - \pL-2\pT~,
\ee
and controls deviations from conformality.  The RTA Boltzmann equation
\rf{eq:Boltzmann_RTA} can now be rewritten as an infinite hierarchy of equations
with \rf{eq:byh} supplemented by 
\begin{equation}
    \frac{\partial\mathcal{M}_n}{\partial\tau}=-\frac{1}{\tau}\left(a_n'\mathcal{M}_n+b_n'\mathcal{M}_{n-1}+c_n'\mathcal{M}_{n+1}\right)
    -\frac{\mathcal{M}_n-\mathcal{M}^{\rm eq}_n}{\tau_R}~,
    \label{eq:M}
\end{equation}
where $\mathcal{M}^{\rm eq}_n$ are the equilibrium moments, and the coefficients read
\begin{equation}
    a_n'=\frac{2(6n^2+3n-1)}{(4n-1)(4n+3)},\quad
    b_n'=\frac{4n^2(2n-1)}{(4n-1)(4n+1)},\quad
    c_n'=-\frac{(2n+1)^2(2n+2)}{(4n+1)(4n+3)}~.
\end{equation}
Note that the $\MM_n$ moments are coupled to the $\mathcal{L}_n$ by the
presence of the $\mathcal{M}^{\rm eq}_n$ terms, and decouple in the
collisionless limit of $\tau_R\rightarrow\infty$. The equations \rf{eq:L}
constitute a closed system which has to be solved first, determining the moments
$\LL_n$.

Using this infinite hierarchy of evolution equations one can identify an
early-time far-from-equilibrium attractor which describes free-streaming. This
is in line with results found in this model in
Refs.~\cite{Chattopadhyay:2021ive,Jaiswal:2021uvv,Chen:2021wwh}.  Out of the
three independent components of the energy-momentum tensor (see \rf{eq:AdefNC}),
one has an attractor while the remaining two carry information about the initial
state to the asymptotic late-time region.  This can be contrasted with conformal
systems, where the attractor appears in the pressure anisotropy, and information
about the initial conditions is carried to asymptotically late times by the
energy density alone. 

In \rfc{Jaiswal:2022udf} the quantity which has an attractor is expressed as 
\begin{equation}
    g_0\equiv
    \frac{\tau}{\LL_0}\frac{\partial\LL_0}{\partial\tau} = 
    -1 - \f{\pL}{\edens}.
    \label{eq:g0def}
\end{equation}
In the conformal case this is trivially related to the pressure anisotropy, but
in the absence of conformal symmetry it differs from it due to
bulk pressure. For this reason the function $g_0$ does not satisfy a single ODE, as it
would in a conformal model; instead, 
the pair of functions $g_0, \edens$ satisfy a coupled set of ODEs which
determine their dynamics.

The attractor solution tends to $g_0=-1$ at early
times, which corresponds to free streaming ($\pL=0$). Its late-time behaviour is
much harder to analyse than in the conformal case, since the equations of state
are more involved. In particular, the velocity of sound tends to zero in this
limit, which leads to rather nontrivial
asymptotics~\cite{Kamata:2022jrc,Kamata:2022ola}. The hydrodynamic region is
approached much more slowly and in way which depends of the mass.  This
behaviour is illustrated in \rff{fig:g0}. 

Truncations of the hierarchy of evolution equations provide workable
approximations which capture essential features of the full dynamics.  The most
straightforward truncation, which accounts for all three independent components
of the energy-momentum tensor, consists of keeping $\LL_0$, $\LL_1$ and $\MM_0$.
The higher modes $\LL_2$, and $\MM_1$ coupled with the three lowest ones have to
be modelled in some way, out of which the most simple, but not unreasonable, is
to set $\LL_2=\MM_1=0$.  This system of equations is hydrodynamic in spirit in
the sense that it is a closed set of three ODEs. However, it is different from
the hydrodynamic model used in
Refs.~\cite{Chattopadhyay:2021ive,Jaiswal:2021uvv,Chen:2021wwh}. The main virtue
of this description is that it captures the early-time attractor identified in
the underlying kinetic theory, as seen in \rff{fig:g0}.  It should however be
noted that it is obtained by truncating the boost-invariant moment hierarchy
\rf{eq:byh}, \rfn{eq:M} and it is not entirely clear how it could be obtained
from a set of covariant hydrodynamic equations such as variants of MIS
theory~\cite{Denicol:2012cn,Baier:2007ix}.


\begin{figure}
\begin{center}
\includegraphics[height = .36\textheight]{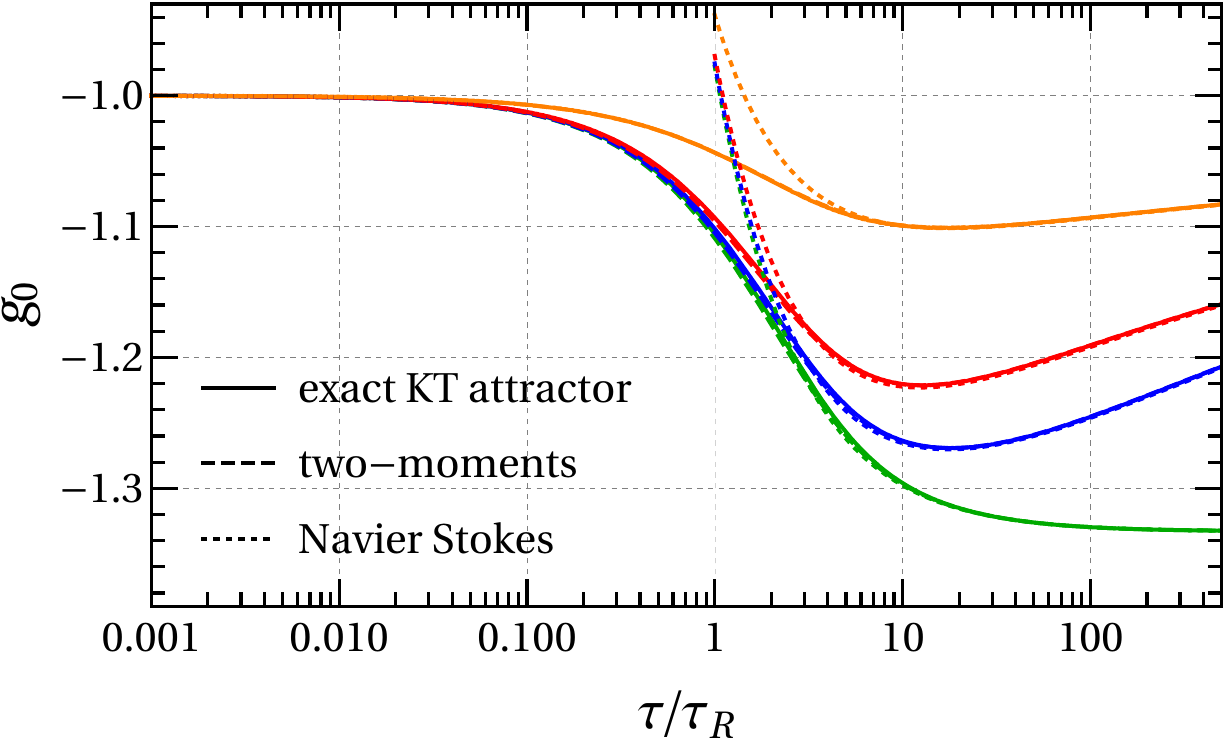} 
\caption{The attractor in the nonconformal kinetic model of
    Ref.~\cite{Jaiswal:2022udf}, together attractors in effective hydrodynamic
    models (recall that $g_0$ is defined in \rf{eq:g0def}). 
The green, blue, red, and orange curves represent $m/T(\tau_R)=0.001,0.5,1$, and $5$. 
``Two-moments'' refers to 
the truncation with $\LL_2=0$. This truncation clearly captures the early-time attraction, in contrast to the Navier-Stokes, which refers to the truncated gradient expansion.  
This plot is taken from Ref.~\cite{Jaiswal:2022udf}.
}
\label{fig:g0}
\end{center}
\end{figure}


\subsection{Incorporating transverse dynamics}

Since the longitudinal expansion is believed to be dominant at early times, the attractors seen
in models of Bjorken flow may retain their relevance even in the presence of
transverse dynamics. The persistence of early-time attractors in such
circumstances was recently studied in
Refs.~\cite{Kurkela:2019set,Kurkela:2020wwb,Ambrus:2021sjg} (see also Ref.~\cite{Borrell:2021cmh}). The first of these papers
considered transverse flow in the case of kinetic theory in the RTA.  The
authors have solved the boost-invariant Boltzmann equation numerically for a
choice of realistic initial transverse profiles. It was found that as long as
the transverse gradients remain negligible compared to the longitudinal ones at
initialisation time, arbitrary initial conditions in 3+1D evolve towards the
1+1D attractor. The late-time evolution does however depend on the transverse
profile of energy and transverse momentum, reflecting the fact that the space of
solutions of boost-invariant perfect fluid hydrodynamics has a higher
dimensionality than what is seen in the case of Bjorken flow.  These results
suggests a degree of robustness of early time attractors in the presence of
transverse dynamics.

Other work on attractors with transverse dynamics includes studies of Gubser
flow~\cite{Denicol:2018pak,Behtash:2017wqg,Behtash:2019qtk,Dash:2020zqx} (this
activity was recently reviewed by Soloviev~\cite{Soloviev:2021lhs}).
The applicability of hydrodynamics, including the effects of
the build-up of transverse dynamics at early times was also the subject of
recent 
Refs.~\cite{Ambrus:2022koq,Ambrus:2022qya,Ambrus:2023oyk}.

%% file: outlook.tex
\section{Outlook}
\label{sec:Out}

In this review we have attempted to present the main ideas behind the hypothesis
that the applicability of fluid dynamics to early phases of QGP dynamics can be
explained by a far-from-equilibrium hydrodynamic attractor. We have emphasised
the role of the kinematic setting specific to heavy-ion collisions, which makes
it plausible that such an attractor occurs also in QCD.  At the conceptual level
this picture is rather compelling. However, Bjorken flow is a very restrictive
setting, and despite some existing applications using the attractor in practical
calculations of phenomenologically interesting observables requires developing
effective methods of working with models with a large number of degrees of
freedom. Progress is likely to come gradually, by learning how to deal with
models of increasing complexity, extending the studies reviewed in Sections
\ref{sec:PhaseSpace} and \ref{sec:beyond}. 

The idea of hydrodynamic attractors has been closely connected with the
divergence of the hydrodynamic gradient expansion. In this review we have not
discussed this connection beyond its utilitarian aspects, but on a conceptual
level there have been some important developments in recent times. This includes
a proof of the generic divergence of the gradient expansion at the linearised
level without any symmetry assumptions, and its connection with the properties
of dispersion relations~\cite{Withers:2018srf,Heller:2020uuy}.  At the nonlinear
level, some of the results for Bjorken flow have been generalised to a much
wider class of flows, called longitudinal flows. In particular, the gradient
expansion has been shown to diverge for this class of
solutions~\cite{Heller:2021oxl}. It was also found that the large order
behaviour of the gradient series can be expressed in terms of new degrees of
freedom, the singulant fields, which track transient
effects~\cite{Heller:2021yjh}. The relevance of these advances to the study of
attractors remains an interesting challenge for the future. 

A number of issues were not addressed in this review. One is the presence of
other degrees of freedom, such as those  connected with chiral symmetry
breaking, and their possible effect on the attractor dynamics of
QGP~\cite{Mitra:2020mei,Mitra:2020hbj} (see also Ref.~\cite{Mitra:2022uhv}).
Another such issue is the role of fluctuations, which has been mostly neglected
in the attractor literature, with the notable exception of
Refs.~\cite{Akamatsu:2016llw,Chen:2022ryi}. 

It would also be very interesting to understand the role of quantum effects in
the early-time dynamics. Of course the hydrodynamic picture implicitly contains
them, but in a rather opaque way. On the other hand, the kinetic theory
description arises from quantum field theory through a series of
approximations~\cite{Mueller:2002gd,Jeon:2004dh} and it should be possible to
study the origin and robustness of the kinetic theory attractor in a framework
which allows for a systematic incorporation of quantum corrections. This is
connected with other approaches to early-time dynamics, including those
involving ideas such as the Color Glass Condensate or non-thermal
attractors~\cite{Berges:2008wm,Mazeliauskas:2018yef,Brewer:2019oha,Brewer:2022ifw,Brewer:2022vkq,Berges:2020fwq}.
It is not yet fully understood how they are related to the ideas reviewed here,
and clarifying this appears to be a very promising avenue for future research. 

Finally, there is the more general question about far-from-equilibrium
attractors in nonequilibrium systems.  In the context of heavy-ion collisions
the specific kinematic circumstances have lead us to consider boost-invariant
expansion where far-from-equilibrium hydrodynamic behaviour was first noted, but
there could be other situations where analogous phenomena might
appear~\cite{Baggioli:2021tzr}, perhaps even in the non-relativistic
domain~\cite{Le:2022ntg}.  Another context where far from equilibrium
hydrodynamic attractors can occur is the dynamics of systems in nontrivial
spacetime backgrounds (see e.g. Ref.~\cite{Vyas:2022hkm}).